\documentclass[prb, twocolumn, superscriptaddress, footinbib, amsmath, amssymb, amsthm]{revtex4-2}

\usepackage[utf8]{inputenc}
\usepackage[T1]{fontenc}
\usepackage[main=english]{babel}

\usepackage{multirow}
\usepackage{graphicx}
\usepackage[version=4]{mhchem}

\usepackage[scr=boondoxo, scrscaled=1.03]{mathalpha}

\usepackage{bm}
\usepackage{upgreek}

\usepackage{mleftright}
\usepackage{hyperref}
\hypersetup{colorlinks=true, allcolors=blue, breaklinks=true, pdfstartview=}

\makeatletter
\newcommand*{\defeq}{\mathrel{\rlap{%
\raisebox{0.3ex}{$\m@th\cdot$}}%
\raisebox{-0.3ex}{$\m@th\cdot$}}=}
\makeatother

\newcommand{\one}{\text{\usefont{U}{bbold}{m}{n}1}}
\MakeRobust{\one}

\newcommand*{\iu}{\mathrm{i}}
\newcommand*{\Elr}{\mathrm{e}}
\newcommand*{\Pauli}{\upsigma}
\newcommand*{\GellMann}{\Uplambda}
\newcommand*{\LCs}{\upepsilon}
\newcommand*{\Kd}{\updelta}
\DeclareMathOperator{\Dd}{\updelta}

\newcommand*{\abs}[1]{\mleft\lvert {#1} \mright\rvert}
\newcommand*{\ev}[1]{\mleft\langle {#1} \mright\rangle}

\newcommand*{\dd}[2][]{\mathop{}\!\mathrm{d}^{#1} {#2}}
\newcommand*{\var}[2][]{\mathop{}\!\delta_{#1} {#2}}
\newcommand*{\fdv}[2]{\frac{\delta #1}{\delta #2}}
\newcommand*{\DD}[1]{\mleft[\dd{#1}\mright]}

\newcommand*{\vdot}{\bm{\cdot}}
\newcommand*{\vcross}{\bm{\times}}
\newcommand*{\vb}[1]{\bm{#1}}
\newcommand*{\vu}[1]{\bm{\hat{#1}}}

\let\Re\relax
\DeclareMathOperator{\Re}{Re}
\let\Im\relax
\DeclareMathOperator{\Im}{Im}
\newcommand*{\Hc}{\mathrm{H.c.}}
\DeclareMathOperator{\tr}{tr}
\DeclareMathOperator{\Tr}{Tr}
\DeclareMathOperator{\diag}{diag}

\newcommand*{\Z}{\mathbb{Z}}
\let\C\relax
\newcommand*{\C}{\mathbb{C}}

\DeclareMathOperator{\Ugp}{U}

\DeclareMathOperator{\Ogp}{O}
\DeclareMathOperator{\SO}{SO}

\newcommand*{\actS}{\mathcal{S}}
\newcommand*{\partZ}{\mathcal{Z}}
\newcommand*{\Haml}{\mathcal{H}}

\usepackage{xcolor}

\begin{document}
\title{Unbiased large-$N$ approach to competing vestigial orders of density-wave and superconducting instabilities}
\author{Grgur Palle}
\email{gpalle@illinois.edu}
\affiliation{Department of Physics, The Grainger College of Engineering, University of Illinois Urbana-Champaign, Urbana, Illinois 61801, USA}
\affiliation{Anthony J.\ Leggett Institute for Condensed Matter Theory, The Grainger College of Engineering, University of Illinois Urbana-Champaign, Urbana, Illinois 61801, USA}
\author{Rafael M.\ Fernandes}
\affiliation{Department of Physics, The Grainger College of Engineering, University of Illinois Urbana-Champaign, Urbana, Illinois 61801, USA}
\affiliation{Anthony J.\ Leggett Institute for Condensed Matter Theory, The Grainger College of Engineering, University of Illinois Urbana-Champaign, Urbana, Illinois 61801, USA}
\date{\today}
\begin{abstract}
When a primary order breaks multiple symmetries, partially ordered phases that only break a subset of those symmetries, known as vestigial phases, may onset at a higher temperature.
This concept has been applied to a wide range of systems, including iron pnictides, cuprates, van der Waals antiferromagnets, doped topological insulators, and twisted bilayer graphene.
In general, a multicomponent primary order parameter (OP) supports multiple vestigial channels, each described by a quadratic (or higher-order) composite OP.
However, the standard large-$N$ approach to the Ginzburg-Landau action of the primary OP has an intrinsic ambiguity in how one decouples the composite OPs, leading to situations in which one can seemingly enhance or eliminate altogether any vestigial instability.
Here, we show that this ambiguity is a direct consequence of redundancy relations, such as Fierz identities, that relate different composite OPs, reflecting the fact that different vestigial channels interfere with each other and thus cannot be treated separately.
To resolve this ambiguity, we propose an unbiased large-$N$ approach that respects both the redundancy relations and the underlying symmetry-group structure, and that gives unique values for the effective interactions of all vestigial channels.
Our analysis reveals the generic existence of regions in the parameter space of quartic Landau coefficients where no vestigial order is stable, in contrast to the standard large-$N$ approach, but in agreement with weak-coupling and variational approaches.
We illustrate our results by analyzing the vestigial orders of charge-density waves, spin-density waves, and multicomponent superconductors in tetragonal, hexagonal, and cubic systems, respectively, revealing the presence of exotic vestigial phases describing spin-quadrupolar, charge-$4e$ superconducting, and altermagnetic orders.
\end{abstract}

\maketitle

\section{Introduction}
A ubiquitous feature of the phase diagrams of many complex systems, such as cuprates~\cite{Fradkin2015, Keimer2015, Proust2019}, iron pnictides~\cite{Fernandes2014, Fernandes2022}, heavy-fermion compounds~\cite{Stewart1984, Lohneysen2007, Pfleiderer2009}, and twisted moiré systems~\cite{Carr2020, Andrei2021}, is the appearance of multiple phases with distinct broken symmetries in the vicinity of each other.
Although there is much debate regarding the interplay of these distinct phases, an appealing perspective proposes that these phases are \emph{intertwined}~\cite{Fradkin2015}.
Unlike competing phases emerging from separate degrees of freedom, intertwined orders can cooperate and mutually affect each other in nontrivial ways, as they emerge from the same primary degrees of freedom.
It is an important question to better theoretically understand how ordered states intertwine, starting from the simplest of examples.

In this regard, one of the simplest, and perhaps most paradigmatic, examples of intertwined orders are vestigial orders~\cite{Nie2014, Fradkin2015, Fernandes2019}.
Given a certain primary or parent order which spontaneously breaks multiple symmetries of the non-ordered system, vestigial phases are orders that spontaneously break only a subset of the symmetries that the parent order breaks.
Following Ref.~\cite{Fernandes2019}, if $\vb{\eta} = (\eta_1, \eta_2, \ldots)$ is the multicomponent order parameter (OP) of the parent order, then vestigial orders correspond to the condensation of composite OPs of the form $\Psi_{\mu} = \vb{\eta}^{\intercal} \Gamma_{\mu} \vb{\eta}$, with some matrix $\Gamma_{\mu}$, that transform nontrivially under the symmetries of the non-ordered system, while the primary order parameter remains uncondensed, i.e., $\ev{\vb{\eta}} = \vb{0}$.
The existence of a multicomponent OP is key here because only then can the structure of the fluctuations around $\ev{\vb{\eta}} = \vb{0}$ cause symmetry-breaking, resulting in a vestigial phase.

\begin{figure*}[t]
\includegraphics[width=\textwidth]{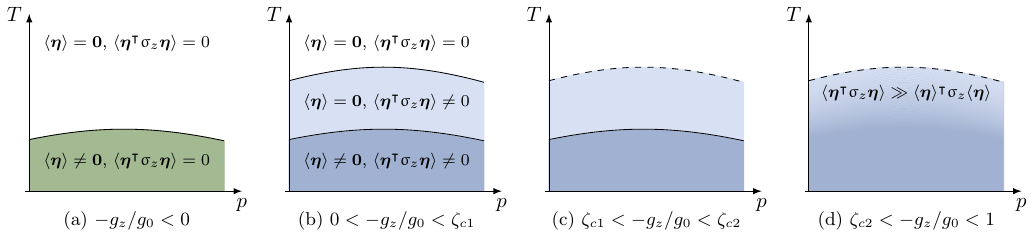}
\caption{The temperature $T$ vs.\ external tuning parameter $p$ phase diagram of the model~\eqref{eq:intro-action} obtained in Ref.~\cite{Fernandes2012} for dimension $2 < d < 3$ and different values of the ratio $g_z / g_0$.
Here $\zeta_{c1} = (6 - 2d) / (6 - d)$ and $\zeta_{c2} = 3 - d$.
In the primary phase (dark green or blue) $\ev{\vb{\eta}}$ is finite with a finite $\ev{\vb{\eta}^{\intercal} \Pauli_z \vb{\eta}}$ when $g_z < 0$ (panels (b),(c),(d)), while in the vestigial phase (light blue) only $\ev{\vb{\eta}^{\intercal} \Pauli_z \vb{\eta}}$ is finite.
In the disordered phase (white) both $\ev{\vb{\eta}}$ and $\ev{\vb{\eta}^{\intercal} \Pauli_z \vb{\eta}}$ vanish.
The blue shading under (d) emphasizes the fact that near the transition $\ev{\vb{\eta}^{\intercal} \Pauli_z \vb{\eta}}$ is much larger than $\ev{\vb{\eta}}^{\intercal} \Pauli_z \ev{\vb{\eta}}$, with an enhanced transition temperature compared to (a).
Solid (dashed) lines indicate second-order (first-order) transitions.}
\label{fig:vestigial-transitions}
\end{figure*}

The idea of vestigial order~\cite{Fradkin2015, Fernandes2019} builds on several seminal studies, most notably of partially ordered phases in liquid crystals~\cite{Stephen1974, Chaikin1995, Kivelson1998} and order-by-disorder phenomena in frustrated magnets~\cite{Villain1977, Fradkin1977, Shender1982, Henley1989, Chandra1990, Korshunov2006}, and one of the earliest works to have discussed the phenomenon that is now identified as vestigial nematic order was by Golubović and Kostić~\cite{Golubovic1988}.
Theoretically, this concept has been applied to a wide variety of electronic parent orders, ranging from charge-density waves~\cite{Nie2014, YuxuanWang2014, Venderbos2016hex1, Nie2017, Fernandes2019} and spin-density waves~\cite{Golubovic1988, Fang2006, Fang2008, Xu2008, Qi2009, Millis2010, Fernandes2012, Chern2012, Venderbos2016hex2, Fernandes2016, Nie2017, Zhang2017, Christensen2018, Fernandes2019} to multicomponent superconductors~\cite{Fischer2016, Hecker2018, Fernandes2019, Jian2021, Fernandes2021, Garaud2022, Hecker2023, How2023, How2024, Poduval2024, Lin2025, Verghis2025, Maccari2025, Zou2025, Gao2026} and pair-density waves~\cite{Himeda2002, Berg2007, Agterberg2008, Berg2009-PRB, Berg2009-Nat, Berg2009, Loder2011, Wang2015-PRL, Fradkin2015, Wu2024, Pan2024, Huecker2026}.
Although the most commonly studied electronic vestigial phases have been nematic ones~\cite{Golubovic1988, Kivelson1998, Fang2006, Fang2008, Xu2008, Qi2009, Millis2010, Fernandes2012, Nie2014, Fernandes2016, Nie2017, Zhang2017, Christensen2018, Borisov2019, OBrien2026}, the range of possible symmetry-breaking patterns that can arise through vestigial ordering is considerably wealthier~\cite{Fradkin2015, Fernandes2019, Fernandes2016, Fischer2016, Hecker2023}, covering charge-density waves, ferromagnets, antiferromagnets, altermagnets, spin loop-currents, and charge-$4e$ superconductors.
There is experimental evidence for vestigial order in cuprates~\cite{Berg2009, Fradkin2015, Mukhopadhyay2019}, iron pnictides~\cite{Fernandes2012, Fernandes2014, Borisov2019, Grinenko2021}, doped \ce{Bi2Se3}~\cite{Yonezawa2019, Cho2020}, heavy-fermion compounds~\cite{Seo2020}, twisted bilayer graphene~\cite{Fernandes2021, Cao2021}, and van der Waals antiferromagnets~\cite{Ni2023, Hwangbo2024, Sun2024}, among other systems~\cite{Zhang2017, Ge2024}.
Moreover, this idea has found applications in diverse settings~\cite{Kivelson1998, Loison2000, Weber2003, Kamiya2011, Bojesen2014, Jeevanesan2015, Roy2015, ZhentaoWang2017, Willa2020, Takahashi2020, Liu2023, Volovik2024, Liu2024, Francini2024, OBrien2026}, ranging from cold atoms~\cite{Gopalakrishnan2017, Yu2025} to neutron stars~\cite{Herland2010}.

Because vestigial order is driven by fluctuations, its theoretical description requires methods that go beyond Ginzburg-Landau theory and that incorporate the fluctuations of the primary order parameter.
The most common one has been the large-$N$ approach~\cite{Golubovic1988, Fang2006, Fang2008, Fernandes2012, Chern2012, Nie2014, YuxuanWang2014, Fernandes2016, Zhang2017, Hecker2018, Willa2020, Jian2021, Hecker2023, Wu2024, Poduval2024, Verghis2025}, but Gaussian variational~\cite{Fischer2016, Nie2017, Hecker2023}, self-consistent Hartree-Fock~\cite{How2023, How2024}, renormalization group~\cite{Qi2009, Millis2010, Fernandes2012}, and Monte Carlo~\cite{Loison2000, Weber2003, Kamiya2011, Jeevanesan2015, Zou2025, Gao2026} methods have also been used in this context.
To illustrate the mechanisms responsible for vestigial order, below we summarize the results of the traditional large-$N$ method, as discussed for instance in Refs.~\cite{Fang2008, Nie2014, Fernandes2012, Fernandes2019, Hecker2023}.

Consider the simplest case of a two-component real-valued OP $\vb{\eta} = (\eta_1, \eta_2)$ whose Euclidean action has the form:
\begin{align}
\actS = \frac{1}{2} \int_q \chi_q^{-1} \vb{\eta}^{\intercal} \vb{\eta} + \frac{1}{8} \int_x \mleft[g_0 (\vb{\eta}^{\intercal} \vb{\eta})^2 + g_z (\vb{\eta}^{\intercal} \Pauli_z \vb{\eta})^2\mright]. \label{eq:intro-action}
\end{align}
Here $q$ is momentum, $x$ is space, $\Pauli_z$ is the $z$ Pauli matrix, $\chi_q$ is the susceptibility (propagator) of the primary OP, and $g_0 > 0$, $g_z > - g_0$ to ensure stability (that $\actS$ is bounded from below as $\abs{\vb{\eta}} \to \infty$).
This action describes, for instance, a degenerate Ising-like stripe spin-density wave (SDW) on the tetragonal lattice with wavevectors $(\pi,0) \sim \eta_1$ and $(0,\pi) \sim \eta_2$~\cite{Christensen2018}.
The vestigial order corresponds to the condensation of the bilinear $\vb{\eta}^{\intercal} \Pauli_z \vb{\eta}$, which in the case of the stripe SDW example implies electronic nematic order.

After introducing a Hubbard-Stratonovich field for $\Psi_z = \vb{\eta}^{\intercal} \Pauli_z \vb{\eta}$, integrating out $\vb{\eta}$, and performing a saddle-point approximation (SPA) for $\Psi_z$ (which becomes exact as $N \to \infty$~\cite{Moshe2003}), the following condition for the condensation of $\Psi_z$ is obtained:
\begin{align}
\frac{1}{g_z} &= - \int_q \chi_{q}^2. \label{eq:intro-vestigial-cond}
\end{align}
Assuming $g_z < 0$ and taking a generic form for the susceptibility $\chi_q^{-1} = \xi^{-2} + q^2$ with $\xi \propto \abs{T - T_c}^{-\nu}$, where $T_c$ is the transition temperature of the primary transition, below $d = 4$ dimensions we find a vestigial transition temperature
\begin{align}
T_v - T_c &\propto \mleft(\frac{- g_z}{4-d}\mright)^{\frac{1}{(4-d)\nu}} \label{eq:intro-vestigial-Tv}
\end{align}
that, evidently, appears for arbitrarily small, negative $g_z$; in $d = 4$, the term on the right is replaced by $\Elr^{1 / (g_z \nu)}$.
Thus, we can interpret $g_z$ as the effective interaction of the nematic vestigial channel.
That $g_z < 0$ corresponds to attraction in the nematic channel is consistent with the fact that $g_z < 0$ favors the $\vb{\eta} = (1,0)/(0,1)$ primary OP configuration.
The existence of $T_v > T_c$ is a necessary, but not sufficient, condition for the stabilization of a distinct vestigial phase.
In particular, if the vestigial transition at $T_v$ is first-order, it can trigger a simultaneous transition of the primary order, leaving no regime for which $\ev{\Psi_z} \neq 0$ but $\ev{\vb{\eta}} = \vb{0}$.
Further analysis that includes the mass renormalization due to the $g_0 (\vb{\eta}^{\intercal} \vb{\eta})^2$ term reveals~\cite{Fernandes2012} that the system undergoes a joint first-order transition for $\abs{g_z} / g_0 > 3 - d \equiv \zeta_{c2}$, while a transition to a distinct vestigial phase takes place only for $\abs{g_z} / g_0 < \zeta_{c2}$, i.e., for 2D or quasi-2D systems.
This is shown in Fig.~\ref{fig:vestigial-transitions}.

As recently observed~\cite{Hecker2023}, difficulties appear in the above analysis if we attempt to consistently allow for all possible competing vestigial channels.
In the model of Eq.~\eqref{eq:intro-action}, this means considering a more general quartic interaction
\begin{align}
g_0 (\vb{\eta}^{\intercal} \vb{\eta})^2 + g_x (\vb{\eta}^{\intercal} \Pauli_x \vb{\eta})^2 + g_z (\vb{\eta}^{\intercal} \Pauli_z \vb{\eta})^2
\end{align}
and introducing an additional Hubbard-Stratonovich field for $\Psi_x = \vb{\eta}^{\intercal} \Pauli_x \vb{\eta}$.
Using the example of the stripe SDW primary order, this vestigial order parameter corresponds to a charge-density wave with wavevector $(\pi,\pi)$, which is nonzero when the primary order condenses in the configuration $\vb{\eta} = (1, \pm 1)$ instead of $\vb{\eta} = (1,0)/(0,1)$.
The main difficulty here is that the different channels feed into each other, as specified by the Fierz identity
\begin{align}
- (\vb{\eta}^{\intercal} \vb{\eta})^2 + (\vb{\eta}^{\intercal} \Pauli_x \vb{\eta})^2 + (\vb{\eta}^{\intercal} \Pauli_z \vb{\eta})^2 = 0. \label{eq:intro-Fierz}
\end{align}
Thus, changing the quartic Landau coefficients $(g_0, g_x, g_z)$ to $(g_0 - \var{g}, g_x + \var{g}, g_z + \var{g})$ does not change the action in Eq.~\eqref{eq:intro-action}.
Yet, after this change of $g_{\mu}$, the saddle-point equation~\eqref{eq:intro-vestigial-cond} can acquire dramatically different solutions.
For instance, by choosing a large enough $\var{g}$,  we can make $g_x$ or $g_z$ positive, thus apparently eliminating altogether the possibility of vestigial order in the respective channels.
Conversely, by rendering $g_x$ and $g_z$ more negative, we can seemingly enhance the vestigial transition temperature $T_v$.
The concurrent change in $g_0$ does not fix this ambiguity.

Although Gaussian variational~\cite{Fischer2016, Nie2017, Hecker2023} and self-consistent Hartree-Fock~\cite{How2023, How2024} approaches to vestigial order do not suffer from this ambiguity, they are either uncontrolled or only apply at weak coupling, i.e., when the dimensionless coupling constants $\xi^{4-d} g_{\mu}$ that control the perturbative expansion are small.
However, according to Eq.~\eqref{eq:intro-vestigial-Tv}, vestigial order takes place when $\xi^{4-d} g_{\mu}$ is of order $1$.
The large-$N$ method discussed here enables one to analytically access the strong-coupling regime in a controlled manner~\cite{Altland2010}.
Importantly, as observed in Ref.~\cite{Hecker2023}, the large-$N$ and variational approaches give qualitatively different results: while in the former any infinitesimal $-g_z$ in Eq.~\eqref{eq:intro-vestigial-cond} yields a vestigial transition, in the latter $-g_z$ needs to overcome a threshold value.
Given the widespread use of the large-$N$ method~\cite{Golubovic1988, Fang2006, Fang2008, Fernandes2012, Chern2012, Nie2014, YuxuanWang2014, Fernandes2016, Zhang2017, Hecker2018, Willa2020, Jian2021, Hecker2023, Wu2024, Poduval2024, Verghis2025}, it is of considerable interest to elucidate how one should modify the large-$N$ approach in a way that respects the intrinsic coupling between the different vestigial channels, as expressed by the Fierz identities.

In this article, we provide a unique prescription for how the large-$N$ analysis must be carried out to respect the Fierz identities.
In fact, as we shall demonstrate, the Fierz identities represent only a special instance of a more general ``composite redundancy structure'' that arises because the number of composite vestigial OPs $\Psi_{\mu} = \vb{\eta}^{\intercal} \Gamma_{\mu} \vb{\eta}$ is larger than the number of primary OP components $\vb{\eta} = (\eta_1, \eta_2, \ldots)$.
We accomplish this by carefully defining a large-$N$ limit of the theory that respects both the initial group structure and the composite redundancy structure.

The main result is that the coupling constants appearing in the saddle-point equations must be symmetrized with respect to the redundancy identities.
As we show in this paper, only then can we expect the infinite-$N$ results to approximate well the original small-$N$ system.
In addition, the consistent treatment of all possible vestigial channels requires that we introduce Hubbard-Stratonovich fields for all of them.
The competition and interference between the different vestigial channels is then correctly accounted for by the symmetrization of the coupling constants.

In the preceding example, described by the action~\eqref{eq:intro-action}, application of our proposed method gives saddle-point equations of the same form as Eq.~\eqref{eq:intro-vestigial-cond}, but with coupling constants that are symmetrized according to
\begin{align}
\begin{aligned}
(\mathscr{S} \circ g)_0 &= \frac{2 g_0 + g_x + g_z}{3}, \\
(\mathscr{S} \circ g)_x &= \frac{g_0 + 2 g_x - g_z}{3}, \\
(\mathscr{S} \circ g)_z &= \frac{g_0 - g_x + 2 g_z}{3}.
\end{aligned}
\end{align}
The key point is that the coupling constants are now invariant under $(g_0, g_x, g_z) \mapsto (g_0 - \var{g}, g_x + \var{g}, g_z + \var{g})$.
As a result, just like the original action, the saddle-point equations now obey the constraints imposed by the Fierz identities.
The leading vestigial order, in this case, is the one with the most negative (i.e., most attractive) $(\mathscr{S} \circ g)_{x,z}$.
The corresponding phase diagrams are the same as in Fig.~\ref{fig:vestigial-transitions}, but with $g_z / g_0$ replaced by $(\mathscr{S} \circ g)_z / (\mathscr{S} \circ g)_0$.
Consequently, even in the parameter range $g_z > 0$ of panel~\ref{fig:vestigial-transitions}(a), $(\mathscr{S} \circ g)_x = (g_0 - g_z) / 3$ can become negative, resulting in $\vb{\eta}^{\intercal} \Pauli_x \vb{\eta}$ vestigial order.
Importantly, an infinitesimal negative $g_z$ is insufficient to ensure $\vb{\eta}^{\intercal} \Pauli_z \vb{\eta}$ vestigial order.
This is a consequence of the mutual interference between vestigial channels, which cannot be treated separately.
As we later prove, systems generally need to be a finite distance away from the point where all channels are degenerate to yield vestigial order.
Moreover, the results obtained by symmetrizing the coupling constant are in quantitative agreement with those obtained from the variational and Hartree-Fock methods, as we also show.

The article is organized into two parts: theoretical framework (Sec.~\ref{sec:general-teo}) and applications (Sec.~\ref{sec:applications}).
We start in Sec.~\ref{sec:general-teo} by writing down a general $M$-component $\Phi^4$ theory, reformulating it in terms of composite fields, and deriving its saddle-point equations.
In Sec.~\ref{sec:Fierz} we then derive the quartic composite redundancy relations that are responsible for the large-$N$ ambiguities.
These relations, which include the Fierz identities, are the lowest order ones of the composite redundancy structure, which is explored more fully in App.~\ref{sec:comp-redundancy}.
The main result is derived in Sec.~\ref{sec:why-symm}, namely, that the large-$N$ limit respects the composite redundancy structure only if the coupling constants are symmetrized.
The need for symmetrization is further substantiated in Sec.~\ref{sec:Hartree-Fock} by showing that symmetrization is necessary for the large-$N$ results to be able to reproduce weak-coupling (Hartree-Fock) results.

In Sec.~\ref{sec:applications}, we illustrate our formalism on three examples: that of a charge-density wave (CDW), spin-density wave (SDW), and superconducting (SC) primary order.
We focus on the case of three-component SC (Sec.~\ref{sec:cubic-SC}), which is possible in cubic systems, but has remained scarcely explored, and show that in this case there are wide parameter ranges in which charge-$4e$ SC and ferromagnetic order are leading vestigial instabilities (Fig.~\ref{fig:3D-SC-phase-diag}).
Although SDWs can give vestigial orders that are spin vectors ($\ell = 1$) and quadrupoles ($\ell = 2$)~\cite{Fernandes2019}, to date the competitiveness of these higher-$\ell$ channels has not been systematically investigated (App.~\ref{sec:spin-like-OPs}).
For $M$-point SDWs in primitive hexagonal systems (Sec.~\ref{sec:M-SDW}), we find that vestigial spin-vector loop-current order can indeed arise as a leading instability (Fig.~\ref{fig:M-SDW-phase-diag}).
In contrast, spin-quadrupole channels are subleading or, at best, degenerate with the leading vestigial channel on fine-tuned lines of the broader parameter space.

\section{General formalism} \label{sec:general-teo}
We assume that the primary order is described by an $M$-component real-valued OP $\vb{\eta} = (\eta_1, \ldots, \eta_M)$ that, at a minimum, has a $\Z_2$ symmetry $\vb{\eta}(x) \mapsto - \vb{\eta}(x)$.
The fluctuations of $\vb{\eta}$ near its second-order transition are then described by the following general Euclidean action
\begin{align}
\begin{aligned}
\actS[\eta] &= \frac{1}{2} \int_{x_1 x_2} \sum_{ab} \eta_a(x_1) \chi_{ab}^{-1}(x_1-x_2) \eta_b(x_2) \\
&\phantom{=}\quad + \frac{1}{8} \int_x \sum_{\mu\nu} \Psi_{\mu}(x) g_{\mu\nu} \Psi_{\nu}(x),
\end{aligned} \label{eq:real-primary-action}
\end{align}
where $\int_x = \int \dd[d]{x}$ is an integral over spatial coordinates, the Latin indices refer to primary OP components $a, b, c, d \in \{1, \ldots, M\}$, and the Greek indices label the possible composite OPs,  $\mu, \nu, \rho, \sigma \in \{0, 1, \ldots, \frac{M (M+1)}{2} - 1\}$.
In line with previous studies~\cite{Fernandes2019}, throughout this paper we focus on classical (finite-temperature) phase transitions for which the imaginary time dependence is unimportant.
Thus $x$ includes only spatial coordinates, while the $\beta = \int_0^{\beta} \dd{\tau} = 1 / (k_B T)$ prefactor has been absorbed in $\chi_{ab}^{-1}$ and $g_{\mu\nu}$.
Here, $\chi_{ab}$ is the susceptibility (bare propagator) of the primary OP, $g_{\mu\nu}$ are the quartic Landau coefficients of the action, and
\begin{align}
\Psi_{\mu}(x) \defeq \vb{\eta}^{\intercal}(x) \Gamma_{\mu} \vb{\eta}(x) = \sum_{ab} \eta_a(x) \Gamma_{\mu,ab} \eta_b(x) \label{eq:real-psi-bilinear}
\end{align}
are the composite OPs whose condensation we are interested in investigating.
Here and throughout, sums go over all indices, unless noted otherwise.
Given that $\eta_a \eta_b = \eta_b \eta_a$, $\Gamma_{\mu}^{\intercal} = \Gamma_{\mu}$ are symmetric.
We choose them so that they are orthonormal ($w > 0$) with $\Gamma_0 \propto \one$:
\begin{align}
\tr \Gamma_{\mu} \Gamma_{\nu} &= w \cdot \Kd_{\mu \nu}, &
\Gamma_{0,ab} &= \sqrt{\frac{w}{M}} \cdot \Kd_{ab}. \label{eq:Gamma-normalization}
\end{align}
This normalization enables us to directly compare the Landau coefficients $g_{\mu\nu}$ of the different $\Psi_{\mu}$.
In total, there are $\frac{M (M+1)}{2}$ possible composite OPs.
Symmetries constrain the form of the susceptibility $\chi_{ab}^{-1}(x) = \chi_{ba}^{-1}(-x)$ and of the couplings $g_{\mu\nu} = g_{\nu\mu}$.
The stability (i.e., that $\actS$ is bounded from below as $\abs{\vb{\eta}} \to \infty$ for all directions) of the action against first-order transitions constrains the quartic term of Eq.~\eqref{eq:real-primary-action} to be positive-definite in $\eta_a$.
However, we shall leave them completely general for now.
Specific examples of such actions are studied in Part~\ref{sec:applications}.

The model of Eq.~\eqref{eq:real-primary-action} is a realization of a multicomponent (Landau-Ginzburg-Wilson) $\Phi^4$ theory~\cite{Kleinert2001, Vicari2008}, also known as the general $n$-vector model~\cite{Brezin1974} or multiscalar theory~\cite{Rychkov2019}.
While the main features of interest are captured by expanding to quartic order, depending on the character of the vestigial transition (Fig.~\ref{fig:vestigial-transitions}), sixth-order terms may also become important, for instance, by lifting accidental degeneracies that are present at quartic order, but absent in the system, or by stabilizing the action against a fluctuation-induced first-order transition.
Including imaginary time dynamics in the formalism is in principle straightforward and should not affect our main conclusions (Sec.~\ref{sec:why-symm}) regarding the need to symmetrize the coupling constants in the large-$N$ limit.
Because we are interested in studying all possible vestigial orders, we are using a complete set of $\tfrac{M (M+1)}{2}$ linearly independent quadratic composite OPs $\Psi_{\mu}$.
Although the quartic $\Psi_{\mu} \Psi_{\nu}$ terms appearing in the interaction of Eq.~\eqref{eq:real-primary-action} are overcomplete (Sec.~\ref{sec:Fierz}), working with the most general redundant $g_{\mu\nu}$ is preferred because different vestigial channels always interfere with each other, even when the corresponding $g_{\mu\nu}$ vanishes.
If we were to focus only on the smallest possible $g_{\mu\nu}$, the large-$N$ limit would become ambiguous, as we explain in Sec.~\ref{sec:why-symm}.

\subsection{Composite field theory and \linebreak the saddle-point treatment of vestigial order} \label{sec:SPA}
To introduce composite fields, we perform a Hubbard-Stratonovich transformation~\cite{Altland2010} by inserting
\begin{gather}
\one = \int \DD{\psi} \prod_{x,\mu} \Dd\mleft(\psi_{\mu}(x) - \Psi_{\mu}(x)\mright) \label{eq:Hubbard-Stratonovich-identity} \\
= \int \DD{\psi} \DD{\phi} \exp\mleft(\frac{1}{2} \int_x \sum_{\mu} \phi_{\mu}(x) \mleft[\psi_{\mu}(x) - \Psi_{\mu}(x)\mright]\mright) \notag
\end{gather}
into the path integral.
Here $\Psi_{\mu}$ is the bilinear of Eq.~\eqref{eq:real-psi-bilinear}, while $\psi_{\mu}$ is the corresponding fluctuating real bosonic field.
$\phi_{\mu}$ is a purely imaginary Lagrange multiplier field.
We may replace $\Psi_{\mu}$ with $\psi_{\mu}$ in the quartic interaction because of the Dirac delta function, which once expressed in terms of $\phi_{\mu}$ enables us to integrate out the primary field $\eta_a$.

The result is the following composite field action
\begin{align}
\begin{aligned}
\actS_f[\psi, \phi] &= \frac{1}{8} \int_{x} \sum_{\mu\nu} \psi_{\mu}(x) g_{\mu\nu} \psi_{\nu}(x) \\
&\phantom{=}\quad - \frac{1}{2} \int_x \sum_{\mu} \phi_{\mu}(x) \psi_{\mu}(x) - \log \partZ_m[\phi],
\end{aligned} \label{eq:real-composite-action}
\end{align}
where $\partZ_m[\phi] = \int \DD{\eta} \Elr^{- \actS_{m, \phi}[\eta]}$ is the mean-field partition function, while the mean-field action is defined as
\begin{align}
\begin{aligned}
\actS_{m, \phi}[\eta] &= \frac{1}{2} \int_{x_1 x_2} \sum_{ab} \eta_a(x_1) \chi_{ab}^{-1}(x_1-x_2) \eta_b(x_2) \\
&\phantom{=}\quad + \frac{1}{2} \int_x \sum_{\mu} \phi_{\mu}(x) \Psi_{\mu}(x).
\end{aligned} \label{eq:real-meanfield-action}
\end{align}
Since $\actS_{m, \phi}$ is Gaussian,
\begin{align}
\log \partZ_m[\phi] = - \frac{1}{2} \Tr \log G_{\phi}^{-1}
\end{align}
where
\begin{align}
G_{\phi}^{-1}(x_1, x_2) = \chi^{-1}(x_1-x_2) + \sum_{\mu} \phi_{\mu}(x_1) \Gamma_{\mu} \Dd(x_1-x_2).
\end{align}

The saddle-point equations of the composite field action are obtained by minimizing the $\actS_f[\psi, \phi]$ of Eq.~\eqref{eq:real-composite-action}:
\begin{align}
\begin{aligned}
\phi_{\mu}(x) &= \frac{1}{2} \sum_{\nu} g_{\mu\nu} \psi_{\nu}(x), \\
\psi_{\mu}(x) &= \ev{\Psi_{\mu}(x)}_{m,\phi},
\end{aligned} \label{eq:real-saddle-point-eq}
\end{align}
where the average is performed relative to $\actS_{m, \phi}$.
They include the fluctuations of the primary field $\eta_a$, while the composite fields $\psi_{\mu}$ and $\phi_{\mu}$ are treated as stationary and semi-classical.
These are the conventional saddle-point equations that have been employed previously to study vestigial order~\cite{Golubovic1988, Fang2006, Fernandes2012, Nie2014, YuxuanWang2014, Hecker2023}.
Their solutions describe a vestigial phase whenever the $\psi_{\mu}$ and $\phi_{\mu}$ are finite in some symmetry nontrivial channel, while the primary OP $\eta_a$ remains uncondensed.
The latter condition is equivalent to having all eigenvalues of $G_{\phi}^{-1}$ positive, since $G_{\phi}$ is the dressed susceptibility of the primary OP.
Notice that the integration contour of $\phi_{\mu}$, which goes along the imaginary axis, needs to be deformed to cross the above saddle point, which is real.

For small $\phi_{\mu}$, the saddle-point equations give
\begin{align}
\begin{aligned}
\phi_{\mu}(x) &= \frac{1}{2} \sum_{\nu} g_{\mu\nu} \Bigg[\ev{\Psi_{\nu}(x)}_{m,0} \\
&- \frac{1}{2} \int_{x'} \sum_{\rho} \ev{\Psi_{\nu}(x) \Psi_{\rho}(x')}_{m,0}^c \phi_{\rho}(x') + \cdots\Bigg],
\end{aligned}
\end{align}
which can be recast as
\begin{gather}
\begin{gathered}
\sum_{\rho} \Bigg[\Kd_{\mu\rho} + \frac{1}{2} \sum_{\nu} g_{\mu\nu} \int_q \tr \Gamma_{\nu} \chi_{q+k} \Gamma_{\rho} \chi_q\Bigg] \phi_{\rho}(k) = \\
= \Kd_{k,0} \cdot \frac{1}{2} \sum_{\nu} g_{\mu\nu} \int_q \tr \Gamma_{\nu} \chi_q
\end{gathered} \label{eq:real-vestigial-cond}
\end{gather}
where $\int_q = \int \frac{\dd[d]{q}}{(2\pi)^d}$ sums over momentum and $\ev{\Psi_{\nu}(x) \Psi_{\rho}(x')}^c \defeq \ev{\Psi_{\nu}(x) \Psi_{\rho}(x')} - \ev{\Psi_{\nu}(x)} \ev{\Psi_{\rho}(x')}$ denotes a connected 2-point correlation function.
Because the $g_{\mu\nu}$ cannot mix symmetry-trivial $\psi_{\mu}$ with symmetry-nontrivial $\psi_{\nu}$, the right-hand side of Eq.~\eqref{eq:real-vestigial-cond} vanishes for a vestigial composite OP (i.e., symmetry-nontrivial) $\phi_{\mu}$.
The condition for the appearance of vestigial order is thus that a zero eigenvalue exists for the matrix in the square bracket of Eq.~\eqref{eq:real-vestigial-cond}.
More precisely, this condition determines when a vestigial saddle-point solution bifurcates away from the disordered solution $\phi_{\mu} = 0$.
This coincides with the vestigial transition when it is second-order, but when it is first-order this condition only gives the lower spinodal.
Both scenarios may take place~\cite{Fernandes2012}, as depicted in Fig.~\ref{fig:vestigial-transitions}.
To determine the character of the vestigial transition, one needs to solve the original nonlinear saddle-point equations~\eqref{eq:real-saddle-point-eq}.

Very often, the symmetries of the system constrain $g_{\mu\nu} = g_{\mu} \Kd_{\mu\nu}$ to be diagonal, while $\chi_q = \tilde{\chi}_q \one + \cdots$ can be taken to be approximately proportional to the identity matrix, with negligible contributions that are not proportional to the identity matrix and that are anisotropic in momentum.
In this case, which is relevant to all examples studied in Sec.~\ref{sec:applications}, the condition~\eqref{eq:real-vestigial-cond} for vestigial order simplifies to (cf.\ Eq.~\eqref{eq:intro-vestigial-cond})
\begin{align}
\frac{1}{g_{\mu}} &= - \frac{w}{2} \int_q \tilde{\chi}_q^2, \label{eq:isotropic-sus-limit}
\end{align}
where we used Eq.~\eqref{eq:Gamma-normalization}; $w > 0$ is a normalization factor.
Because the right-hand side is the same for all channels, it follows that vestigial order takes place in the channel with the most negative coupling constant $g_{\mu}$.

\subsection{Ambiguities in the saddle-point equations and quartic composite redundancy relations} \label{sec:Fierz}
As noted in the introduction, the $g_{\mu\nu}$ that enter the saddle-point equations~\eqref{eq:real-saddle-point-eq} are not unique, which implies a large ambiguity in the predictions of the saddle-point approximation.
As we explain in this section, the ambiguity in the values of the coupling constants $g_{\mu\nu}$ reflects the redundancy inherent to the composite field formulation of the model.

To account for all possible vestigial channels, the $\Gamma_{\mu}$ that we use to construct the bilinears $\Psi_{\mu}$ (Eq.~\eqref{eq:real-psi-bilinear}) must constitute a complete basis of symmetric $M \times M$ matrices.
Hence not only is $\Psi_{\mu} = \sum_{ab} \eta_a \Gamma_{\mu,ab} \eta_b$, but also
\begin{align}
\eta_a \eta_b &= \frac{1}{w} \sum_{\mu} \Gamma_{\mu,ba} \Psi_{\mu}. \label{eq:Gamma-completeness}
\end{align}
Here we used the normalization of Eq.~\eqref{eq:Gamma-normalization}.
The crucial point here is that there are, in total, $\frac{M (M + 1)}{2}$ distinct composite fields $\Psi_{\mu}$, while the primary field $\eta_a$ has only $M$ components.
Any theory based on composite fields will therefore possess a ``composite redundancy structure'' that encodes the fact that the composite fields are not fully independent but interconnected.

The most familiar relation that encodes this fact is the Fierz identity. To obtain the Fierz identity, we rewrite the left-hand side of Eq.~\eqref{eq:Gamma-completeness} in terms of Kronecker deltas and substitute $\Psi_{\mu} = \sum_{cd} \eta_c \Gamma_{\mu,cd} \eta_d$ into the right-hand side, obtaining:
\begin{align}
\begin{aligned}
\eta_a \eta_b &= \sum_{cd} \frac{1}{2} \mleft[\Kd_{ad} \Kd_{bc} + \Kd_{ac} \Kd_{bd}\mright] \eta_c \eta_d \\
&= \frac{1}{w} \sum_{\mu} \sum_{cd} \Gamma_{\mu,ba}  \Gamma_{\mu,cd} \eta_c \eta_d.
\end{aligned}
\end{align}
Since $\Gamma_{\mu,ab} = \Gamma_{\mu,ba}$ are symmetric matrices, we find:
\begin{align}
\begin{aligned}
\sum_{\mu=0}^{\frac{M(M+1)}{2}-1} \Gamma_{\mu, ab} \Gamma_{\mu, cd} = \frac{w}{2} \mleft[\Kd_{ad} \Kd_{bc} + \Kd_{ac} \Kd_{bd}\mright]\phantom{.}& \label{eq:Fierz-identity} \\
= \frac{M}{2} \mleft[\Gamma_{0, ad} \Gamma_{0, bc} + \Gamma_{0, ac} \Gamma_{0, bd}\mright].&
\end{aligned}
\end{align}
By contracting this identity with $\eta_a \eta_b \eta_c \eta_d$, we obtain
\begin{align}
\sum_{\mu\nu} \mleft(- M \Kd_{\mu0} \Kd_{\nu0} + \Kd_{\mu\nu}\mright) \Psi_{\mu} \Psi_{\nu} &= 0. \label{eq:Fierz-redundancy}
\end{align}

Hence, shifting the coupling constants according to $g_{\mu\nu} \mapsto g_{\mu\nu} + \mleft(- M \Kd_{\mu0} \Kd_{\nu0} + \Kd_{\mu\nu}\mright) \var{g}$ does not change the action.
However, within approximate schemes such as the SPA, a shift in $g_{\mu\nu}$ can fundamentally change the results.
For instance, if $g_{\mu\nu} = g_{\mu} \Kd_{\mu\nu}$ is diagonal, then the Fierz identity can be exploited to make any coupling constant $g_1$, \ldots, $g_{M(M+1)/2-1}$ positive, apparently precluding vestigial order, or even more negative, apparently enhancing the transition temperature.
To better understand why the SPA is sensitive to these changes in $g_{\mu\nu}$, one needs to study the limit that controls the SPA, which we do in the next section.

The general procedure for deriving redundancy relations among composite OPs, of which Eq.~\eqref{eq:Fierz-redundancy} is the simplest one, is the following.
Consider a general term that is quartic in $\vb{\eta}$:
\begin{align}
\sum_{abcd} R_{abcd} \eta_a \eta_b \eta_c \eta_d.
\end{align}
The product $\eta_a \eta_b \eta_c \eta_d$ has the key property that it is fully symmetric under all permutations of $a, b, c, d$.
Thus, if we decompose $R$ into a fully symmetric part $\mathscr{S} \circ R$, given by
\begin{align}
(\mathscr{S} \circ R)_{abcd} \defeq \frac{1}{4!} \sum_{\pi(a, b, c, d)} R_{\pi_1 \pi_2 \pi_3 \pi_4} \label{eq:aux1}
\end{align}
where $\pi$ are permutations of $(a, b, c, d)$, and into a partially symmetric remainder $\mathscr{P}_S \circ R$, given by
\begin{align}
(\mathscr{P}_S \circ R)_{abcd} \defeq R_{abcd} - (\mathscr{S} \circ R)_{abcd}, \label{eq:aux2}
\end{align}
it follows that $\sum_{abcd} (\mathscr{P}_S \circ R)_{abcd} \eta_a \eta_b \eta_c \eta_d = 0$.
Applying Eq.~\eqref{eq:Gamma-completeness} on the $\eta_a \eta_b$ and $\eta_c \eta_d$, we obtain the quartic composite redundancy relations
\begin{align}
\sum_{\mu\nu} (\mathscr{P}_S \circ R)_{\mu\nu} \Psi_{\mu} \Psi_{\nu} &= 0, \label{eq:quartic-redundancy}
\end{align}
where $(\mathscr{P}_S \circ R)_{\mu\nu} \defeq w^{-2} \sum_{abcd} (\mathscr{P}_S \circ R)_{abcd} \Gamma_{\mu,ba} \Gamma_{\nu,dc}$ for an arbitrary tensor $R_{abcd}$.

The relations~\eqref{eq:quartic-redundancy} are the most general ones that express the redundancy of the composite OPs.
They provide us with the most general matrices that we can add to $g_{\mu\nu}$ without affecting the action.
In particular, for $M \geq 3$ we find additional relationships besides the Fierz one expressed by the matrix of Eq.~\eqref{eq:Fierz-redundancy} -- for more details, see Appendix~\ref{sec:comp-redundancy}.
More significantly, this derivation suggests that the solution to the SPA ambiguity problem is to fully symmetrize the coupling constants $g_{\mu\nu}$, given that $\mathscr{S} \circ g$ is invariant under the addition of $\mathscr{P}_S \circ R$ to $g$ because of the identity $\mathscr{S} \circ \mathscr{P}_S = 0$.

Redundancy relations of the type~\eqref{eq:quartic-redundancy} exist for all orders in $\vb{\eta}$ higher than two and are derived analogously (App.~\ref{sec:comp-redundancy}).
This hierarchy of redundancy relations, which we call the composite redundancy structure, expresses the fact that the $\Psi_{\mu}$ are composites of $\eta_a$.

\subsection{A large-$N$ approach that has no saddle-point ambiguities} \label{sec:why-symm}
Having identified the key structure -- the composite redundancy structure -- that underlies the transformation from the primary-field action~\eqref{eq:real-primary-action} to the composite-field action~\eqref{eq:real-composite-action}, we now develop a procedure that preserves this structure within the saddle-point approximation (SPA).

The SPA becomes exact when there is a large prefactor multiplying the action~\cite{Altland2010}.
For the composite action~\eqref{eq:real-composite-action} there is no such naturally occurring prefactor.
The standard way of dealing with this is to formally extend the number of  components of the primary field appearing in the theory to $N$, while rescaling the coupling constants, so that the action becomes overall proportional to $N$~\cite{Moshe2003}.
At the same time, the symmetry is enlarged so as to keep the number of symmetry-allowed terms in the action finite as $N \to \infty$~\cite{Moshe2003}.
The SPA then becomes exact for large $N$ ($N \to \infty$) and corrections to it can be systematically computed in powers of $1/N$.

The action of Eq.~\eqref{eq:real-primary-action} is well known in the literature~\cite{Kleinert2001, Vicari2008, Brezin1974, Rychkov2019} and, as such, it has been studied via several methods besides large-$N$, including $4-\epsilon$ renormalization group (RG) and Hartree-Fock.
Here, we focus on the large-$N$ approach because it applies in arbitrary dimensions from weak to strong coupling and, relatedly, because it is the most common method used to analyze vestigial order~\cite{Golubovic1988, Fang2006, Fang2008, Fernandes2012, Chern2012, Nie2014, YuxuanWang2014, Fernandes2016, Zhang2017, Hecker2018, Willa2020, Jian2021, Hecker2023, Wu2024, Poduval2024, Verghis2025}.
A variety of other problems have also been addressed by the large-$N$ method~\cite{Moshe2003, Affleck1985, Bickers1987, Affleck1988, Marston1989, Sachdev1990, Read1991, Sachdev1993, Vojta2000-PRB, Chubukov2005, DTSon2007, Flint2009, Flint2012, Esterlis2019, Esterlis2021}.

In the context of vestigial order, the most direct large-$N$ limit would be to extend the number of components of the primary OP $M$ to infinity.
However, as we showed in the previous section, the redundancy relations are fundamentally different for distinct values of $M$, which makes defining a large-$M$ limit that preserves them challenging.
Moreover, the set of possible vestigial channels is also altered if we change the dimension of the primary OP, the irreducible representation (irrep) under which the primary OP transforms, or the corresponding symmetry group.
In some models, such as the $\Ogp(N)$ vector model~\cite{Moshe2003}, the $\C P^{N-1}$ model~\cite{Moshe2003}, or the $\Ogp(N_1) \times \Ogp(N_2)$ spin model~\cite{Pelissetto2001}, there is a natural way of generalizing the symmetry group, the primary field, and its irrep from $N = 2, 3$ to arbitrary $N$.
A similar extension, however, is not generally possible for models whose primary OP transforms under an irrep that is unique to a given point group or space group.

To circumvent these problems, we propose the following large-$N$ limit.
Its primary goal is to preserve the vestigial channels and redundancy relations that are intrinsic to the irrep according to which the OP transforms.
Suppose we are given a multicomponent $\Phi^4$ model whose symmetry group is $G$.
Since $\SO(1)$ is isomorphic to the trivial group, we can also express the symmetry group as $G \times \SO(1)$.
We now enlarge the symmetry group from $G \times \SO(1)$ to $G \times \SO(N)$, with $N = 1$ recovering the original model.
This effectively adds to the primary OP $\eta_a$ an internal index $n \in \{1, \ldots, N\}$ such that $\eta_{a,n} \mapsto \sum_{n'} R_{nn'} \eta_{a,n'}$ under $R_{nn'} \in \SO(N)$.
Importantly, this does not alter the structure of the original symmetry group $G$ and all vestigial channels are preserved without modification if we focus on those channels that are trivial (scalar, $\propto \Kd_{nn'}$) in the internal space.
Hence no vestigial channels are omitted or modified in the $N \to \infty$ limit when compared to the original $N = 1$ model.
Moreover, all terms that are symmetry-allowed by $G$ are also symmetry-allowed by $G \times \SO(N)$ as long as $G$ contains the $\Z_2$ symmetry $\vb{\eta} \mapsto - \vb{\eta}$ characteristic of multicomponent $\Phi^4$ theories.

In previous large-$N$ approaches used to study vestigial order~\cite{Golubovic1988, Fang2006, Fernandes2012, Nie2014, YuxuanWang2014, Hecker2023}, it was a continuous subgroup of $G$, usually $\SO(2)$ or $\SO(3)$ related to magnetic degrees of freedom, that was extended to $\SO(N)$.
While this extension can leave the leading vestigial channel invariant, it generally cannot leave all vestigial channels invariant, as we show in App.~\ref{sec:spin-like-OPs}.
Given that the redundancy relations couple different vestigial OPs, this type of large-$N$ approach will not in general preserve them, resulting in ambiguities in the SPA.
Another difference is that within our large-$N$ approach we need to extrapolate down to $N = 1$ instead of the usual $N = 2$ or $3$.
This, however, does not have obvious implications for the reliability of one large-$N$ approach over the other because both expansions are asymptotic in $1/N$~\cite{Marino2014, Note1} and neither the optimal truncation order, the associated error, nor the Borel summability are known.
An alternative test of reliability is to compare the large-$N$ results against weak-coupling expansions.
We do this in the next Sec.~\ref{sec:Hartree-Fock}, where we show that our proposed large-$N$ approach correctly reproduces the competition between the various vestigial channels present in the Hartree-Fock approximation, in contrast to the traditional large-$N$ approach.

\footnotetext[1]{ 
The expansion in $1/N$ is asymptotic because the theory is ill-defined for arbitrarily small negative $1/N$, which implies that the convergence radius is zero~\cite{Marino2014}.
Few studies have examined the Borel summability of general multicomponent $\Phi^4$ theories in $1/N$.
Ref.~\cite{Ferdinand2024} has proven the Borel summability of the vector $\Ogp(N)$ model in $1/N$ in zero dimensions.
See also Ref.~\cite{Hikami1979} and the discussion in Ref.~\cite{Moshe2003}.
}

We are now in a position to explicitly construct our large-$N$ extension of the action~\eqref{eq:real-primary-action}.
First, we rewrite it as
\begin{align}
\begin{aligned}
\actS[\eta] &= \frac{1}{2} \int_{x_1 x_2} \sum_{ab} \eta_a(x_1) \chi_{ab}^{-1}(x_1-x_2) \eta_b(x_2) \\
&\phantom{=}\quad + \frac{1}{8} \int_x \sum_{abcd} g_{abcd} \eta_a(x) \eta_b(x) \eta_c(x) \eta_d(x),
\end{aligned} \label{eq:real-primary-action-v2}
\end{align}
where
\begin{align}
g_{abcd} &\defeq \sum_{\mu\nu} g_{\mu\nu} \Gamma_{\mu,ab} \Gamma_{\nu,cd}, \label{eq:gabcd-gmunu-rel} \\
g_{\mu\nu} &= \frac{1}{w^2} \sum_{abcd} \Gamma_{\mu,ba} \Gamma_{\nu,dc} g_{abcd}.
\end{align}
Given that $g_{\mu\nu} = g_{\nu\mu}$ and $\Gamma_{\mu}^{\intercal} = \Gamma_{\mu}$, $g_{abcd}$ satisfies
\begin{gather}
\begin{gathered}
g_{abcd} = g_{cdab}, \\
g_{abcd} = g_{bacd}.
\end{gathered}
\end{gather}
For reasons that will soon become apparent, we next decompose $g$ into a fully symmetric part $\mathscr{S} \circ g$ and a partially symmetric part $\mathscr{P}_S \circ g$ according to Eqs.~\eqref{eq:aux1}--\eqref{eq:aux2}:
\begin{align}
g &= \mathscr{S} \circ g + \mathscr{P}_S \circ g,
\end{align}
where
\begin{align}
(\mathscr{S} \circ g)_{abcd} &= \frac{1}{3} \mleft(g_{abcd} + g_{acbd} + g_{adbc}\mright), \\
(\mathscr{P}_S \circ g)_{abcd} &= \frac{1}{3} \mleft(2 g_{abcd} - g_{acbd} - g_{adbc}\mright).
\end{align}
These two decompositions satisfy
\begin{gather}
\begin{gathered}
(\mathscr{S} \circ g)_{abcd} = (\mathscr{S} \circ g)_{cdab}, \\
(\mathscr{S} \circ g)_{abcd} = (\mathscr{S} \circ g)_{bacd}, \\
(\mathscr{S} \circ g)_{abcd} = (\mathscr{S} \circ g)_{acbd} = (\mathscr{S} \circ g)_{adbc},
\end{gathered}
\end{gather}
and
\begin{gather}
\begin{gathered}
(\mathscr{P}_S \circ g)_{abcd} = (\mathscr{P}_S \circ g)_{cdab}, \\
(\mathscr{P}_S \circ g)_{abcd} = (\mathscr{P}_S \circ g)_{bacd}, \\
(\mathscr{P}_S \circ g)_{abcd} + (\mathscr{P}_S \circ g)_{acdb} + (\mathscr{P}_S \circ g)_{adbc} = 0.
\end{gathered}
\end{gather}
In the composite-field basis they become:
\begin{align}
(\mathscr{S} \circ g)_{\mu\nu} &= \frac{1}{3} g_{\mu\nu} + \frac{2}{3} (\mathscr{F} \circ g)_{\mu\nu}, \label{eq:symmetrized-gmunu} \\
(\mathscr{P}_S \circ g)_{\mu\nu} &= \frac{2}{3} g_{\mu\nu} - \frac{2}{3} (\mathscr{F} \circ g)_{\mu\nu}, \label{eq:partsymmetrized-gmunu}
\end{align}
where
\begin{align}
(\mathscr{F} \circ g)_{\mu\nu} &\defeq \frac{1}{w^2} \sum_{\rho\sigma} \tr \Gamma_{\mu} \Gamma_{\rho} \Gamma_{\nu} \Gamma_{\sigma} \cdot g_{\rho\sigma}.
\end{align}
In addition, we can also construct a partially antisymmetric part
\begin{align}
(\mathscr{P}_A \circ g)_{abcd} &= \frac{1}{2} \mleft(g_{acbd} - g_{adbc}\mright)
\end{align}
that satisfies
\begin{gather}
\begin{gathered}
(\mathscr{P}_A \circ g)_{abcd} = (\mathscr{P}_A \circ g)_{cdab}, \\
(\mathscr{P}_A \circ g)_{abcd} = - (\mathscr{P}_A \circ g)_{bacd}, \\
(\mathscr{P}_A \circ g)_{abcd} + (\mathscr{P}_A \circ g)_{acdb} - (\mathscr{P}_A \circ g)_{adbc} = 0.
\end{gathered}
\end{gather}
The corresponding composite-field matrix vanishes identically, $(\mathscr{P}_A \circ g)_{\mu\nu} = 0$.

To proceed, we note that $\eta_a \eta_b \eta_c \eta_d$ is fully symmetric under index permutations, as already discussed in the previous Sec.~\ref{sec:Fierz}.
Therefore:
\begin{align}
\sum_{abcd} (\mathscr{P}_S \circ g)_{abcd} \eta_a \eta_b \eta_c \eta_d &= 0, \\
\sum_{abcd} (\mathscr{P}_A \circ g)_{abcd} \eta_a \eta_b \eta_c \eta_d &= 0.
\end{align}
Because $(\mathscr{P}_A \circ g)_{\mu\nu} = 0$, only the former relation gives rise to quadratic composite redundancy relations.

It is now straightforward to extend the action~\eqref{eq:real-primary-action-v2}, which is invariant under the symmetry group $G$, to a new action that is invariant under the symmetry group $G \times \SO(N)$.
The only rank-$4$ tensor that is invariant under $\SO(N)$, for general $N \neq 2, 4$, is $\Kd_{n_1 n_2} \Kd_{n_3 n_4}$, where $n_1, n_2, n_3, n_4 \in \{1, \ldots, N\}$ are indices internal to $\SO(N)$.
Because $\eta_{a, n_1} \eta_{b, n_2} \eta_{c, n_3} \eta_{d, n_4}$ is fully symmetric under permutations of $(a,n_1), (b,n_2), \ldots$ index pairs, only tensors that share this property survive after index contraction.
The most general such tensors are constructed by decomposing both the original quartic coefficients $g_{abcd}$ and $\Kd_{n_1 n_2} \Kd_{n_3 n_4}$ with $\mathscr{S}$, $\mathscr{P}_S$, and $\mathscr{P}_A$ and then combining them~\cite{Note2}.
\begin{widetext}
All in all, the most general extension of the original action of Eq.~\eqref{eq:real-primary-action-v2} that is $\SO(N)$-symmetric in the internal $n$-index space is the following:
\begin{align}
\begin{aligned}
\actS_N[\eta] &= \frac{1}{2} \int_{x_1 x_2} \sum_{ab} \sum_{n_1 n_2} \chi_{ab}^{-1}(x_1-x_2) \Kd_{n_1 n_2} \eta_{a, n_1}(x_1) \eta_{b, n_2}(x_2) \\
&+ \frac{1}{8 N} \int_x \sum_{abcd} \sum_{n_1 n_2 n_3 n_4} (\mathscr{S} \circ g)_{abcd} \frac{\Kd_{n_1 n_2} \Kd_{n_3 n_4} + \Kd_{n_1 n_3} \Kd_{n_2 n_4} + \Kd_{n_1 n_4} \Kd_{n_2 n_3}}{3} \eta_{a, n_1}(x) \eta_{b, n_2}(x) \eta_{c, n_3}(x) \eta_{d, n_4}(x) \\
&+ \frac{1}{8 N} \int_x \sum_{abcd} \sum_{n_1 n_2 n_3 n_4} (\mathscr{P}_S \circ g)_{abcd} \frac{2 \Kd_{n_1 n_2} \Kd_{n_3 n_4} - \Kd_{n_1 n_3} \Kd_{n_2 n_4} - \Kd_{n_1 n_4} \Kd_{n_2 n_3}}{3} \eta_{a, n_1}(x) \eta_{b, n_2}(x) \eta_{c, n_3}(x) \eta_{d, n_4}(x) \\
&+ \frac{1}{8 N} \int_x \sum_{abcd} \sum_{n_1 n_2 n_3 n_4} (\mathscr{P}_A \circ g)_{abcd} \frac{\Kd_{n_1 n_3} \Kd_{n_2 n_4} - \Kd_{n_1 n_4} \Kd_{n_2 n_3}}{2} \eta_{a, n_1}(x) \eta_{b, n_2}(x) \eta_{c, n_3}(x) \eta_{d, n_4}(x).
\end{aligned} \label{eq:real-primary-action-large-N}
\end{align}
\end{widetext}

\footnotetext[2]{ 
Notice that $\mathscr{S}$, $\mathscr{P}_S$, and $\mathscr{P}_A$ decompose rank-$4$ tensors into irreps of the permutation group of degree $4$~\cite{Hamermesh1989}.
Only direct products of irreps with themselves give a trivial irrep, i.e., one that is fully symmetric under permutations of $(a,n_1), (b,n_2), \ldots$ index pairs.
A more pedestrian way of constructing all invariants is to notice that $g_{abcd}$ can be combined with either $\Kd_{n_1 n_2} \Kd_{n_3 n_4}$, $\Kd_{n_1 n_3} \Kd_{n_2 n_4}$, or $\Kd_{n_1 n_4} \Kd_{n_2 n_3}$.
}

Although the terms with partially symmetric and partially antisymmetric quartic coefficients, $\mathscr{P}_S \circ g$ and $\mathscr{P}_A \circ g$, vanish identically in Eq.~\eqref{eq:real-primary-action-large-N} for $N = 1$, at large $N$ they are not only finite, but of the same order in $N$ as the fully symmetric $\mathscr{S} \circ g$ term.
Thus, for a given action of the form~\eqref{eq:real-primary-action-v2}, there is a family of inequivalent large-$N$ extensions that differ in their $\mathscr{P}_S \circ g$ and $\mathscr{P}_A \circ g$ coupling constants.
Because these inequivalent large-$N$ actions give inequivalent saddle-point equations, the corresponding large-$N$ SPA results are different.
This is the reason why the SPA results changed for equivalent choices of coupling constants $g_{\mu\nu}$, $g_{\mu\nu} \mapsto g_{\mu\nu} + (\mathscr{P}_S \circ R)_{\mu\nu}$, as first noticed in Ref.~\cite{Hecker2023}.

Symmetries constrain the form of $g_{\mu\nu}$ and, for $N > 1$, only those $\mathscr{P}_S \circ g$ and $\mathscr{P}_A \circ g$ that respect the symmetries of $G$ are allowed in the large-$N$ extension~\eqref{eq:real-primary-action-large-N}.
Depending on how large and restrictive the symmetry group $G$ is, there can be only one symmetry-allowed $\mathscr{P}_S \circ g$, namely the one associated with the Fierz identity~\eqref{eq:Fierz-redundancy}, or several.
Hence an ambiguity of the large-$N$ limit always exists.
Note that the Fierz $(\mathscr{P}_S \circ g)_{\mu\nu} = \mleft(- M \Kd_{\mu0} \Kd_{\nu0} + \Kd_{\mu\nu}\mright) \var{g}$ always respects all symmetries.
This holds because $G$ acts on the primary fields $\eta_a$ through a representation, $\eta_a \mapsto \sum_b \varrho_{ab}(h) \eta_b$ for $h \in G$, that can always be made orthogonal~\cite{Hamermesh1989, Dresselhaus2007}, implying that $\Psi_0 \propto \vb{\eta}^{\intercal} \vb{\eta}$ transforms trivially, while the invariance of $\sum_{\mu} \Psi_{\mu}^2$ follows from orthogonality of the $\varrho \otimes \varrho$ representation under which $\Psi_{\mu}$ transform.

Having identified the origin of the SPA ambiguity, we now eliminate it by keeping only the fully symmetric combination $g = \mathscr{S} \circ g$ in the large-$N$ action.
For $N = 1$, this corresponds to switching from $g$ to the equivalent $g - \mathscr{P}_S \circ g$ quartic coefficients.
To see why, recall that the initial motivation for constructing this large-$N$ extension was to construct a solvable limit that preserves as much of the symmetry group structure encoded in the original action as possible, because only then will the $N \to \infty$ results extrapolate reasonably well down to $N = 1$.
Our large-$N$ limit manifestly preserves the group structure.
However, the relationship between the primary and composite fields is disrupted if we only allow for composite OPs that are scalars in $\SO(N)$, i.e, $\propto \Kd_{n_1 n_2}$.
For instance, the quartic redundancy relation~\eqref{eq:quartic-redundancy} is violated:
\begin{align}
\sum_{\mu\nu} (\mathscr{P}_S \circ R)_{\mu\nu} \Psi_{N,\mu} \Psi_{N,\nu} &\neq 0
\end{align}
for $\Psi_{N,\mu} \defeq N^{-1} \sum_{abn} \eta_{a, n} \Gamma_{\mu,ab} \eta_{b, n}$.
Thus, in order to preserve the composite redundancy structure, we choose to eliminate the $\mathscr{P}_S \circ g$ and $\mathscr{P}_A \circ g$ terms from the action~\eqref{eq:real-primary-action-large-N}.
Since they vanish for $N = 1$, it is natural to impose this condition for $N \geq 2$.

Further support for using $g = \mathscr{S} \circ g$ is provided in the next section, where we compare our large-$N$ SPA with the weak-coupling self-consistent Hartree-Fock approximation, taking advantage of the fact that weak coupling is one of the few limits where we have analytic control.
The main finding is that the two approaches agree, up to an overall constant, only if $g = \mathscr{S} \circ g$.
Incidentally, this provides a physical interpretation for the symmetrization of the coupling constants $g_{\mu\nu}$.
Given the interaction $\sum_{abcd} \sum_{\mu\nu} g_{\mu\nu} (\eta_a \Gamma_{\mu,ab} \eta_b) \cdot (\eta_c \Gamma_{\nu,cd} \eta_d)$, we can apply Eq.~\eqref{eq:Gamma-completeness} not only on $(\eta_a \eta_b)$ and $(\eta_c \eta_d)$, yielding $g_{\mu\nu}$, but also on $(\eta_a \eta_c)$ and $(\eta_b \eta_d)$, or on $(\eta_a \eta_d)$ and $(\eta_b \eta_c)$, yielding $(\mathscr{F} \circ g)_{\mu\nu}$.
Thus what Eq.~\eqref{eq:symmetrized-gmunu} does is, essentially, average $g_{\mu\nu}$ over the Hartree ($\propto g_{\mu\nu}$) and the two Fock ($\propto (\mathscr{F} \circ g)_{\mu\nu}$) contractions of the interaction.

The construction of the composite field action for the large-$N$ extension~\eqref{eq:real-primary-action-large-N} with $\mathscr{P}_S \circ g = \mathscr{P}_A \circ g = 0$ proceeds as in Sec.~\ref{sec:SPA}.
Because of index permutation symmetry, $(\mathscr{S} \circ g)_{abcd} \frac{1}{3} (\Kd_{n_1 n_2} \Kd_{n_3 n_4} + \Kd_{n_1 n_3} \Kd_{n_2 n_4} + \Kd_{n_1 n_4} \Kd_{n_2 n_3})$ is equivalent to $(\mathscr{S} \circ g)_{abcd} \Kd_{n_1 n_2} \Kd_{n_3 n_4}$.
The interaction can therefore be expressed in terms of $\SO(N)$-scalar $\Psi_{N,\mu} \defeq N^{-1} \sum_{abn} \eta_{a, n} \Gamma_{\mu,ab} \eta_{b, n}$.
Next, one inserts Eq.~\eqref{eq:Hubbard-Stratonovich-identity} with $\Psi_{\mu} \to \Psi_{N,\mu}$ and $\phi_{\mu} \to N \phi_{\mu}$ into the path integral.
The end result is the composite field action~\eqref{eq:real-composite-action} multiplied by an overall $N$ prefactor, $\actS_{N,f}[\psi, \phi] = N \actS_f[\psi, \phi]$, as advertised.
The definitions of $\partZ_m[\phi]$ and $\actS_{m, \phi}[\eta]$ appearing in this action are identical to the $N = 1$ case.
Since the Hubbard-Stratonovich transformation is exact, the partition function $\partZ = \int \DD{\psi} \DD{\phi} \Elr^{- \actS_f[\psi, \phi]}$ is originally invariant under $g \mapsto g + \mathscr{P}_S \circ R$.
Once the $N$ prefactor is added to $\actS_f$ this is no longer the case, as we explicitly see from the dependence of the large-$N$ action~\eqref{eq:real-primary-action-large-N} on $\mathscr{P}_S \circ g$.

For the common situation of diagonal $g_{\mu\nu} = g_{\mu} \Kd_{\mu\nu}$, discussed at the end of Sec~\ref{sec:SPA}, we can demonstrate one important consequence of symmetrization: the threshold effect.
Given a $\tilde{g}_{\mu\nu} = g_0 \Kd_{\mu 0} \Kd_{\nu 0}$ whose vestigial channels are all degenerate and vanishing, $g_{\mu \geq 1} = 0$, is an infinitesimally small negative $g_{\mu \geq 1}$ sufficient for vestigial ordering, as indicated by Eq.~\eqref{eq:isotropic-sus-limit}?
According to our prescription, the answer is negative because the corresponding fully symmetrized coupling matrix equals
\begin{align}
(\mathscr{S} \circ \tilde{g})_{\mu\nu} &= \frac{1 + 2/M}{3} g_0 \Kd_{\mu 0} \Kd_{\nu 0} + \frac{2}{3M} g_0 (\Kd_{\mu\nu} - \Kd_{\mu 0} \Kd_{\nu 0}), \label{eq:threshold-effect}
\end{align}
where $g_0 > 0$ to ensure stability.
The threshold effect is the simple observation that all vestigial coupling constants are positive here and that a finite distance away from this point is needed for vestigial order to appear.

In summary, the source of the SPA ambiguity is the existence of multiple inequivalent large-$N$ limits.
The one that extrapolates best to $N = 1$ is the one in which the coupling constants $g_{\mu\nu}$ are fully symmetrized, i.e., the one with $g_{\mu\nu} = (\mathscr{S} \circ g)_{\mu\nu}$, as given by Eq.~\eqref{eq:symmetrized-gmunu}.

For convenience, in Appendices~\ref{sec:complex-OPs} and~\ref{sec:spin-like-OPs}, we explicitly apply this formalism, respectively, to the case of vestigial superconducting phases, where the primary OP is complex and the symmetry group is $G = G_0 \times \Ugp(1)$, with space group $G_0$, and the case of vestigial magnetic phases, where the primary OPs are isotropic in spin space and the symmetry group is $G_0 \times \SO(3)$.
As we show in these Appendices, some minor modifications of the formalism make the calculations more straightforward.
The corresponding symmetrization of the coupling constants is provided in Eqs.~\eqref{eq:symmetrized-U1-gmunu} and~\eqref{eq:symmetrized-spin-gmunu}, respectively.

\subsection{Comparison to the self-consistent Hartree-Fock approximation} \label{sec:Hartree-Fock}
The self-consistent Hartree-Fock approximation, sometimes formulated in variational terms (App.~\ref{sec:comp-variational}), has been used to study vestigial order on several occasions~\cite{Fischer2016, Nie2017, Hecker2023, How2023, How2024}.
Unlike SPA, it gives unambiguous results, as there is no need to take a large-$N$ limit.
However, this approach applies only at weak coupling, whereas vestigial order can also be a strong-coupling effect.
Nonetheless, it is instructive to compare the predictions of the two approaches, particularly to further validate the large-$N$ approach that we propose.

To derive the Hartree-Fock self-consistency equations, we start from the Baym-Kadanoff functional~\cite{Baym1961, Bickers2004}, which for the real bosonic model~\eqref{eq:real-primary-action} has the form
\begin{align}
\begin{aligned}
\beta F_{\text{BK}}[G] &= \frac{1}{2} \Tr \log G^{-1} - \frac{1}{2} \Tr G \mleft(G^{-1} - \chi^{-1}\mright) \\
&\phantom{=}\quad + \Upphi_{\text{LW}}[G].
\end{aligned} \label{eq:Baym-Kadanoff}
\end{align}
When evaluated at the exact Green's function, $\beta F_{\text{BK}}$ gives the (exact) free energy $\beta F = - \log \partZ$.
The Luttinger-Ward functional $\Upphi_{\text{LW}}$~\cite{Luttinger1960, Bickers2004, Potthoff2006} is defined as the sum over connected skeleton diagrams of $\ev{- \Elr^{- \actS_{\text{int}}[\eta]}}$, where $\actS_{\text{int}}[\eta] = \frac{1}{8} \int_x \sum_{\mu\nu} \Psi_{\mu}(x) g_{\mu\nu} \Psi_{\nu}(x)$.
Within Hartree-Fock, $\Upphi_{\text{LW}}$ is replaced by the lowest-order term $\ev{\actS_{\text{int}}[\eta]}$, which is readily evaluated using Wick's theorem:
\begin{align}
\begin{aligned}
\ev{\actS_{\text{int}}[\eta]} &= \frac{1}{8} \int_x \sum_{\mu\nu} g_{\mu\nu} \ev{\Psi_{\mu}(x)} \ev{\Psi_{\nu}(x)} \\
&\phantom{=}\quad + \frac{1}{4} \int_x \sum_{\mu\nu} g_{\mu\nu} \tr \Gamma_{\mu} G(x,x) \Gamma_{\nu} G(x,x).
\end{aligned}
\end{align}
Here the first term is the Hartree term and the second term is the Fock term.
The latter can be related to the former by exploiting the completeness of the $\Gamma_{\mu}$ matrices.
Since $\ev{\Psi_{\mu}(x)} = \tr \Gamma_{\mu} G(x,x)$ and $\tr \Gamma_{\mu} \Gamma_{\nu} = w \Kd_{\mu \nu}$, it follows that $G(x,x) = w^{-1} \sum_{\mu} \Gamma_{\mu} \ev{\Psi_{\mu}(x)}$ and therefore
\begin{align}
\ev{\actS_{\text{int}}[\eta]} &= \frac{1}{8} \int_x \sum_{\mu\nu} 3 (\mathscr{S} \circ g)_{\mu\nu} \ev{\Psi_{\mu}(x)} \ev{\Psi_{\nu}(x)},
\end{align}
where $(\mathscr{S} \circ g)_{\mu\nu}$ is defined in Eq.~\eqref{eq:symmetrized-gmunu}.
By minimizing $F_{\text{BK}}$ with $\Upphi_{\text{LW}} = \ev{\actS_{\text{int}}[\eta]}$ with respect to $G$, we obtain the Hartree-Fock expression for the self-energy:
\begin{gather}
\begin{gathered}
G^{-1}(x_1, x_2) - \chi^{-1}(x_1, x_2) = \\
= \frac{1}{2} \sum_{\mu\nu} 3 (\mathscr{S} \circ g)_{\mu\nu} \ev{\Psi_{\mu}(x_1)} \Gamma_{\nu} \Dd(x_1 - x_2).
\end{gathered} \label{eq:Hartree-Fock-self-energy}
\end{gather}
Due to the proportionality to $\Dd(x_1 - x_2)$, we may write $G^{-1}(x_1, x_2) = \chi^{-1}(x_1-x_2) + \sum_{\mu} \phi_{\mu}(x_1) \Gamma_{\mu} \Dd(x_1-x_2)$ to recast the above as a self-consistency equation for $\phi_{\mu}$:
\begin{align}
\phi_{\mu}(x) &= \frac{1}{2} \sum_{\nu} 3 (\mathscr{S} \circ g)_{\mu\nu} \ev{\Psi_{\nu}(x)}_{m, \phi}. \label{eq:Hartree-Fock-equation}
\end{align}
This is identical to the saddle-point equations~\eqref{eq:real-saddle-point-eq}, except that the coupling constants $g_{\mu\nu}$ are now replaced by $3 (\mathscr{S} \circ g)_{\mu\nu}$.
The symmetrization naturally arose from the summation over Wick contractions, rendering the coupling constants unique.
The relative factor of $3$ between the two approaches appears because the two Fock contractions are formally of order $1/N$ and thus drop out as $N \to \infty$.
At order $1/N$, there are in addition polarization bubble diagrams that make the large-$N$ expansion sharply distinct from the weak-coupling (Hartree-Fock) expansion.

Which vestigial channel prevails and what is the character of the vestigial transition both depend only on the relative magnitudes of the coupling constants.
Hence, the weak-coupling results agree with SPA only if we symmetrize the $g_{\mu\nu}$ appearing in the saddle-point equations, as we prescribed in the preceding Sec.~\ref{sec:why-symm}.
This weak-coupling cross-verification further supports that our large-$N$ results are informative for strong coupling.
In contrast, in the traditional large-$N$ approach of extending, e.g., the spin $\SO(3)$ subgroup to $\SO(N)$, quartic redundancy relations are employed to make $g_{\mu\nu}$ vanish in spin-nontrivial channels.
Then not only are the $g_{\mu\nu}$ different from the symmetrized ones, biasing the results, but we also systematically neglect the possibility of spin-nontrivial vestigial order, as we further discuss in Appendix~\ref{sec:spin-like-OPs}.

Given that the symmetrized saddle-point equation~\eqref{eq:real-saddle-point-eq} is, up to a factor, identical to the Hartree-Fock self-consistency equation~\eqref{eq:Hartree-Fock-equation}, one may wonder what is the advantage of using large-$N$ over Hartree-Fock.
The key point is that vestigial order generally requires the coupling $g_{\mu\nu}$ to be larger than a threshold value and $N$ to be finite.
Thus, in both methods, we must extrapolate beyond where the method is exact (small coupling or $N \to \infty$) to access the regime of vestigial order.
However, vestigial order is present at infinite $N$, but can be absent at weak coupling because it is not guaranteed for infinitesimal couplings.
This suggests that it is advantageous to start from $N \to \infty$ and then extrapolate to finite $N$.
We also emphasize that the agreement between large-$N$ and Hartree-Fock is not a feature inherent to any large-$N$ method~\cite{Hecker2023}, but rather a result of our proposed unbiased large-$N$ approach.

\section{Applications to density-wave and superconducting instabilities} \label{sec:applications}
As reviewed in the introduction, there have been many theoretical studies of vestigial order~\cite{Fradkin2015, Fernandes2019,
Nie2014, YuxuanWang2014, Venderbos2016hex1, Nie2017, Fernandes2019,
Golubovic1988, Fang2006, Fang2008, Xu2008, Qi2009, Millis2010, Fernandes2012, Chern2012, Venderbos2016hex2, Fernandes2016, Nie2017, Zhang2017, Christensen2018, Fernandes2019,
Fischer2016, Hecker2018, Fernandes2019, Jian2021, Fernandes2021, Garaud2022, Hecker2023, How2023, How2024, Poduval2024, Lin2025, Verghis2025, Maccari2025, Zou2025, Gao2026,
Himeda2002, Berg2007, Agterberg2008, Berg2009-PRB, Berg2009-Nat, Berg2009, Loder2011, Wang2015-PRL, Fradkin2015, Wu2024, Pan2024, Huecker2026,
Golubovic1988, Kivelson1998, Fang2006, Fang2008, Xu2008, Qi2009, Millis2010, Fernandes2012, Nie2014, Fernandes2016, Nie2017, Zhang2017, Christensen2018, Borisov2019, OBrien2026,
Kivelson1998, Loison2000, Weber2003, Kamiya2011, Bojesen2014, Jeevanesan2015, Roy2015, ZhentaoWang2017, Willa2020, Takahashi2020, Liu2023, Volovik2024, Liu2024, Francini2024,
Gopalakrishnan2017, Yu2025,
Herland2010}.
The most commonly studied primary orders have been CDWs~\cite{Nie2014, YuxuanWang2014, Venderbos2016hex1, Nie2017, Fernandes2019}, SDWs~\cite{Golubovic1988, Fang2006, Fang2008, Xu2008, Qi2009, Millis2010, Fernandes2012, Chern2012, Venderbos2016hex2, Fernandes2016, Nie2017, Zhang2017, Christensen2018, Fernandes2019}, multicomponent SCs~\cite{Fischer2016, Hecker2018, Fernandes2019, Jian2021, Fernandes2021, Garaud2022, Hecker2023, How2023, How2024, Poduval2024, Lin2025, Verghis2025, Maccari2025, Zou2025, Gao2026}, and pair-density waves~\cite{Himeda2002, Berg2007, Agterberg2008, Berg2009-PRB, Berg2009-Nat, Berg2009, Loder2011, Wang2015-PRL, Fradkin2015, Wu2024, Pan2024, Huecker2026}, with nematic order being one of the most frequently studied vestigial instabilities~\cite{Golubovic1988, Kivelson1998, Fang2006, Fang2008, Xu2008, Qi2009, Millis2010, Fernandes2012, Nie2014, Fernandes2016, Nie2017, Zhang2017, Christensen2018, Borisov2019, OBrien2026}.
Yet, vestigial order as a phenomenon is applicable to any parent order with a multicomponent order parameter (OP), including, e.g., ferroelectrics, loop currents, and altermagnets.
Moreover, a primary multicomponent OP in general supports multiple competing vestigial orders associated with unconventional symmetry-breaking phases, such as charge-$4e$ order~\cite{Berg2009-Nat} or spin-current order~\cite{Fernandes2016}.

Our goal in this section is to apply the modified large-$N$ approach, developed in Sec.~\ref{sec:general-teo}, to several cases of interest in order to not only to showcase its features, but, more importantly, to shed new light on the fate of systems with multiple competing vestigial instabilities.
This regime of competing vestigial orders is where the traditional large-$N$ approach faces difficulties.
The outcome of these analyses, as anticipated in the previous section, is that the saddle-point equations obtained from the traditional large-$N$ approach still apply, provided that we replace all coupling constants with their symmetrized variants.
In light of this, some of the results obtained here with the modified large-$N$ approach recover results previously obtained via the traditional large-$N$ approach, with one of the main differences being that the condition for vestigial order is more stringent in the former, as there is a threshold value that the coupling constants must meet in order for vestigial order to take place; see Eq.~\eqref{eq:threshold-effect}.

We focus on three distinct types of primary instabilities.
The first, a two-component commensurate CDW, is chosen for its simplicity in order to illustrate the formalism and the aforementioned threshold effect.
The second example of multicomponent SC is motivated by the fact that cubic systems allow for three-component pairing, whose vestigial orders have been relatively unexplored.
Interestingly, we show that cubic three-component SC is the first known case to give vestigial charge-$4e$ SC as the leading instability (Fig.~\ref{fig:3D-SC-phase-diag}).
Previously, vestigial charge-$4e$ SC was at best found to be degenerate with nematic order~\cite{Fernandes2021}.
The third and last example is that of vestigial order due to $M$-point SDWs in hexagonal systems.
Although this problem was analyzed before using the traditional large-$N$ approach~\cite{Chern2012, Venderbos2016hex2, Fernandes2019}, the possibility of vestigial order in the spin-quadrupole channels has not been fully investigated.
We find that spin-vector loop-current vestigial order can arise (Fig.~\ref{fig:M-SDW-phase-diag}), whereas the spin-quadrupole instabilities are generally subleading.

We extensively employ group theory~\cite{Dresselhaus2007} throughout this section.
Space group irreducible representations (irreps) are labeled according to their high-symmetry momenta, following Miller and Love~\cite{MillerLove1967, CDML1979}, with $\pm$ superscripts denoting parity under space-inversion and the $m$ prefix, if present, denoting oddness under time-reversal (e.g., $m\Gamma_2^{+}$ is odd under time-reversal, even under spatial-inversion, and has momentum $\vb{k} = \vb{0}$).
We use the irrep bases of the \texttt{ISOTROPY} software package~\cite{ISOTROPY}, which conveniently has modules for constructing transformation matrices and invariants.
In what follows, the lattice constants of the direct primitive lattice are set to unity, $(x, y, z)$ and $(k_x, k_y, k_z)$ denote Cartesian coordinates of the real-space position and momentum, respectively, and in Miller index notation $[hkl] \equiv h \vb{a}_1 + k \vb{a}_2 + l \vb{a}_3$ and $(hkl) \equiv h \vb{b}_1 + k \vb{b}_2 + l \vb{b}_3$ are always relative to the primitive direct ($\vb{a}_i$) and primitive reciprocal ($\vb{b}_i$) lattice vectors ($\vb{a}_i \vdot \vb{b}_j = 2 \pi \Kd_{ij}$).
For simplicity, we work in the regime where the susceptibility is proportional to the identity in the multi-dimensional space of the primary OP irrep, as in Eq.~\eqref{eq:isotropic-sus-limit}, although for completeness we also list the more general forms of the anisotropic susceptibilities.

\subsection{\boldmath{$X$}-point charge-density waves in tetragonal systems} \label{sec:X-CDW}
\begin{figure}
\includegraphics[width=\columnwidth]{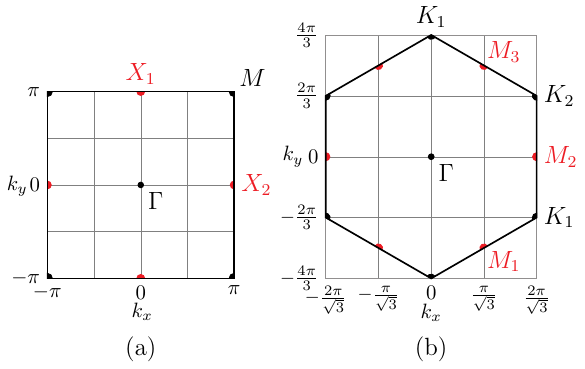}
\caption{The $k_z = 0$ cross-sections of the primitive tetragonal (a) and hexagonal (b) Brillouin zones.
The direct lattice constants have been set to unity.
The primitive basis vectors of the hexagonal lattice are the conventional $[100] \equiv (0,-1,0)$, $[010] \equiv (\sqrt{3}/2,1/2,0)$, and $[001] \equiv (0,0,1)$~\cite{Dresselhaus2007, BradleyCracknell2009, CDML1979}.
The momenta highlighted in red correspond to the wavevectors of the primary orders studied in Sec.~\ref{sec:X-CDW}, for panel (a), and in Sec.~\ref{sec:M-SDW}, for panel (b).} \label{fig:Brillouin}
\end{figure}

We start by studying one of the simplest manifestations of vestigial order, namely, the one arising from CDW instabilities on the tetragonal lattice whose wavevectors are related by a $90^{\circ}$ rotation.
This situation is realized in a broad range of systems, such as cuprates~\cite{Nie2014, YuxuanWang2014, Nie2017}, nickel arsenides~\cite{Souliou2022, Lacmann2023, Collini2023}, and rare-earth tritellurides~\cite{Hu2014}.
For concreteness, we consider a simple tetragonal lattice with space group $P4/mmm$ (\#123).
The corresponding point group is $4/mmm$ or $D_{4h}$, and it is generated by fourfold rotations around the (principal) $z$ axis, twofold rotations around the $x$ axis, twofold rotations around the diagonal $x+y$, and space-inversion (parity).
Its Brillouin zone is shown in Fig.~\ref{fig:Brillouin}(a).

For simplicity, we focus on CDWs with commensurate wavevectors $\vb{k}_{X_1} = (0, \pi, 0)$ and $\vb{k}_{X_2} = (\pi, 0, 0)$.
They correspond to the $X$-point of the Brillouin zone of Fig.~\ref{fig:Brillouin}(a) and, as such, the primary CDW OP $\vb{\eta} = (\eta_1, \eta_2)$ must transform according to an $X$-point irrep of the space group.
Given the commensurability relations $2 \vb{k}_{X_1} \cong 2 \vb{k}_{X_2} \cong \vb{0} \equiv \vb{k}_{\Gamma}$ and $\vb{k}_{X_1} \pm \vb{k}_{X_2} \cong (\pi, \pi, 0) \equiv \vb{k}_{M}$, the bilinear composite vestigial order parameters must transform as space group irreps of the $\Gamma$ and $M$ points.
A list of $\Gamma$-, $X$-, and $M$-point irreps and their basis functions is provided in Table~\ref{tab:tetragonal-irreps}.
Physically, the $X$-point CDW with a two-component OP $\vb{\eta} = (\eta_1, \eta_2)$ corresponds to a modulation of the charge density of the form
\begin{align}
\var{\rho(\vb{x})} &= \eta_1 f_1(\vb{x}) + \eta_2 f_2(\vb{x}),
\end{align}
where the pair of functions $(f_1(\vb{x}) | f_2(\vb{x}))$ transforms according to the same $X$-point irrep as $\vb{\eta}$; see Table~\ref{tab:tetragonal-irreps} for examples of such functions.

\begin{table}
\caption{The irreducible representations (irreps) of the primitive tetragonal space group $P4/mmm$ (\#123) for the high-symmetry points $\Gamma = (0, 0, 0)$, $X \in \{(0\tfrac{1}{2}0) \equiv (0, \pi, 0), \, (\tfrac{1}{2}00) \equiv (\pi, 0, 0)\}$, and $M = (\tfrac{1}{2}\tfrac{1}{2}0) \equiv (\pi, \pi, 0)$, together with the corresponding basis functions.
For the $\Gamma$ irreps, we provide the Mulliken names~\cite{Mulliken1933} in addition to the Miller-Love ones~\cite{MillerLove1967, CDML1979}.
The 2D irrep basis functions have been chosen so that they transform under the physically-irreducible matrices of \texttt{ISOTROPY}~\cite{ISOTROPY}.}
{\renewcommand{\arraystretch}{1.3}
\renewcommand{\tabcolsep}{1.7pt}
\begin{tabular}{lc|lc}
\hline \hline
\multicolumn{4}{c}{tetragonal $P4/mmm$ (\#123) space group} \tabularnewline
\multicolumn{2}{c|}{even-parity irreps} & \multicolumn{2}{c}{odd-parity irreps} \tabularnewline
\hline
$\Gamma_1^{+} (A_{1g})$ & $1$, $x^2 + y^2$, $z^2$ & $\Gamma_1^{-} (A_{1u})$ & $x y z (x^2 - y^2)$ \tabularnewline
$\Gamma_2^{+} (B_{1g})$ & $x^2 - y^2$ & $\Gamma_2^{-} (B_{1u})$ & $x y z$ \tabularnewline
$\Gamma_3^{+} (A_{2g})$ & $x y (x^2 - y^2)$ & $\Gamma_3^{-} (A_{2u})$ & $z$ \tabularnewline
$\Gamma_4^{+} (B_{2g})$ & $x y$ & $\Gamma_4^{-} (B_{2u})$ & $z (x^2 - y^2)$ \tabularnewline
$\Gamma_5^{+} (E_g)$ & $(y z | - x z)$ & $\Gamma_5^{-} (E_u)$ & $(x | y)$ \tabularnewline
\hline
$X_1^{+}$ & $(c_y | c_x)$, $(s_y y | s_x x)$ & $X_1^{-}$ & $(s_y x z | - s_x y z)$ \tabularnewline
$X_2^{+}$ & $(s_y x | s_x y)$ & $X_2^{-}$ & $(c_y z | - c_x z)$ \tabularnewline
$X_3^{+}$ & $(s_y z | s_x z)$ & $X_3^{-}$ & $(c_y x | - c_x y)$ \tabularnewline
$X_4^{+}$ & $(c_y x z | c_x y z)$ & $X_4^{-}$ & $(s_y | - s_x)$, $(c_y y | - c_x x)$ \tabularnewline
\hline
$M_1^{+}$ & $c_x c_y$ & $M_1^{-}$ & $s_x s_y z (x^2 - y^2)$ \tabularnewline
$M_2^{+}$ & $c_x c_y (x^2 - y^2)$ & $M_2^{-}$ & $s_x s_y z$ \tabularnewline
$M_3^{+}$ & $s_x s_y (x^2 - y^2)$ & $M_3^{-}$ & $c_x c_y z$ \tabularnewline
$M_4^{+}$ & $s_x s_y$ & $M_4^{-}$ & $c_x c_y z (x^2 - y^2)$ \tabularnewline
$M_5^{+}$ & $(s_{d-} z | s_{d+} z)$ & $M_5^{-}$ & $(s_{d-} | - s_{d+})$ \tabularnewline
\hline
\multicolumn{4}{c}{\footnotesize Abbreviations: $c_x = \cos(\pi x)$, $s_x = \sin(\pi x)$, $c_y = \cos(\pi y)$,} \tabularnewline[-2pt]
\multicolumn{4}{c}{\footnotesize $s_y = \sin(\pi y)$, $s_{d\pm} = \sin\mleft(\pi (x \pm y)\mright) = s_x c_y \pm c_x s_y$.} \tabularnewline
\hline \hline
\end{tabular}}
\label{tab:tetragonal-irreps}
\end{table}

For a CDW belonging to any of the eight $X$-point irreps, the most general symmetry-allowed action has the form
\begin{align}
\actS = \frac{1}{2} \int_q \vb{\eta}^{\intercal} \chi_q^{-1} \vb{\eta} + \frac{1}{8} \int_x \sum_{\mu = x, z} g_{\mu} (\vb{\eta}^{\intercal} \Pauli_{\mu} \vb{\eta})^2,
\end{align}
where $\Pauli_{\mu}$ are Pauli matrices and
\begin{align}
\chi_q^{-1} &= (r + q^2 + \kappa_z q_z^2) \one + \var{\kappa_{\perp}} (q_x^2 - q_y^2) \Pauli_z.
\end{align}
Cubic terms are forbidden by the $\vb{a}_1 + \vb{a}_2$ translation symmetry which acts like $\vb{\eta} \mapsto - \vb{\eta}$ on $X$-point irreps.
A quartic $(\vb{\eta}^{\intercal} \vb{\eta})^2$ term is symmetry-allowed, but we have eliminated it without loss of generality by using the Fierz identity~\eqref{eq:intro-Fierz}.
If we assume an isotropic susceptibility $\var{\kappa_{\perp}} = \kappa_z = 0$ and exploit the Fierz identity~\eqref{eq:intro-Fierz} to replace $(\vb{\eta}^{\intercal} \Pauli_x \vb{\eta})^2$ with $(\vb{\eta}^{\intercal} \vb{\eta})^2$, we recover the action~\eqref{eq:intro-action} discussed in the introduction.
The action is stable (bounded from below) and describes a second-order CDW transition only when $g_x$ and $g_z$ are positive, which we henceforth assume.

There are three possible composite bilinears for this two-component real OP, as also discussed in Ref.~\cite{Fernandes2019}:
the trivial one, $\vb{\eta}^{\intercal} \Pauli_0 \vb{\eta} \sim \Gamma_1^{+}$, which is always finite since $\ev{\vb{\eta}^{\intercal} \Pauli_0 \vb{\eta}} = \ev{\eta_1^2 + \eta_2^2} > 0$ and therefore does not give rise to a vestigial order, and two nontrivial channels associated with symmetry-distinct vestigial phases.
The bilinear $\vb{\eta}^{\intercal} \Pauli_z \vb{\eta}$ transforms as $\Gamma_2^{+}$ for all $X$-point irreps and it therefore describes a $d_{x^2-y^2}$-wave nematic vestigial phase.
On the other hand, $\vb{\eta}^{\intercal} \Pauli_x \vb{\eta}$ transforms as $M_1^{+}$ for $\vb{\eta} \sim X_{1,2}^{\pm}$ and as $M_4^{+}$ for $\vb{\eta} \sim X_{3,4}^{\pm}$.
The $\vb{\eta}^{\intercal} \Pauli_x \vb{\eta}$ vestigial order is thus an $M$-point CDW that may ($M_4^{+}$) or may not ($M_1^{+}$) break the twofold rotation symmetry around $x$ and $y$.
Either way, the degree of translation symmetry-breaking (doubling of the unit cell) is smaller than in the corresponding parent $X$-point double-$Q$ CDW, which quadruples the unit cell.

Assuming an isotropic susceptibility, $\var{\kappa_{\perp}} = \kappa_z = 0$, we recover Eq.~\eqref{eq:isotropic-sus-limit} of Sec.~\ref{sec:SPA} with $w = 2$.
As a result, vestigial order takes place in the channel with the most negative coupling constant $g_{\mu}$.
Moreover, as we showed in Sec.~\ref{sec:why-symm}, this is determined by the \emph{symmetrized} coupling constants.
By setting $g_{\mu\nu} = \diag(0, g_x, g_z)$ and $\Gamma_{\mu} = (\Pauli_0, \Pauli_x, \Pauli_z)$ in Eq.~\eqref{eq:symmetrized-gmunu}, we obtain the symmetrized coupling constants
\begin{align}
\begin{aligned}
(\mathscr{S} \circ g)_x^{\text{CDW}} &= \frac{2 g_x - g_z}{3} = (\mathscr{S} \circ g)_0 \cdot \mleft(\tfrac{1}{2} + \lambda\mright), \\
(\mathscr{S} \circ g)_z^{\text{nem}} &= \frac{- g_x + 2 g_z}{3} = (\mathscr{S} \circ g)_0 \cdot \mleft(\tfrac{1}{2} - \lambda\mright),
\end{aligned}
\end{align}
where we have expressed them in terms of $(\mathscr{S} \circ g)_0 = (g_x + g_z) / 3 > 0$ and a dimensionless ratio $\lambda = 3 (g_x - g_z) / [2 (g_x + g_z)]$.
The stable parameter space corresponds to $g_x > 0$ and $g_z > 0$, that is $\lambda \in \langle - \tfrac{3}{2}, \tfrac{3}{2} \rangle$.

\begin{figure}
\includegraphics[width=\columnwidth]{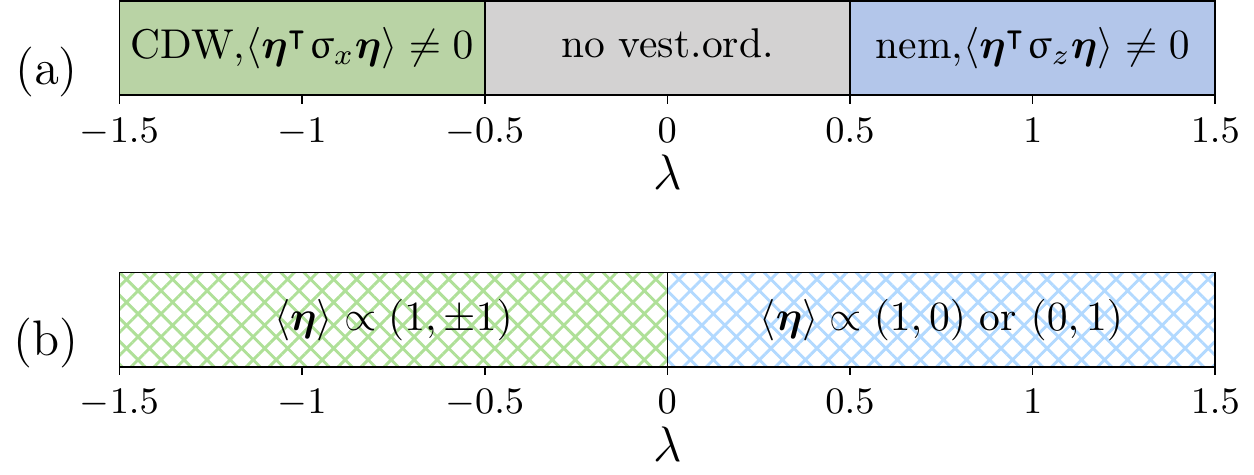}
\caption{Phase diagrams for $X$-point CDWs on the tetragonal lattice showing (a) the vestigial order (if any) that can onset above $T_c$ ($r > 0$) and (b) the primary OP configuration that minimizes the mean-field free energy below $T_c$ ($r < 0$), both as a function of the dimensionless ratio $\lambda = 3 (g_x - g_z) / [2 (g_x + g_z)] \in \langle - \tfrac{3}{2}, \tfrac{3}{2} \rangle$.
$\vb{\eta}^{\intercal} \Pauli_z \vb{\eta} \in \Gamma_2^{+}$ for all $X$-point irreps, while $\vb{\eta}^{\intercal} \Pauli_x \vb{\eta}$ belongs to $M_1^{+}$ for $\vb{\eta} \sim X_{1,2}^{\pm}$ and to $M_4^{+}$ for $\vb{\eta} \sim X_{3,4}^{\pm}$.
The former describes a $d_{x^2-y^2}$-wave nematic vestigial order (nem in panel~(a)), while the latter describes an $M$-point CDW vestigial order (CDW in panel~(a)).} \label{fig:X-CDW-phase-diag}
\end{figure}

In Fig.~\ref{fig:X-CDW-phase-diag}, we show the phase diagram of both the leading vestigial instability as a function of $\lambda$, in panel~(a), as well as the phase diagram of the primary CDW order, in panel~(b).
The primary CDW OP condensed into either the double-$Q$ $\ev{\vb{\eta}} \propto (1, \pm 1)$ or the single-$Q$ $\ev{\vb{\eta}} \propto (1, 0)/(0, 1)$ configuration.
These two phase diagrams are consistent in the sense that $\ev{\vb{\eta}} \propto (1, \pm 1)$ is characterized by $\ev{\vb{\eta}}^{\intercal} \Pauli_x \ev{\vb{\eta}} \neq 0$ and $\ev{\vb{\eta}}^{\intercal} \Pauli_z \ev{\vb{\eta}} = 0$, whereas $\ev{\vb{\eta}} \propto (1, 0)$ or $(0, 1)$ supports $\ev{\vb{\eta}}^{\intercal} \Pauli_x \ev{\vb{\eta}} = 0$ and $\ev{\vb{\eta}}^{\intercal} \Pauli_z \ev{\vb{\eta}} \neq 0$.
For $\lambda = 0$, the $\Pauli_x$ and $\Pauli_z$ channels are degenerate and repulsive.
Although an infinitesimal $\lambda$ is enough to choose the free-energy minimum of the primary CDW order below $T_c$, vestigial order is only induced if $\abs{\lambda} > \tfrac{1}{2}$.
This is the threshold effect we discussed in Eq.~\eqref{eq:threshold-effect}.
Thus we see that near the degeneracy point of the primary OP phase diagram, the competition between different vestigial instabilities is strong enough to completely suppress the condensation of any vestigial OP.
This is in contrast to the traditional large-$N$ approach, for which vestigial order exists for any infinitesimal $\lambda$.

We emphasize that the regions highlighted in the vestigial phase diagram of Fig.~\ref{fig:X-CDW-phase-diag}(a) allow for vestigial order, but do not necessarily imply it~\cite{Fernandes2012}.
This is because, as explained in the introduction, if vestigial order onsets as a first-order transition, it may simultaneously triggers a finite primary OP.
To determine what precisely happens within these regions of Fig.~\ref{fig:X-CDW-phase-diag}(a), one needs to go beyond Eq.~\eqref{eq:isotropic-sus-limit} and analyze the full nonlinear saddle-point equations.
For this particular form of the action, one can take advantage of the results of Ref.~\cite{Fernandes2012} with the bare coupling constants replaced by the symmetrized coupling constants.
Thus, the results shown in Fig.~\ref{fig:vestigial-transitions} still apply if we replace $g_z / g_0$ with $(\mathscr{S} \circ g)_x / (\mathscr{S} \circ g)_0$ or $(\mathscr{S} \circ g)_z / (\mathscr{S} \circ g)_0$, depending on the region.
Including the susceptibility anisotropies $\var{\kappa_{\perp}}, \kappa_z$ does not change the results qualitatively, unless they are large enough to change the effective dimensionality of the system~\cite{Fernandes2012}.

\subsection{Multicomponent superconductivity in cubic systems} \label{sec:cubic-SC}
Vestigial superconducting orders have been primarily investigated for pairing states described by two-component OPs.
Such pairing states can be realized by singlet or triplet unconventional pairing in hexagonal and rhombohedral systems like twisted bilayer graphene~\cite{Fernandes2021, Jian2021, Poduval2024} and doped \ce{Bi2Se3}~\cite{Hecker2018}, or by triplet pairing in tetragonal systems~\cite{Fischer2016, Fernandes2019}.
Here, we apply our modified large-$N$ approach to a less explored situation: vestigial superconducting orders stemming from unconventional singlet or triplet pairing in cubic systems, which support three-component SC OPs~\cite{SigristUeda1991}.
Interestingly, experiments have recently reported signatures consistent with multi-component superconductivity in the cubic topological semimetal \ce{CaSn3}~\cite{Siddiquee2022}.

\begin{table*}
\caption{The irreducible representations (irreps) of the cubic point group $O_h$ ($m\bar{3}m$) and their basis functions as constructed from momentum $(k_x | k_y | k_z) \sim \Gamma_4^{-}$ and spin $(\Pauli_x | \Pauli_y | \Pauli_z) \sim \Gamma_4^{+}$.
Parity is denoted by the $\pm$ superscripts in the Miller-Love notation~\cite{MillerLove1967, CDML1979} and by the $g/u$ subscripts in Mulliken notation~\cite{Mulliken1933}.
The two-dimensional and three-dimensional irrep basis functions have been chosen so that they transform under the physically-irreducible matrices of \texttt{ISOTROPY}~\cite{ISOTROPY}.}
{\renewcommand{\arraystretch}{1.3}
\renewcommand{\tabcolsep}{3.0pt}
\begin{tabular}{lc|lc}
\hline \hline
\multicolumn{4}{c}{cubic $O_h$ ($m\bar{3}m$) point group} \tabularnewline
\multicolumn{2}{c|}{even-parity irreps} & \multicolumn{2}{c}{odd-parity irreps} \tabularnewline
\hline
$\Gamma_1^{+} (A_{1g})$ & $1$, $k_x^2 + k_y^2 + k_z^2$ & $\Gamma_1^{-} (A_{1u})$ & $k_x \Pauli_x + k_y \Pauli_y + k_z \Pauli_z$ 
\tabularnewline
$\Gamma_2^{+} (A_{2g})$ & $k_x^4 (k_y^2 - k_z^2) + k_y^4 (k_z^2 - k_x^2) + k_z^4 (k_x^2 - k_y^2)$ & $\Gamma_2^{-} (A_{2u})$ & {\footnotesize $k_x k_y (k_x \Pauli_y - k_y \Pauli_x) + k_y k_z (k_y \Pauli_z - k_z \Pauli_y) + k_z k_x (k_z \Pauli_x - k_x \Pauli_z)$} 
\tabularnewline
$\Gamma_3^{+} (E_g)$ & $\big(k_x^2 + k_y^2 - 2 k_z^2 \big| \sqrt{3} (k_x^2 - k_y^2)\big)$ & $\Gamma_3^{-} (E_u)$ & $\big(\sqrt{3} (k_x \Pauli_x - k_y \Pauli_y) \big| 2 k_z \Pauli_z - k_x \Pauli_x - k_y \Pauli_y\big)$ 
\tabularnewline
$\Gamma_4^{+} (T_{1g})$ & $\big(k_y k_z (k_y^2 - k_z^2) \big| k_z k_x (k_z^2 - k_x^2) \big| k_x k_y (k_x^2 - k_y^2)\big)$ & $\Gamma_4^{-} (T_{1u})$ & $(k_x | k_y | k_z)$, $(k_y \Pauli_z - k_z \Pauli_y | k_z \Pauli_x - k_x \Pauli_z | k_x \Pauli_y - k_y \Pauli_x)$ \tabularnewline
$\Gamma_5^{+} (T_{2g})$ & $(k_x k_y | k_y k_z | k_z k_x)$ & $\Gamma_5^{-} (T_{2u})$ & $(k_x \Pauli_y + k_y \Pauli_x | k_y \Pauli_z + k_z \Pauli_y | k_z \Pauli_x + k_x \Pauli_z)$ 
\tabularnewline
\hline \hline
\end{tabular}}
\label{tab:cubic-irreps}
\end{table*}

Since SC is a homogeneous zero-momentum order, we only need to specify the point group, which we assume to be the full cubic group $O_h$ (i.e., $m\bar{3}m$).
This group is generated by fourfold rotations around the $z$ axis, threefold rotations around the space diagonal $x+y+z$, twofold rotations around the in-plane diagonal $x+y$, and space-inversion (parity).
As shown in Table~\ref{tab:cubic-irreps}, this group allows multi-component SC OPs that transform as one of two possible two-dimensional irreps $\Gamma_3^{\pm}$ or as one of four possible three-dimensional irreps $\Gamma_4^{\pm}$ and $\Gamma_5^{\pm}$.
For concreteness, we focus on $d$-wave $\Gamma_3^{+}$ and $p$-wave $\Gamma_4^{-}$ Cooper pairing.
The analysis for $\Gamma_3^{-}$ pairing is identical to that for $\Gamma_3^{+}$, as is the analysis for $\Gamma_4^{+}$ and $\Gamma_5^{\pm}$ pairing identical to that for $\Gamma_4^{-}$, in the sense that the symmetry-allowed actions agree between these irreps and that the possible vestigial channels have the same symmetries.
That said, if one employs the transformation matrices of \texttt{ISOTROPY}~\cite{ISOTROPY}, a change of basis is needed to cast the action and vestigial OPs in the same form.
In the case of odd-parity spin-triplet pairing, we shall assume that the direction of the $\vb{d}$-vector is pinned by spin-orbit coupling.
Thus, the odd-parity irreps specify the transformation rules under operations that act simultaneously in orbital and spin space.
Physically, SC with a multi-component OP $\vb{\eta} = (\eta_1, \eta_2, \ldots)$ entails a pairing term in the Bogoliubov-de~Gennes Hamiltonian of the form
\begin{align}
\Haml_{\Delta} &= \sum_{\vb{k} a} c_{\vb{k}}^{\dag} \eta_a f_a(\vb{k}) (\iu \uptau_y) \big[c_{-\vb{k}}^{\dag}\big]^{\intercal} + \Hc,
\end{align}
where the functions $(f_1(\vb{k}) | f_2(\vb{k}) | \cdots)$ transform according to the same irrep as $\vb{\eta}$.
Examples of such functions are given in Table~\ref{tab:cubic-irreps}.
Here $c_{\vb{k}}^{\dag} = (c_{\vb{k} \uparrow}^{\dag}, c_{\vb{k} \downarrow}^{\dag})$ are fermionic creation operators and Pauli matrices $\uptau_{\mu}$ act on spins.

\subsubsection{Two-component superconductivity} \label{sec:2D-SC}
The most general symmetry-allowed action for a $(d_{z^2}, d_{x^2-y^2})$-wave SC on the cubic lattice whose order parameter $\vb{\eta} = (\eta_1, \eta_2)$ belongs to the $\Gamma_3^{+}$ irrep is
\begin{align}
\actS &= \int_q \vb{\eta}^{\dag} \chi_q^{-1} \vb{\eta} + \frac{1}{2} \int_x \sum_{\mu = x, y, z} g_{\mu} (\vb{\eta}^{\dag} \Pauli_{\mu} \vb{\eta})^2,
\end{align}
where $g_x = g_z$ due to the cubic symmetry and
\begin{align}
\begin{aligned}
\chi_q^{-1} &= (r + q^2) \one \\
&\phantom{=} + \kappa \mleft[(q_x^2 + q_y^2 - 2 q_z^2) \Pauli_z - \sqrt{3} (q_x^2 - q_y^2) \Pauli_x\mright].
\end{aligned}
\end{align}
Here a quartic $\propto (\vb{\eta}^{\dag} \vb{\eta})^2$ term is also allowed by symmetry, but we can always eliminate it by exploiting the Fierz identity $(\vb{\eta}^{\dag} \vb{\eta})^2 = \sum_{\mu = x, y, z} (\vb{\eta}^{\dag} \Pauli_{\mu} \vb{\eta})^2$.
Although $g_x = g_z$, it is instructive to consider the more general $g_x \neq g_z$ case, which can be used to describe two-component SC in tetragonal systems~\cite{Hecker2023}.
To ensure stability (i.e., that the action is bounded from below), all three $g_{\mu}$ must be positive.

In total, there are four possible nontrivial vestigial channels.
The particle-hole bilinear $\vb{\eta}^{\dag} \Pauli_y \vb{\eta}$ transforms as $m\Gamma_2^{+}$ and it therefore describes a $d$-wave altermagnetic vestigial phase~\cite{Smejkal2022, Smejkal2022-emerging, Fernandes2024} that is invariant under a combination of time-reversal symmetry and a twofold rotation around an in-plane diagonal.
The vestigial order associated with the pair of particle-hole bilinears $(\vb{\eta}^{\dag} \Pauli_z \vb{\eta} | - \vb{\eta}^{\dag} \Pauli_x \vb{\eta}) \sim \Gamma_3^{+}$ is a $d$-wave nematic.
In the particle-particle sector, $\vb{\eta}^{\intercal} \Pauli_0 \vb{\eta}$ and its complex conjugate transform trivially under all point-group operations.
They thus represent $s$-wave charge-$4e$ order, since they carry twice the charge of the parent SC.
The pair of particle-particle bilinears $(\vb{\eta}^{\intercal} \Pauli_z \vb{\eta} | - \vb{\eta}^{\intercal} \Pauli_x \vb{\eta}) \sim \Gamma_3^{+}$, on the other hand, describe a more exotic vestigial $d$-wave charge-$4e$ order, similar to those that emerge in the case of two-component SC in hexagonal systems~\cite{Hecker2023}.

\begin{figure*}
\includegraphics[width=\textwidth]{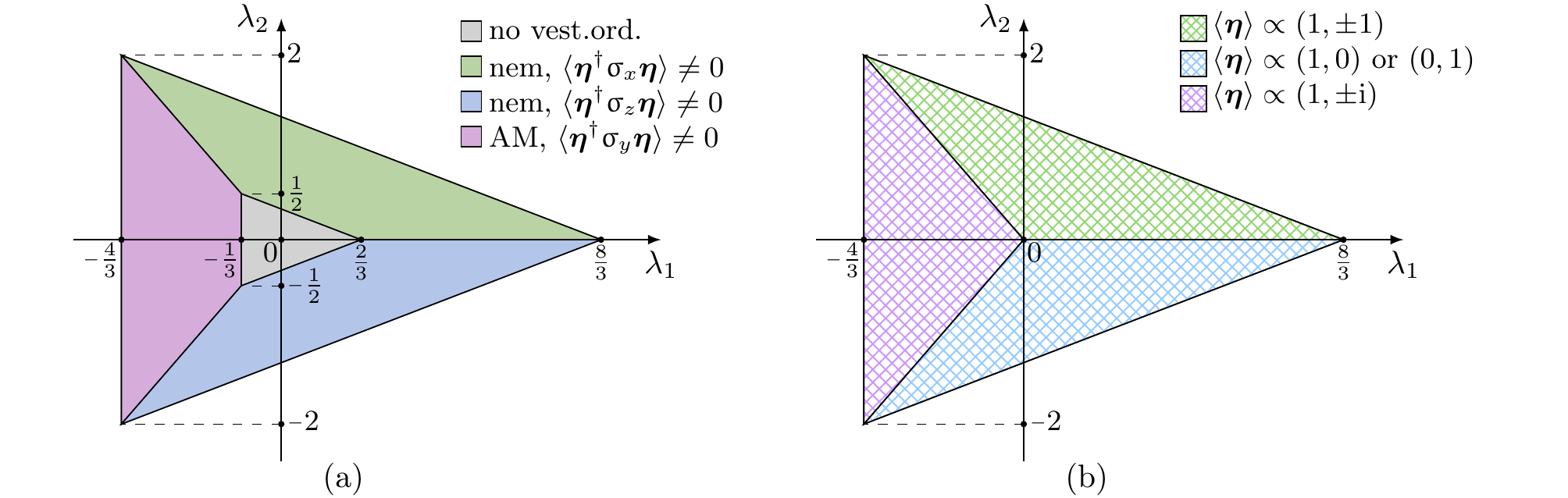}
\caption{Phase diagrams for a two-component $(d_{z^2}, d_{x^2-y^2})$-wave SC state in a cubic system showing (a) the leading vestigial instability and (b) the SC OP configuration in the primary ordered phase, both as a function of the dimensionless ratios $\lambda_1 = 4 (2 g_y - g_x - g_z) / [3 (g_x + g_y + g_z)]$ and $\lambda_2 = 2 (g_z - g_x) / (g_x + g_y + g_z)$.
The case of cubic systems corresponds to $\lambda_2 = 0$.
Outside of the colored triangles, the action is unbounded from below and thus unstable.
Physically, $(\vb{\eta}^{\dag} \Pauli_z \vb{\eta} | - \vb{\eta}^{\dag} \Pauli_x \vb{\eta}) \sim \Gamma_3^{+}$ is a $d$-wave nematic (nem in panel~(a)) whose components are degenerate only when $\lambda_2 = 0$, whereas $\vb{\eta}^{\dag} \Pauli_y \vb{\eta} \sim m\Gamma_2^{+}$ is a $d$-wave altermagnet (AM in panel~(a)).} \label{fig:2D-SC-phase-diag}
\end{figure*}

The extension of the formalism of Sec.~\ref{sec:general-teo} to $\Ugp(1)$-symmetric complex OPs is presented in detail in App.~\ref{sec:complex-OPs}.
To determine the symmetrized vestigial coupling constants, we plug $g_{\mu\nu} = \diag(0, g_x, g_z, g_y)$ and $\gamma_{\mu} = (\Pauli_0, \Pauli_x, \Pauli_z, \Pauli_y)$ into Eq.~\eqref{eq:symmetrized-U1-gmunu} of App.~\ref{sec:complex-OPs}, finding that
\begin{align}
\begin{aligned}
\frac{(\mathscr{S} \circ \tilde{g})_{x}^{\text{nem}}}{(\mathscr{S} \circ \tilde{g})_{0}} &= \frac{3 g_x - g_y - g_z}{6 (\mathscr{S} \circ \tilde{g})_{0}} = \tfrac{1}{3} - \tfrac{\lambda_1}{2} - \lambda_2, \\
\frac{(\mathscr{S} \circ \tilde{g})_{z}^{\text{nem}}}{(\mathscr{S} \circ \tilde{g})_{0}} &= \frac{- g_x - g_y + 3 g_z}{6 (\mathscr{S} \circ \tilde{g})_{0}} = \tfrac{1}{3} - \tfrac{\lambda_1}{2} + \lambda_2, \\
\frac{(\mathscr{S} \circ \tilde{g})_{y}^{\text{AM}}}{(\mathscr{S} \circ \tilde{g})_{0}} &= \frac{- g_x + 3 g_y - g_z}{6 (\mathscr{S} \circ \tilde{g})_{0}} = \tfrac{1}{3} + \lambda_1, \\
\frac{(\mathscr{S} \circ \tilde{g})_{0}^{s\text{-}4e}}{(\mathscr{S} \circ \tilde{g})_{0}} &= \frac{g_x - g_y + g_z}{6 (\mathscr{S} \circ \tilde{g})_{0}} = \tfrac{1}{3} - \tfrac{\lambda_1}{2}, \\
\frac{(\mathscr{S} \circ \tilde{g})_{x}^{d\text{-}4e}}{(\mathscr{S} \circ \tilde{g})_{0}} &= \frac{g_x + g_y - g_z}{6 (\mathscr{S} \circ \tilde{g})_{0}} =  \tfrac{1}{3} + \tfrac{\lambda_1}{4} - \tfrac{\lambda_2}{2}, \\
\frac{(\mathscr{S} \circ \tilde{g})_{z}^{d\text{-}4e}}{(\mathscr{S} \circ \tilde{g})_{0}} &= \frac{- g_x + g_y + g_z}{6 (\mathscr{S} \circ \tilde{g})_{0}} = \tfrac{1}{3} + \tfrac{\lambda_1}{4} + \tfrac{\lambda_2}{2},
\end{aligned} \label{eq:aux_cubic}
\end{align}
where $(\mathscr{S} \circ \tilde{g})_{0} = (g_x + g_y + g_z) / 6$ and the dimensionless ratios are given by:
\begin{align}
\begin{aligned}
\lambda_1 &= \frac{4 (2 g_y - g_x - g_z)}{3 (g_x + g_y + g_z)}, \\
\lambda_2 &= \frac{2 (g_z - g_x)}{g_x + g_y + g_z}.
\end{aligned}
\end{align}
Here, the superscript nem stands for nematic, AM for altermagnetic, and $4e$ for charge-$4e$ superconductivity, which can be either $s$-wave ($\Gamma_1^{+}$) or $d$-wave ($\Gamma_3^{+}$).
The action is stable when $\lambda_1 > - 4/3$ and $\abs{\lambda_2} < (8 - 3 \lambda_1) / 6$, whereas $\lambda_2 = 0$ represents the cubic $g_x = g_z$ subspace.

The phase diagram of the vestigial orders, shown in Fig.~\ref{fig:2D-SC-phase-diag}(a), is determined in a straightforward way by finding the most negative coupling constant as a function of $\lambda_1$ and $\lambda_2$.
Evidently, from Eqs.~\eqref{eq:aux_cubic} we see that for generic $\lambda_1$ and $\lambda_2$, only the nematic and altermagnetic vestigial instabilities are leading instabilities.
$s$-wave and $d$-wave charge-$4e$ order appear only as subleading instabilities.
However, on the $\lambda_2 = 0$ line, there is a hidden symmetry that enforces the $s$-wave charge-$4e$ channel to become exactly degenerate with the nematic channel, similarly to what was obtained previously for the case of two-component pairing in hexagonal lattices~\cite{Fernandes2021}.
Analogous statements hold on the lines that connect the origin to the other two vertices of the outer triangle of Fig.~\ref{fig:2D-SC-phase-diag}.
As in the previous case of CDW order (Sec.~\ref{sec:X-CDW}), the primary SC OP configurations that minimize the mean-field action in the $(\lambda_1, \lambda_2)$ parameter space, shown in Fig.~\ref{fig:2D-SC-phase-diag}(b), are consistent with the leading vestigial instability in the same parameter region, shown in Fig.~\ref{fig:2D-SC-phase-diag}(a), since we have $\vb{\eta}^{\dag} \Pauli_{\mu} \vb{\eta}$ as a vestigial order only when $\ev{\vb{\eta}}^{\dag} \Pauli_{\mu} \ev{\vb{\eta}} \neq 0$ in the ordered state.
Lastly, we emphasize again the threshold effect: in order for a vestigial instability to occur, the parameters need to be a finite distance away from the origin $\lambda_1 = \lambda_2 = 0$, where all vestigial channels are degenerate.

\begin{figure*}
\includegraphics[width=\textwidth]{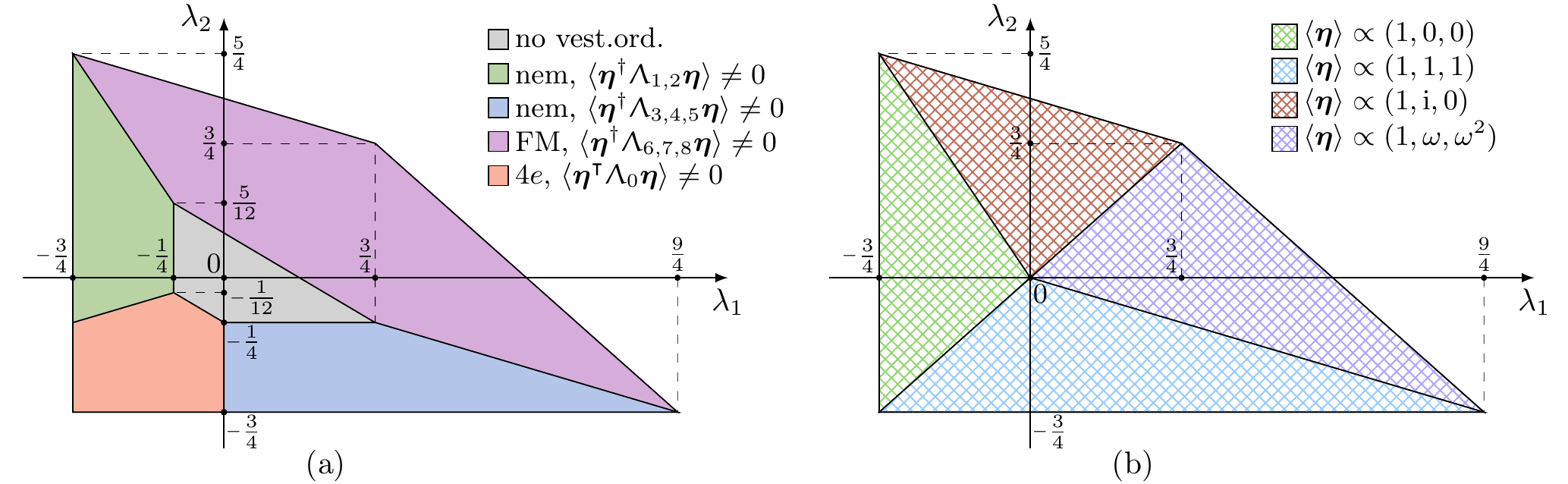}
\caption{Phase diagrams for the $(p_x, p_y, p_z)$-wave three-component SC order on the cubic lattice showing (a) the leading vestigial instability and (b) the SC OP configuration in the primary ordered phase, both as a function of $\lambda_1 = 9 (2 g_{\Gamma_3} - g_{\Gamma_4} - g_{\Gamma_5}) / [4 (2 g_{\Gamma_3} + 3 g_{\Gamma_4} + 3 g_{\Gamma_5})]$ and $\lambda_2 = - 3 (2 g_{\Gamma_3} + 3 g_{\Gamma_4} - 5 g_{\Gamma_5}) / [4 (2 g_{\Gamma_3} + 3 g_{\Gamma_4} + 3 g_{\Gamma_5})]$.
Outside of the colored regions, the action is unstable.
Physically, $\vb{\eta}^{\dag} \GellMann_{1,2} \vb{\eta} \sim \Gamma_3^{+}$ and $\vb{\eta}^{\dag} \GellMann_{3,4,5} \vb{\eta} \sim \Gamma_5^{+}$ are $d$-wave nematics (nem in panel~(a)), $\vb{\eta}^{\dag} \GellMann_{6,7,8} \vb{\eta} \sim m\Gamma_4^{+}$ is a ferromagnet (FM in panel~(a)), and $\vb{\eta}^{\intercal} \GellMann_{0} \vb{\eta} \sim \Gamma_1^{+}$ is an $s$-wave charge-$4e$ order ($4e$ in panel~(a)).
In the legend of panel (b), only one of many symmetry-related superpositions is provided, with $\omega \equiv \Elr^{\iu 2 \pi / 3}$.} \label{fig:3D-SC-phase-diag}
\end{figure*}

The phase diagram of Fig.~\ref{fig:2D-SC-phase-diag} assumes $\kappa = 0$, that is, complete isotropy on the quadratic level.
Allowing for cubic warping slightly changes the phase boundaries, but not the structure of the phase diagram.
Whether the transition into the vestigial phase is first-order or second-order depends on the ratio of the leading symmetrized coupling constant to $(\mathscr{S} \circ \tilde{g})_{0}$ precisely as shown in Fig.~\ref{fig:vestigial-transitions}.
This follows from the fact that the full nonlinear saddle-point equations (with $\kappa = 0$) have the same form as those studied in Refs.~\cite{Fernandes2012, How2023, How2024}.

\subsubsection{Three-component superconductivity} \label{sec:3D-SC}
The most general symmetry-allowed action for a $(p_x, p_y, p_z)$-wave SC whose order parameter $\vb{\eta} = (\eta_1, \eta_2, \eta_3)$ transforms according to the $\Gamma_4^{-}$ irrep is
\begin{align}
\actS &= \int_q \vb{\eta}^{\dag} \chi_q^{-1} \vb{\eta} + \frac{1}{2} \int_x \bigg[g_{\Gamma_3} \sum_{\mu = 1, 2} (\vb{\eta}^{\dag} \GellMann_{\mu} \vb{\eta})^2 \\
&\phantom{=} + g_{\Gamma_5} \sum_{\mu = 3, 4, 5} (\vb{\eta}^{\dag} \GellMann_{\mu} \vb{\eta})^2 + g_{\Gamma_4} \sum_{\mu = 6, 7, 8} (\vb{\eta}^{\dag} \GellMann_{\mu} \vb{\eta})^2\bigg], \notag
\end{align}
where $\GellMann_{\mu}$ are the Gell-Mann matrices defined in App.~\ref{sec:GM-mat} and the susceptibility is given by
\begin{align}
\begin{aligned}
\chi_q^{-1} &= (r + q^2) \one \\
&\phantom{=} + \kappa_{\Gamma_3} \mleft[(q_x^2 + q_y^2 - 2 q_z^2) \GellMann_1 + \sqrt{3} (q_x^2 - q_y^2) \GellMann_2\mright] \\
&\phantom{=} + \kappa_{\Gamma_5} \mleft[q_y q_z \GellMann_3 + q_z q_x \GellMann_4 + q_x q_y \GellMann_5\mright].
\end{aligned}
\end{align}
Here we once again eliminated $(\vb{\eta}^{\dag} \vb{\eta})^2$ from the action by exploiting the Fierz identity $(\vb{\eta}^{\dag} \vb{\eta})^2 = \frac{1}{2} \sum_{\mu = 1}^8 (\vb{\eta}^{\dag} \GellMann_{\mu} \vb{\eta})^2$.
For fixed $\vb{\eta}^{\dag} \vb{\eta} = 1$, the extrema of the quartic terms correspond to the representative OP configurations $\vb{\eta} = (1, 0, 0)$, $(1, 1, 1) / \sqrt{3}$, $(1, \iu, 0) / \sqrt{2}$, and $(1, \omega, \omega^2) / \sqrt{3}$ with energies $g_{\Gamma_3}$, $g_{\Gamma_5}$, $(g_{\Gamma_3} + 3 g_{\Gamma_4}) / 4$, and $(3 g_{\Gamma_4} + g_{\Gamma_5}) / 4$, respectively.
Here, $\omega \equiv \Elr^{\iu 2 \pi / 3}$ and for each representative OP configuration there are additional symmetry-equivalent ones.
Hence, the action is bounded from below and stable when $g_{\Gamma_3} > 0$, $g_{\Gamma_5} > 0$, and $g_{\Gamma_4} > - \frac{1}{3} \min\{g_{\Gamma_3}, g_{\Gamma_5}\}$.

There are six possible (nontrivial) vestigial channels for three-component $\Gamma_4^{-}$ SC.
These same channels arise for the other three-component pairings that are allowed on the cubic lattice, corresponding to the irreps $\Gamma_4^{+}$ and $\Gamma_5^{\pm}$.
The particle-hole composite OPs formed out of symmetric Gell-Mann matrices $(\vb{\eta}^{\dag} \GellMann_1 \vb{\eta} | \vb{\eta}^{\dag} \GellMann_2 \vb{\eta}) \sim \Gamma_3^{+}$ and $(\vb{\eta}^{\dag} \GellMann_5 \vb{\eta} | \vb{\eta}^{\dag} \GellMann_3 \vb{\eta} | \vb{\eta}^{\dag} \GellMann_4 \vb{\eta}) \sim \Gamma_5^{+}$ describe $d$-wave nematic vestigial orders.
In contrast, the condensation of bilinears formed out of antisymmetric Gell-Mann matrices, $(\vb{\eta}^{\dag} \GellMann_6 \vb{\eta} | \vb{\eta}^{\dag} \GellMann_7 \vb{\eta} | \vb{\eta}^{\dag} \GellMann_8 \vb{\eta}) \sim m\Gamma_4^{+}$, is associated with ferromagnetic vestigial order.
In the particle-particle sector, $\vb{\eta}^{\intercal} \GellMann_0 \vb{\eta} \sim \Gamma_1^{+}$ corresponds to $s$-wave charge-$4e$ order, whereas $(\vb{\eta}^{\intercal} \GellMann_1 \vb{\eta} | \vb{\eta}^{\intercal} \GellMann_2 \vb{\eta}) \sim \Gamma_3^{+}$ and $(\vb{\eta}^{\intercal} \GellMann_5 \vb{\eta} | \vb{\eta}^{\intercal} \GellMann_3 \vb{\eta} | \vb{\eta}^{\intercal} \GellMann_4 \vb{\eta}) \sim \Gamma_5^{+}$ correspond to the two different sets of $d$-wave charge-$4e$ orders.

To determine the leading vestigial instability, we insert $g_{\mu\nu} = \diag(0, g_{\Gamma_3}, g_{\Gamma_3}, g_{\Gamma_5}, g_{\Gamma_5}, g_{\Gamma_5}, g_{\Gamma_4},$ $g_{\Gamma_4}, g_{\Gamma_4})$ and $\gamma_{\mu} = \GellMann_{\mu}$ into Eq.~\eqref{eq:symmetrized-U1-gmunu}.
This gives
\begin{align}
\frac{(\mathscr{S} \circ \tilde{g})_{\Gamma_3}^{\text{nem}}}{(\mathscr{S} \circ \tilde{g})_{0}} &= \frac{10 g_{\Gamma_3} - 3 g_{\Gamma_4} - 3 g_{\Gamma_5}}{18 (\mathscr{S} \circ \tilde{g})_{0}} = \tfrac{1}{4} + \lambda_1, \notag \\
\frac{(\mathscr{S} \circ \tilde{g})_{\Gamma_5}^{\text{nem}}}{(\mathscr{S} \circ \tilde{g})_{0}} &= \frac{- 2 g_{\Gamma_3} - 3 g_{\Gamma_4} + 9 g_{\Gamma_5}}{18 (\mathscr{S} \circ \tilde{g})_{0}} = \tfrac{1}{4} + \lambda_2, \notag \\
\frac{(\mathscr{S} \circ \tilde{g})_{\Gamma_4}^{\text{FM}}}{(\mathscr{S} \circ \tilde{g})_{0}} &= \frac{- 2 g_{\Gamma_3} + 9 g_{\Gamma_4} - 3 g_{\Gamma_5}}{18 (\mathscr{S} \circ \tilde{g})_{0}} = \tfrac{1}{4} - \tfrac{2}{3} \lambda_1 - \lambda_2, \notag \\
\frac{(\mathscr{S} \circ \tilde{g})_{0}^{4e}}{(\mathscr{S} \circ \tilde{g})_{0}} &= \frac{2 g_{\Gamma_3} - 3 g_{\Gamma_4} + 3 g_{\Gamma_5}}{9 (\mathscr{S} \circ \tilde{g})_{0}} = \tfrac{1}{4} + \tfrac{2}{3} \lambda_1 + \lambda_2, \notag \\
\frac{(\mathscr{S} \circ \tilde{g})_{\Gamma_3}^{4e}}{(\mathscr{S} \circ \tilde{g})_{0}} &= \frac{4 g_{\Gamma_3} + 3 g_{\Gamma_4} - 3 g_{\Gamma_5}}{18 (\mathscr{S} \circ \tilde{g})_{0}} = \tfrac{1}{4} + \tfrac{1}{6} \lambda_1 - \tfrac{1}{2} \lambda_2, \notag \\
\frac{(\mathscr{S} \circ \tilde{g})_{\Gamma_5}^{4e}}{(\mathscr{S} \circ \tilde{g})_{0}} &= \frac{- 2 g_{\Gamma_3} + 3 g_{\Gamma_4} + 3 g_{\Gamma_5}}{18 (\mathscr{S} \circ \tilde{g})_{0}} = \tfrac{1}{4} - \tfrac{1}{3} \lambda_1,
\end{align}
where $(\mathscr{S} \circ \tilde{g})_{0} = (2 g_{\Gamma_3} + 3 g_{\Gamma_4} + 3 g_{\Gamma_5}) / 9$ and the dimensionless ratios are defined as
\begin{align}
\begin{aligned}
\lambda_1 &= \frac{9 (2 g_{\Gamma_3} - g_{\Gamma_4} - g_{\Gamma_5})}{4 (2 g_{\Gamma_3} + 3 g_{\Gamma_4} + 3 g_{\Gamma_5})}, \\
\lambda_2 &= - \frac{3 (2 g_{\Gamma_3} + 3 g_{\Gamma_4} - 5 g_{\Gamma_5})}{4 (2 g_{\Gamma_3} + 3 g_{\Gamma_4} + 3 g_{\Gamma_5})}.
\end{aligned}
\end{align}
The action is stable provided that $\lambda_1 > - 3/4$, $\lambda_2 > -3/4$, $\lambda_2 < 1 - \lambda_1 / 3$, and $\lambda_2 < 3/2 - \lambda_1$.

The most negative symmetrized coupling constant determines the leading vestigial channel in the $(\lambda_1, \lambda_2)$ parameter space, as shown in Fig.~\ref{fig:3D-SC-phase-diag}(a).
For simplicity, in this phase diagram we have assumed an isotropic susceptibility $\kappa_{\Gamma_3} = \kappa_{\Gamma_5} = 0$.
An important difference when compared to the two-component SC case (Fig.~\ref{fig:2D-SC-phase-diag}(a)) is the appearance of a region (red) where charge-$4e$ order is the leading vestigial ordering channel.
Previously~\cite{Fernandes2021}, charge-$4e$ order was at best found to be degenerate with the leading particle-hole vestigial channel, as is the case for two-component SC in hexagonal and cubic systems (Sec.~\ref{sec:2D-SC}).
It is interesting to note that the region with leading charge-$4e$ vestigial order appears around the line where the two nematic SC OP configurations, namely $(1,0,0)$ and $(1,1,1)$, are degenerate, as shown in Fig.~\ref{fig:3D-SC-phase-diag}(b).
This indicates that the charge-$4e$ order arises because of the strong competition between the two symmetry-distinct nematic vestigial instabilities near this line.
This is similar to what happens for two-component SC, albeit with the notable difference that the charge-$4e$ region is extended in the current case.
The threshold effect is, once again, clearly visible in Fig.~\ref{fig:3D-SC-phase-diag}(a).
The nature of the transition into the vestigial phase depends on the ratio of the leading symmetrized coupling constant to $(\mathscr{S} \circ \tilde{g})_{0}$, qualitatively as depicted in Fig.~\ref{fig:vestigial-transitions}.

\subsection{\boldmath{$M$}-point spin-density waves in hexagonal systems} \label{sec:M-SDW}
SDWs that admit multiple wavevectors related by rotational symmetries of the lattice are ideal candidates for realizing nematic and other vestigial orders.
In tetragonal lattices, such as those present in iron-based superconductors~\cite{Fernandes2012, Fernandes2014, Borisov2019, Grinenko2021}, there are generally two wavevectors related by $90^{\circ}$ rotations.
The analysis of the corresponding SDW vestigial instabilities is thus similar to the case of $X$-point CDWs studied in Sec.~\ref{sec:X-CDW}.
In hexagonal systems, however, there are generally three wavevectors related by $60^{\circ}$ rotations.
Materials that exhibit these types of SDWs include the intercalated \ce{Fe_{1/3}NbS2}~\cite{Little2020, Li2026} and the family of van der Waals transition metal phosphorus trichalcogenides antiferromagnets~\cite{Ni2023, Hwangbo2024, Sun2024}.
Most of the previous analyses of vestigial orders in these systems focused on vestigial nematicity.
Here, our goal is to apply our modified large-$N$ formalism and investigate whether more complex vestigial orders can emerge in these systems that involve simultaneous broken symmetries in real-space and spin-space (i.e., spin-quadrupolar order).

\begin{table*}
\caption{The irreducible representations (irreps) of the primitive hexagonal space group $P6/mmm$ (\#191) at the high-symmetry points $\Gamma = (0, 0, 0)$ and $M \in \{(\tfrac{1}{2}00) \equiv (\pi/\sqrt{3}, -\pi, 0), \, (0\tfrac{1}{2}0) \equiv (2\pi/\sqrt{3}, 0, 0), \, (\tfrac{1}{2}\tfrac{1}{2}0) \cong (\pi/\sqrt{3}, \pi, 0)\}$, together with the corresponding basis functions.
The primitive basis vectors are $[100] \equiv (0,-1,0)$, $[010] \equiv (\sqrt{3}/2,1/2,0)$, and $[001] \equiv (0,0,1)$~\cite{Dresselhaus2007, BradleyCracknell2009, CDML1979}.
The two-dimensional and three-dimensional irrep basis functions transform under the physically-irreducible matrices of \texttt{ISOTROPY}~\cite{ISOTROPY}.}
{\renewcommand{\arraystretch}{1.3}
\renewcommand{\tabcolsep}{28.0pt}
\begin{tabular}{lc|lc}
\hline \hline
\multicolumn{4}{c}{hexagonal $P6/mmm$ (\#191) space group} \tabularnewline
\multicolumn{2}{c|}{even-parity irreps} & \multicolumn{2}{c}{odd-parity irreps} \tabularnewline
\hline
$\Gamma_1^{+} (A_{1g})$ & $1$, $x^2 + y^2$, $z^2$ & $\Gamma_1^{-} (A_{1u})$ & $z (x^3 - 3 x y^2) (3 x^2 y - y^3)$ \tabularnewline
$\Gamma_2^{+} (A_{2g})$ & $(x^3 - 3 x y^2) (3 x^2 y - y^3)$ & $\Gamma_2^{-} (A_{2u})$ & $z$ \tabularnewline
$\Gamma_3^{+} (B_{2g})$ & $z (x^3 - 3 x y^2)$ & $\Gamma_3^{-} (B_{2u})$ & $3 x^2 y - y^3$ \tabularnewline
$\Gamma_4^{+} (B_{1g})$ & $z (3 x^2 y - y^3)$ & $\Gamma_4^{-} (B_{1u})$ & $x^3 - 3 x y^2$ \tabularnewline
$\Gamma_5^{+} (E_{2g})$ & $(x^2 - y^2 | - 2 x y)$ & $\Gamma_5^{-} (E_{2u})$ & $\big(2 x_1 y_1 z | - z (x_1^2 - y_1^2)\big)$ \tabularnewline
$\Gamma_6^{+} (E_{1g})$ & $(y z | - x z)$ & $\Gamma_6^{-} (E_{1u})$ & $(x | y)$ \tabularnewline
\hline
$M_1^{+}$ & $(c_1 | c_2 | c_3)$, $(s_1 x_1 | s_2 x_2 | s_3 x_3)$ & $M_1^{-}$ & $\big(s_1 z y_1 | s_2 z y_2 | s_3 z y_3\big)$ \tabularnewline
$M_2^{+}$ & $\big(s_1 y_1 | s_2 y_2 | s_3 y_3\big)$ & $M_2^{-}$ & $(c_1 z | c_2 z | c_3 z)$ \tabularnewline
$M_3^{+}$ & $(s_1 z | s_2 z | s_3 z)$ & $M_3^{-}$ & $\big(c_1 y_1 | c_2 y_2 | c_3 y_3\big)$ \tabularnewline
$M_4^{+}$ & $\big(c_1 z y_1 | c_2 z y_2 | c_3 z y_3\big)$ & $M_4^{-}$ & $(s_1 | s_2 | s_3)$, $(c_1 x_1 | c_2 x_2 | c_3 x_3)$ \tabularnewline
\hline
\multicolumn{4}{c}{\footnotesize Abbreviations: $x_1 = \tfrac{1}{2} x - \tfrac{\sqrt{3}}{2} y$, $x_2 = x$, $x_3 = \tfrac{1}{2} x + \tfrac{\sqrt{3}}{2} y$, $y_1 = \tfrac{\sqrt{3}}{2} x + \tfrac{1}{2} y$,} \tabularnewline[-2pt]
\multicolumn{4}{c}{\footnotesize $y_2 = y$, $y_3 = - \tfrac{\sqrt{3}}{2} x + \tfrac{1}{2} y$, $c_n = \cos\!\big(2 \pi x_n / \sqrt{3}\big)$, $s_n = \sin\!\big(2 \pi x_n / \sqrt{3}\big)$.} \tabularnewline
\hline \hline
\end{tabular}}
\label{tab:hexagonal-irreps}
\end{table*}

We consider a simple hexagonal lattice with space group $P6/mmm$ (\#191).
The corresponding point group $6/mmm$, that is $D_{6h}$, is generated by sixfold rotations around the (principal) $z$ axis, twofold rotations around the $x$ axis, twofold rotations around the $y$ axis, and space-inversion (parity).
Its Brillouin zone is shown in Fig.~\ref{fig:Brillouin}(b).

We focus on SDWs with wavevectors at the three $M$-points $\vb{k}_{M_1} = (\pi/\sqrt{3}, -\pi, 0)$, $\vb{k}_{M_2} = (2\pi/\sqrt{3}, 0, 0)$, and $\vb{k}_{M_3} = (\pi/\sqrt{3}, \pi, 0)$.
We also assume no spin-orbit coupling, which implies that orbital and spin degrees of freedom transform separately.
The SDW transformation rules under lattice operations are thus set by one of the $M$-point irreps, which are listed in Table~\ref{tab:hexagonal-irreps}.
On the other hand, under spin rotations, the SDWs by definition transform like a time-reversal-odd vector, i.e., they have angular momentum $\ell = 1$ in the $\SO(3)$ space of spin rotations.
Due to the commensurability relations $2 \vb{k}_{M_{1,2,3}} \cong \vb{0} \equiv \vb{k}_{\Gamma}$ and $\vb{k}_{M_1} \pm \vb{k}_{M_2} \cong \vb{k}_{M_3}$, the bilinear vestigial orders transform, with respect to lattice operations, as space group irreps of the $\Gamma$ and $M$ points.
Meanwhile, with respect to spin rotation operations, they can be either time-reversal-even scalars, vectors, or quadrupoles, as follows from $(\ell=1) \otimes (\ell=1) = (\ell=0) \oplus (\ell=1) \oplus (\ell=2)$.
The nine-component OP $\vb{\eta} = (\vb{\eta}_1, \vb{\eta}_2, \vb{\eta}_3)$ of an $M$-point SDW is therefore made of three spin-vectors $\vb{\eta}_a = (\eta_{a,x}, \eta_{a,y}, \eta_{a,z})$, one for each $M$-point.
Physically, it corresponds to a modulation of the spin density of the form
\begin{align}
\var{\vb{S}(\vb{x})} &= \vb{\eta}_1 f_1(\vb{x}) + \vb{\eta}_2 f_2(\vb{x}) + \vb{\eta}_3 f_3(\vb{x}),
\end{align}
where the triplet of functions $(f_1(\vb{x}) | f_2(\vb{x}) | f_3(\vb{x}))$ transforms, under lattice operations, according to the same $M$-point irrep as $\eta_{a,i}$.
Examples of such functions are provided in Table~\ref{tab:hexagonal-irreps}.

For a SDW which transforms under lattice operations as any of the eight $M$-point irreps, the most general symmetry-allowed action has the form
\begin{align}
\begin{aligned}
\actS &= \frac{1}{2} \int_q \vb{\eta}^{\intercal} \chi_q^{-1} \vb{\eta} + \frac{1}{8} \int_x \bigg[g_{0} (\vb{\eta}^{\intercal} \GellMann_{0} \GellMann^s_{0} \vb{\eta})^2 \\
&\phantom{=} + g_{\Gamma_5} \sum_{\mu = 1, 2} (\vb{\eta}^{\intercal} \GellMann_{\mu} \GellMann^s_{0} \vb{\eta})^2 + g_{M_1} \sum_{\mu = 3, 4, 5} (\vb{\eta}^{\intercal} \GellMann_{\mu} \GellMann^s_{0} \vb{\eta})^2\bigg].
\end{aligned}
\end{align}
The Gell-Mann matrices $\GellMann_{\mu}$ are defined in Appendix~\ref{sec:GM-mat} and when they have $s$ superscripts, this means that they act in spin space only, i.e., $\GellMann_{\mu} \GellMann^s_{\nu} \equiv \GellMann_{\mu} \otimes \GellMann^s_{\nu}$.
The SDW susceptibility is given by
\begin{align}
\chi_q^{-1} &= (r + q^2) \GellMann_{0} \GellMann^s_{0} + \kappa [(q_x^2 - q_y^2) \tilde{\GellMann}_{1} - 2 q_x q_y \tilde{\GellMann}_{2}] \GellMann^s_{0},
\end{align}
where the rotated Gell-Mann matrices $\tilde{\GellMann}_{\mu}$ are also defined in App.~\ref{sec:GM-mat}.
Cubic terms, such as $\vb{\eta}_1 \vdot (\vb{\eta}_2 \vcross \vb{\eta}_3)$, are precluded by time-reversal symmetry.
For a fixed order parameter magnitude $\vb{\eta}^{\intercal} \vb{\eta}$, the quartic terms are extremized by the following SDW OP configurations: $\vb{\eta} = (\vu{e}_z, \vb{0}, \vb{0})$, corresponding to a collinear single-$\mathbf{Q}$ phase with energy $(g_0 + 2 g_{\Gamma_5}) / 8$; $(\vu{e}_z, \vu{e}_z, \vu{e}_z) / \sqrt{3}$, corresponding to a collinear triple-$\mathbf{Q}$ phase with energy $(g_0 + 2 g_{M_1}) / 8$; and $(\vu{e}_x, \vu{e}_y, \vu{e}_z) / \sqrt{3}$, corresponding to a noncoplanar triple-$\mathbf{Q}$ phase with energy $g_0 / 8$.
The stability conditions for the action are thus $g_0 > 0$, $g_{\Gamma_5} > - g_0 / 2$, and $g_{M_1} > - g_0 / 2$.

\begin{figure*}
\includegraphics[width=\textwidth]{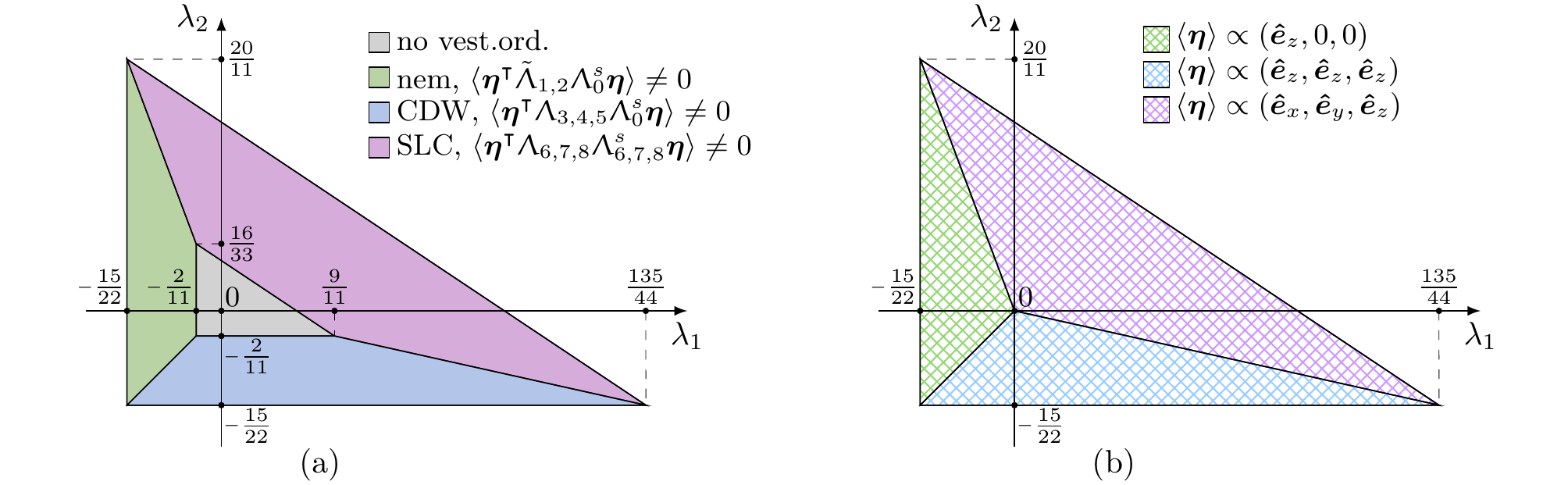}
\caption{Phase diagrams for $M$-point SDW order on the hexagonal lattice showing (a) the leading vestigial instability and (b) the SDW OP configuration in the primary ordered phase, both as a function of the dimensionless ratios $\lambda_1 = - 45 (g_{M_1} - 3 g_{\Gamma_5}) / [11 (11 g_0 + 6 g_{M_1} + 4 g_{\Gamma_5})]$ and $\lambda_2 = 30 (4 g_{M_1} - g_{\Gamma_5}) / [11 (11 g_0 + 6 g_{M_1} + 4 g_{\Gamma_5})]$.
The action is not bounded from below in the region outside the colored triangles.
Physically, $\vb{\eta}^{\intercal} \tilde{\GellMann}_{1,2} \GellMann^s_{0} \vb{\eta} \sim \Gamma_5^{+} (\ell = 0)$ is a $d$-wave nematic (nem in panel~(a)), $\vb{\eta}^{\intercal} \GellMann_{3,4,5} \GellMann^s_{0} \vb{\eta} \sim M_1^{+} (\ell = 0)$ is an $M$-point CDW (CDW in panel~(a)), and $\vb{\eta}^{\intercal} \GellMann_{6,7,8} \GellMann^s_{6,7,8} \vb{\eta} \sim M_2^{+} (\ell = 1)$ is an $M$-point spin loop-current order (SLC in panel~(a)).
In the legend of panel (b), only one of many symmetry-related superpositions is given for each option.} \label{fig:M-SDW-phase-diag}
\end{figure*}

Using group theory, we find six distinct vestigial channels.
Two of them transform as spin-scalars and correspond to the composites $(\vb{\eta}^{\intercal} \tilde{\GellMann}_{1} \GellMann^s_{0} \vb{\eta} | \vb{\eta}^{\intercal} \tilde{\GellMann}_{2} \GellMann^s_{0} \vb{\eta}) \sim \Gamma_5^{+} (\ell = 0)$ and $(\vb{\eta}^{\intercal} {\GellMann}_{3} \GellMann^s_{0} \vb{\eta} | s_{\eta} \vb{\eta}^{\intercal} {\GellMann}_{4} \GellMann^s_{0} \vb{\eta} | \vb{\eta}^{\intercal} {\GellMann}_{5} \GellMann^s_{0} \vb{\eta}) \sim M_1^{+} (\ell = 0)$.
The former is a $d$-wave nematic, while the latter is an $M$-point CDW.
Hereafter, we use the notation $s_{\eta} = +$ when $\vb{\eta} \sim mM_{1,2}^{\pm} (\ell = 1)$ and $s_{\eta} = -$ when $\vb{\eta} \sim mM_{3,4}^{\pm} (\ell = 1)$.
Recall that $\pm$ superscripts denote parity, while the $m$ prefix indicates that the irrep is odd under time-reversal.
Only one of the vestigial phases has a composite order parameter that transforms as a vector in spin space: $(\vb{\eta}^{\intercal} {\GellMann}_{6} \GellMann^s_{\mu} \vb{\eta} | s_{\eta} \vb{\eta}^{\intercal} {\GellMann}_{7} \GellMann^s_{\mu} \vb{\eta} | \vb{\eta}^{\intercal} {\GellMann}_{8} \GellMann^s_{\mu} \vb{\eta}) \sim M_2^{+} (\ell = 1)$ that is constructed from the spin-antisymmetric $\mu \in \{6, 7, 8\}$ Gell-Mann matrices.
This order corresponds to an $M$-point spin loop-current order, which is even with respect to time-reversal symmetry.

Finally, three of the six nontrivial vestigial composites correspond to spin-quadrupolar orders, namely: $\vb{\eta}^{\intercal} \GellMann_0 \GellMann^s_{\mu} \vb{\eta} \sim \Gamma_1^{+} (\ell = 2)$, associated with pure quadrupolar order in spin space; $(\vb{\eta}^{\intercal} \tilde{\GellMann}_{1} \GellMann^s_{\mu} \vb{\eta} | \vb{\eta}^{\intercal} \tilde{\GellMann}_{2} \GellMann^s_{\mu} \vb{\eta}) \sim \Gamma_5^{+} (\ell = 2)$, corresponding to spin-quadrupolar order accompanied by real-space $d$-wave nematic order; and $(\vb{\eta}^{\intercal} {\GellMann}_{3} \GellMann^s_{\mu} \vb{\eta} | s_{\eta} \vb{\eta}^{\intercal} {\GellMann}_{4} \GellMann^s_{\mu} \vb{\eta} | \vb{\eta}^{\intercal} {\GellMann}_{5} \GellMann^s_{\mu} \vb{\eta}) \sim M_1^{+} (\ell = 2)$, corresponding to a spin quadrupolar $M$-point density wave.
In all these case, $\mu$ spans the values $\{1, 2, 3, 4, 5\}$, corresponding to spin-symmetric Gell-Mann matrices.
Physically, they can be interpreted as generalizations of the widely studied spin-nematic orders~\cite{Andreev1984, Chandra1990, Chubukov1990, Chandra1991, Chubukov1991}, corresponding to the spontaneous breaking of rotational symmetry in spin space, and here described by the composite $\vb{\eta}^{\intercal} \GellMann_0 \GellMann^s_{\mu} \vb{\eta}$.

The extension of the modified large-$N$ formalism to treat vestigial orders originating from primary SDW OPs is described in Appendix~\ref{sec:spin-like-OPs}.
Here, we use the results derived in that Appendix.
The $\gamma_{\mu}$ matrices of Sec.~\ref{sec:spin-like-OPs} are the Gell-Mann matrices $\GellMann_{\mu}$ and $g_{\mu\nu} = \diag(g_0, g_{\Gamma_5}, g_{\Gamma_5}, g_{M_1}, g_{M_1}, g_{M_1})$.
Thus, we obtain via Eq.~\eqref{eq:symmetrized-spin-gmunu} the symmetrized coupling constants:
\begin{align}
\begin{aligned}
\frac{(\mathscr{S} \circ \tilde{g})_{\Gamma_5}^{\ell=0}}{(\mathscr{S} \circ \tilde{g})_{0}} &= \frac{2 g_0 - 3 g_{M_1} + 13 g_{\Gamma_5}}{27 (\mathscr{S} \circ \tilde{g})_{0}} = \tfrac{2}{11} + \lambda_1, \\
\frac{(\mathscr{S} \circ \tilde{g})_{M_1}^{\ell=0}}{(\mathscr{S} \circ \tilde{g})_{0}} &= \frac{2 g_0 + 12 g_{M_1} - 2 g_{\Gamma_5}}{27 (\mathscr{S} \circ \tilde{g})_{0}} = \tfrac{2}{11} + \lambda_2, \\
\frac{(\mathscr{S} \circ \tilde{g})_{\Gamma_1}^{\ell=2}}{(\mathscr{S} \circ \tilde{g})_{0}} &= \frac{2 g_0 + 6 g_{M_1} + 4 g_{\Gamma_5}}{27 (\mathscr{S} \circ \tilde{g})_{0}} = \tfrac{2}{11} + \tfrac{2}{5} \lambda_1 + \tfrac{3}{5} \lambda_2, \\
\frac{(\mathscr{S} \circ \tilde{g})_{\Gamma_5}^{\ell=2}}{(\mathscr{S} \circ \tilde{g})_{0}} &= \frac{2 g_0 - 3 g_{M_1} + 4 g_{\Gamma_5}}{27 (\mathscr{S} \circ \tilde{g})_{0}} = \tfrac{2}{11} + \tfrac{1}{5} \lambda_1 - \tfrac{3}{10} \lambda_2, \\
\frac{(\mathscr{S} \circ \tilde{g})_{M_1}^{\ell=2}}{(\mathscr{S} \circ \tilde{g})_{0}} &= \frac{2 g_0 + 3 g_{M_1} - 2 g_{\Gamma_5}}{27 (\mathscr{S} \circ \tilde{g})_{0}} = \tfrac{2}{11} - \tfrac{1}{5} \lambda_1 + \tfrac{1}{10} \lambda_2, \\
\frac{(\mathscr{S} \circ \tilde{g})_{M_2}^{\ell=1}}{(\mathscr{S} \circ \tilde{g})_{0}} &= \frac{2 g_0 - 3 g_{M_1} - 2 g_{\Gamma_5}}{27 (\mathscr{S} \circ \tilde{g})_{0}} = \tfrac{2}{11} - \tfrac{1}{3} \lambda_1 - \tfrac{1}{2} \lambda_2,
\end{aligned}
\end{align}
where $(\mathscr{S} \circ \tilde{g})_{0} = (11 g_0 + 6 g_{M_1} + 4 g_{\Gamma_5}) / 27$ and
\begin{align}
\begin{aligned}
\lambda_1 &= - \frac{45 (g_{M_1} - 3 g_{\Gamma_5})}{11 (11 g_0 + 6 g_{M_1} + 4 g_{\Gamma_5})}, \\
\lambda_2 &= \frac{30 (4 g_{M_1} - g_{\Gamma_5})}{11 (11 g_0 + 6 g_{M_1} + 4 g_{\Gamma_5})}
\end{aligned}
\end{align}
are dimensionless ratios.
In terms of these ratios, the region where the action is stable corresponds to $\lambda_1 > - 15/22$, $\lambda_2 > - 15/22$, and $\lambda_2 < (45 - 22 \lambda_1) / 33$.

As in the previous examples, we consider an isotropic ($\kappa = 0$) primary OP susceptibility and then determine the leading vestigial instability in the $(\lambda_1, \lambda_2)$ parameter space by finding the most negative symmetrized coupling constant; see Eq.~\eqref{eq:isotropic-sus-limit}.
The result is the phase diagram shown in Fig.~\ref{fig:M-SDW-phase-diag}(a).
Similarly, the mean-field phase diagram of the primary order of Fig.~\ref{fig:M-SDW-phase-diag}(b) is determined by finding the direction of the primary OP $\eta_{a,i}$ along which the quartic term acquires its smallest value.
For arbitrary $\lambda_1$ and $\lambda_2$, we find that only nematic, CDW, and spin loop-current vestigial instabilities can be the leading ones.
Hence the symmetry that differentiates between the vestigial and primary phase can either be translations, spin rotations, or time-reversal, respectively.
The more exotic spin-quadrupolar orders do not arise as leading vestigial instabilities for general $\lambda_1$ and $\lambda_2$.
However, it turns out that on each line in Fig.~\ref{fig:M-SDW-phase-diag}(a) where two leading vestigial channels become degenerate, one of the three spin quadrupolar $\ell = 2$ vestigial channels also becomes degenerate with them.
Thus the $\ell = 2$ vestigial channels can at best become degenerate with the leading vestigial channels along fine-tuned lines in the $(\lambda_1, \lambda_2)$ parameter space.
This is reminiscent of the behavior found for charge-$4e$ order in two-component superconductors in Sec.~\ref{sec:2D-SC}.
As in the previous two examples, there is no stable vestigial order near the degeneracy point $\lambda_1 = \lambda_2 = 0$ due to the interference between the strongly competing channels.

\section{Conclusions}
In this work, we proposed a modified large-$N$ approach that treats all vestigial orders arising from the same primary order on an equal footing.
While group theory allows one to readily obtain all possible bilinear (or higher-order) composite OPs associated with a multi-component primary OP, as well as relate them to the symmetry-allowed primary OP configurations that minimize the action, only a microscopic calculation can establish which, if any, vestigial phase emerges in the space spanned by the parameters of the low-energy action.
Such calculations must necessarily go beyond the mean-field approximation, as vestigial phases are intrinsically fluctuation-driven ordered states.
The standard large-$N$ approach, in which a composite Hubbard-Stratonovich field is introduced only for the single vestigial channel that one wishes to investigate, gives a saddle-point equation for the corresponding composite OP that provides crucial insight into the behavior of the vestigial phase.
However, this approach becomes fundamentally problematic once composite Hubbard-Stratonovich fields are introduced to describe more than one vestigial phase at the same time, which is unavoidable in the systems of interest.

Here, we showed that the origin of this issue with the standard large-$N$ approach lies in the overcomplete nature of the composite fields.
This property gives rise to a rich composite redundancy structure, of which the well-known Fierz identities are just one example.
Indeed, other types of identities also follow from the overcomplete nature of the composite fields, as we demonstrated in this paper.
Physically, these equations that relate different composite fields imply that different vestigial channels necessarily interfere with each other and therefore cannot be treated separately.
Since multi-component primary OPs generically support more than one vestigial composite OP, an alternative large-$N$ approach is required to consistently treat all vestigial channels on an equal footing.

The main result of our paper is the derivation of a modified large-$N$ approach that respects the composite redundancy structure.
Within it, the symmetry group $G$ is extended to $G \times \SO(N)$, effectively adding an internal flavor to the primary OP, without altering the structure of the original symmetry group $G$ or or the irreducible representation of the primary OP.
We argue that these features are crucial, since both the composite redundancy relations and the vestigial channels are specific to a given symmetry group and irrep.
The resulting saddle-point equations, which become exact as $N \to \infty$, are structurally identical to the saddle-point equations of the traditional large-$N$ approach, with the key difference being that the bare vestigial coupling constants (i.e., the coefficients of the quartic terms in the action) are replaced by the unique symmetrized coupling constants.
Physically, these symmetrized coupling constants correspond to averaging over Hartree and Fock contractions and they are invariant under reparameterizations of the quartic coefficients that are possible due to the Fierz and other identities arising from the composite redundancy structure.
Thus the results derived in this paper provide a prescription on how to obtain the unbiased effective interactions corresponding to different vestigial channels, which in turn allows one to consistently compare the different vestigial instabilities and determine the leading one.

Of course, it is an open question the extent to which the large-$N$ results capture the behavior of the original system, which in our case always corresponds to $N = 1$.
To shed light on this issue, we compared the results of the modified large-$N$ approach with those obtained from a weak-coupling self-consistent Hartree-Fock treatment of vestigial order and with those that follow from a Gaussian variational approach.
While the agreement between these methods provides further support for the validity of the modified large-$N$ results, it would be interesting to compare them with unbiased numerical methods, such as Monte Carlo.
Another interesting direction is to investigate $1/N$ corrections, since this is the lowest order at which we expect to see qualitative differences between our proposed unbiased large-$N$ treatment and the self-consistent Hartree-Fock treatment.

An immediate consequence of the modified large-$N$ results is that a given vestigial order can only appear when the corresponding bare coupling constant overcomes a threshold value.
Thus, in contrast to the standard large-$N$ approach, there are wide regions in the parameter space spanned by the action's coefficients where no vestigial order is stable.
We showed that these regions are located around the point where all vestigial orders are degenerate.
This threshold behavior reflects a strong competition between symmetry-distinct vestigial orders, a competition that is promoted by the coupling between composite order parameters enabled through Fierz-like identities.

This central result was illustrated by applying the modified large-$N$ approach to three different types of primary order that are known to support vestigial order: two-component CDW on the tetragonal lattice, two- and three-component SC on the cubic lattice, and three-component SDW on the hexagonal lattice.
In the last two cases, the symmetry group consists of a direct product between the crystalline space group and an internal continuous group, namely, global-phase $\Ugp(1)$ for SC and spin-rotational $\SO(3)$ for SDW.
Our analyses also showed that the competition between almost degenerate vestigial instabilities sometimes enables composite orders that transform nontrivially with respect to the internal groups (i.e., charge-$4e$ order in the case of $\Ugp(1)$ symmetry and spin-current or spin-quadrupolar order in the case of $\SO(3)$ symmetry) to become the leading vestigial instabilities over certain regions of the parameter space.
Specifically, charge-$4e$ order is stabilized in the case of three-component SC on the cubic lattice, whereas spin-current order is stabilized in the case of $M$-point SDW on the hexagonal lattice.
In addition, we found that the more exotic spin-quadrupolar composite order, of which spin-nematics are one example, cannot be a leading vestigial instability, but it is degenerate with other vestigial phases along fine-tuned lines of the parameter space.
Finally, we found that both two- and three-component SC orders on the cubic lattice support a magnetic vestigial phase.

Besides shedding light on the leading vestigial instabilities of a given primary order, and on the absence of any stable vestigial orders near points of degeneracy, the modified large-$N$ approach can also be employed in future investigations to address other interesting questions regarding vestigial phases.
For instance, it would be interesting to extend the linearized saddle-point equations to the full nonlinear regime in order to assess under which conditions a subleading vestigial instability could occur below the leading vestigial order transition, but above the primary order transition.
Previous investigations of this possibility were carried out in a limited setting through a variational approach, but suggested that interesting behaviors are possible~\cite{Hecker2024_local}.
In a similar vein, it would be interesting to develop a theory of collective vestigial modes inside the primary state to elucidate whether non-condensed vestigial orders with attractive interactions could emerge as collective modes with finite excitation energies. \\

\textit{Note added.} Recently, a Monte Carlo study of vestigial phases of cubic superconductors appeared~\cite{Gao2026} whose findings are consistent with our results from Sec.~\ref{sec:cubic-SC}. Note that the limit of isotropic susceptibility that we focused on corresponds to the regime where the relative and absolute phase stiffnesses of Ref.~\cite{Gao2026} are equal.

\begin{acknowledgments}
We thank M.\ Hecker, R.\ Willa, J.\ Schmalian, A.\ R.\ Chakraborty, M.\ C.\ O'Brien, E.\ Fradkin, and A.\ V.\ Chubukov for useful discussions.
During revision of this work, R.M.F. was supported by the U.S. Department of Energy, Office of Science, Office of Basic Energy Sciences, under Award Number DE-SC0026471.

Disclaimer: This report was prepared as an account of work sponsored by an agency of the United States Government.
Neither the United States Government nor any agency thereof, nor any of their employees, makes any warranty, express or implied, or assumes any legal liability or responsibility for the accuracy, completeness, or usefulness of any information, apparatus, product, or process disclosed, or represents that its use would not infringe privately owned rights.
Reference herein to any specific commercial product, process, or service by trade name, trademark, manufacturer, or otherwise does not necessarily constitute or imply its endorsement, recommendation, or favoring by the United States Government or any agency thereof.
The views and opinions of authors expressed herein do not necessarily state or reflect those of the United States Government or any agency thereof.
\end{acknowledgments}

\appendix

\section{The composite redundancy structure} \label{sec:comp-redundancy}
In Sec.~\ref{sec:Fierz} we derived the quartic composite redundancy relations~\eqref{eq:quartic-redundancy}.
Here we discuss the higher-order redundancy relations, classify them, and give examples.

To count the number of distinct quartic redundancy relations, note that $\sum_{\mu\nu} R_{\mu\nu} \psi_{\mu} \psi_{\nu} = 0$ vanishes trivially even for noncomposite $\psi_{\mu}$ when $R_{\mu\nu} = - R_{\nu\mu}$ so we may assume that $R_{\mu\nu} = R_{\nu\mu}$ is symmetric.
The symmetric $R_{\mu\nu}$ can be divide into those that are $\propto (\mathscr{S} \circ R)_{\mu\nu}$ and those that are $\propto (\mathscr{P}_S \circ R)_{\mu\nu}$.
The latter give redundancy relations.
The number of $\propto (\mathscr{P}_S \circ R)_{\mu\nu}$ matrices equals the number of symmetric $R_{\mu\nu}$ minus the number of fully symmetric $R_{abcd}$.
Given that $\mu, \nu \in \{0, 1, \ldots, \frac{M(M+1)}{2} - 1\}$, the number of symmetric $R_{\mu\nu}$ is
\begin{align*}
\frac{\frac{M(M+1)}{2} \mleft(\frac{M(M+1)}{2}+1\mright)}{2} = \frac{M (M+1) \mleft[M^2 + M + 2\mright]}{2 \cdot 4},
\end{align*}
while the number of fully symmetric $R_{abcd}$ can be counted using the combinatorial ``stars and bars'' representation:
\begin{align*}
\binom{M+4-1}{4} = \frac{M(M+1)(M+2)(M+3)}{4!}.
\end{align*}
The difference between these two gives the number of nontrivial quartic composite redundancy relations:
\begin{align}
\mathcal{N}_{\eta^4\text{redun.rel.}} &= \frac{(M-1) M^2 (M+1)}{3 \cdot 4}.
\end{align}

For $M = 2$, $\mathcal{N}_{\eta^4\text{redun.rel.}} = 1$ and the only redundancy relation is the Fierz identity~\eqref{eq:intro-Fierz} from the introduction.
It corresponds to the $(\mathscr{P}_S \circ R)_{\mu\nu}$:
\begin{align}
\begin{pmatrix}
 -1 & 0 & 0\\
 0 & 1 & 0 \\
 0 & 0 & 1
\end{pmatrix}.
\end{align}

For $M \geq 3$ there are additional relations.
For instance, three-component OPs ($M = 3$) have $\mathcal{N}_{\eta^4\text{redun.rel.}} = 6$.
If we use the Gell-Mann matrices $\GellMann_{\mu}$ of App.~\ref{sec:GM-mat} for the $\Gamma_{\mu}$, then Eq.~\eqref{eq:partsymmetrized-gmunu} readily generates all six $(\mathscr{P}_S \circ R)_{\mu\nu}$:
\begin{gather}
\begin{gathered}
\begin{pmatrix}
 -2 & 0 & 0 & 0 & 0 & 0 \\
 0 & 1 & 0 & 0 & 0 & 0 \\
 0 & 0 & 1 & 0 & 0 & 0 \\
 0 & 0 & 0 & 1 & 0 & 0 \\
 0 & 0 & 0 & 0 & 1 & 0 \\
 0 & 0 & 0 & 0 & 0 & 1 \\
\end{pmatrix}, \\
\begin{pmatrix}
 0 & -1 & 0 & 0 & 0 & 0 \\
 -1 & -\sqrt{2} & 0 & 0 & 0 & 0 \\
 0 & 0 & \sqrt{2} & 0 & 0 & 0 \\
 0 & 0 & 0 & -\frac{1}{\sqrt{2}} & 0 & 0 \\
 0 & 0 & 0 & 0 & -\frac{1}{\sqrt{2}} & 0 \\
 0 & 0 & 0 & 0 & 0 & \sqrt{2} \\
\end{pmatrix}, \\
\begin{pmatrix}
 0 & 0 & -1 & 0 & 0 & 0 \\
 0 & 0 & \sqrt{2} & 0 & 0 & 0 \\
 -1 & \sqrt{2} & 0 & 0 & 0 & 0 \\
 0 & 0 & 0 & -\sqrt{\frac{3}{2}} & 0 & 0 \\
 0 & 0 & 0 & 0 & \sqrt{\frac{3}{2}} & 0 \\
 0 & 0 & 0 & 0 & 0 & 0 \\
\end{pmatrix}, \\
\begin{pmatrix}
 0 & 0 & 0 & -1 & 0 & 0 \\
 0 & 0 & 0 & -\frac{1}{\sqrt{2}} & 0 & 0 \\
 0 & 0 & 0 & -\sqrt{\frac{3}{2}} & 0 & 0 \\
 -1 & -\frac{1}{\sqrt{2}} & -\sqrt{\frac{3}{2}} & 0 & 0 & 0 \\
 0 & 0 & 0 & 0 & 0 & \sqrt{\frac{3}{2}} \\
 0 & 0 & 0 & 0 & \sqrt{\frac{3}{2}} & 0 \\
\end{pmatrix},
\end{gathered}
\end{gather}
and
\begin{gather}
\begin{gathered}
\begin{pmatrix}
 0 & 0 & 0 & 0 & -1 & 0 \\
 0 & 0 & 0 & 0 & -\frac{1}{\sqrt{2}} & 0 \\
 0 & 0 & 0 & 0 & \sqrt{\frac{3}{2}} & 0 \\
 0 & 0 & 0 & 0 & 0 & \sqrt{\frac{3}{2}} \\
 -1 & -\frac{1}{\sqrt{2}} & \sqrt{\frac{3}{2}} & 0 & 0 & 0 \\
 0 & 0 & 0 & \sqrt{\frac{3}{2}} & 0 & 0 \\
\end{pmatrix}, \\
\begin{pmatrix}
 0 & 0 & 0 & 0 & 0 & -1 \\
 0 & 0 & 0 & 0 & 0 & \sqrt{2} \\
 0 & 0 & 0 & 0 & 0 & 0 \\
 0 & 0 & 0 & 0 & \sqrt{\frac{3}{2}} & 0 \\
 0 & 0 & 0 & \sqrt{\frac{3}{2}} & 0 & 0 \\
 -1 & \sqrt{2} & 0 & 0 & 0 & 0 \\
\end{pmatrix}.
\end{gathered}
\end{gather}

Very often symmetries constrain $g_{\mu\nu} = g_{\mu} \Kd_{\mu\nu}$ to be completely diagonal.
In such cases, the $(\mathscr{P}_S \circ R)_{\mu\nu}$ of most interest are those that are likewise diagonal because only such $\mathscr{P}_S \circ g$ terms are allowed by symmetry to appear in the large-$N$ action of Eq.~\eqref{eq:real-primary-action-large-N}.
For $M$ going up to $10$ ($\mathcal{N}_{\eta^4\text{redun.rel.}} = 825$), we have verified that in a generalized Gell-Mann basis the only completely diagonal $(\mathscr{P}_S \circ R)_{\mu\nu}$ is the Fierz one $\mleft(- M \Kd_{\mu0} \Kd_{\nu0} + \Kd_{\mu\nu}\mright)$ of Eq.~\eqref{eq:Fierz-redundancy}.
This does not necessarily hold for other bases.

Redundancy relations also appear for higher-order terms in the action.
For instance, in the sixth-order term
\begin{align}
\sum_{a_i} R_{a_1 a_2 a_3 a_4 a_5 a_6} \eta_{a_1} \eta_{a_2} \eta_{a_3} \eta_{a_4} \eta_{a_5} \eta_{a_6}
\end{align}
only the fully symmetric $R_{a_1 a_2 a_3 a_4 a_5 a_6}$ survives for the same reason as before (Sec.~\ref{sec:Fierz}).
This gives the sixth-order composite redundancy relations
\begin{align}
\sum_{\mu\nu\rho} (\mathscr{P}_S \circ R)_{\mu\nu\rho} \Psi_{\mu} \Psi_{\nu} \Psi_{\rho} &= 0,
\end{align}
where $\mathscr{P}_S \circ R \defeq R - \mathscr{S} \circ R$ and
\begin{align}
\begin{aligned}
(\mathscr{P}_S \circ R)_{\mu\nu\rho} \defeq \frac{1}{w^3} \sum_{a_i} &(\mathscr{P}_S \circ R)_{a_1 a_2 a_3 a_4 a_5 a_6} \\[-4pt]
&\times \Gamma_{\mu,a_2 a_1} \Gamma_{\nu,a_4 a_3} \Gamma_{\rho,a_6 a_5}.
\end{aligned}
\end{align}
Such composite redundancy relations exist at all even orders and their number is given by the difference between the number of fully symmetric $R_{\mu_1 \cdots \mu_n}$ and fully symmetric $R_{a_1 \cdots a_{2n}}$, both of which can be counted using the combinatorial ``stars and bars'' representation:
\begin{align}
\mathcal{N}_{\eta^{2n}\text{redun.rel.}} &= \binom{\frac{M(M+1)}{2}+n-1}{n} - \binom{M+2n-1}{2n}.
\end{align}

In addition, there are mixed redundancy relations for odd orders.
For instance, cubic composite redundancy relations have the form
\begin{align}
\sum_{a\mu} R_{a\mu} \eta_a \Psi_{\mu} &= 0,
\end{align}
where $R_{a\mu} \defeq \frac{1}{w} \sum_{bc} (\mathscr{P}_S \circ R)_{abc} \Gamma_{\mu,cb}$.
Their general form is
\begin{align}
\sum_{a\mu\nu\cdots} R_{a\mu\nu\cdots} \eta_a \Psi_{\mu} \Psi_{\nu} \cdots &= 0,
\end{align}
i.e., the first index contracts with the primary $\eta_a$, while all the other ones contract with composite $\Psi_{\mu}$.
Their number is given by the number of $R_{a\mu\nu\cdots}$ which do not vanish identically when contracted with noncomposite $\psi_{\mu} \psi_{\nu} \cdots$ minus the total number of fully symmetric $R_{abc\cdots}$:
\begin{align}
\mathcal{N}_{\eta^{2n+1}\text{redun.rel.}} &= M \binom{\frac{M(M+1)}{2}+n-1}{n} - \binom{M+2n}{2n+1}.
\end{align}

In summary, we have found a hierarchy of redundancy relations that convey how the composite $\Psi_{\mu}$ are made from an underlying primary field $\eta_a$.
This hierarchy is what we dub the composite redundancy structure.

\section{Complex primary order parameters \linebreak with $\mathrm{U}(1)$ phase-rotation symmetry} \label{sec:complex-OPs}
OPs possessing $\Ugp(1)$ symmetry often arise.
For example, they describe superconductors, superfluids, and incommensurate charge-density waves.
Although a complex $M$-component OP can be treated as a real $2M$-component OP, when the OP has a $\Ugp(1)$ phase rotation symmetry, it is beneficial to employ a formalism specifically suited for complex OPs.
Such a formalism is developed in this Appendix.

Let us assume that the primary order is described by an $M$-component complex-valued OP $\vb{\eta} = (\eta_1, \ldots, \eta_M)$ that transforms under $\Ugp(1)$ as $\vb{\eta} \mapsto \Elr^{\iu \varphi} \vb{\eta}$.
The fluctuations of $\vb{\eta}$ near the transition are then described by the following general Euclidean action:
\begin{align}
\begin{aligned}
\actS[\eta] &= \int_{x_1 x_2} \sum_{ab} \eta_a^{*}(x_1) \chi_{ab}^{-1}(x_1-x_2) \eta_b(x_2) \\
&\phantom{=}\quad + \frac{1}{2} \int_x \sum_{abcd} g_{abcd} \eta_a^{*}(x) \eta_b(x) \eta_c^{*}(x) \eta_d(x) \\
&= \int_{x_1 x_2} \sum_{ab} \eta_a^{*}(x_1) \chi_{ab}^{-1}(x_1-x_2) \eta_b(x_2) \\
&\phantom{=}\quad + \frac{1}{8} \int_x \sum_{\mu\nu} \Psi_{\mu}(x) g_{\mu \nu} \Psi_{\nu}(x),
\end{aligned} \label{eq:complex-primary-action}
\end{align}
where we have expressed the quartic self-interaction as a coupling between $\Ugp(1)$-invariant composite OPs:
\begin{align}
\Psi_{\mu}(x) \defeq 2 \, \vb{\eta}^{\dag}(x) \gamma_{\mu} \vb{\eta}(x).
\end{align}
There is no loss of generality in doing so.
That said, although these $\Psi_{\mu}$ are sufficient for writing down the most general symmetry-allowed action, to cover all vestigial channels we shall soon introduce additional $\Ugp(1)$-charged composite OPs.
Here $g_{\mu\nu}^* = g_{\mu\nu} = g_{\nu\mu}$ and the factor of $2$ has been inserted for later convenience, as well as to ensure that (cf.\ Eq.~\eqref{eq:gabcd-gmunu-rel})
\begin{align}
g_{abcd} = \sum_{\mu \nu} g_{\mu \nu} \gamma_{\mu, ab} \gamma_{\nu, cd}.
\end{align}
Here, $\gamma_{\mu}$ are Hermitian $M \times M$ matrices, $\gamma_{\mu}^{\dag} = \gamma_{\mu}$.
We order the $M^2$ Hermitian $\gamma_{\mu}$ matrices so that the first $\frac{M(M+1)}{2}$ matrices are symmetric, while the last $\frac{M(M-1)}{2}$ ones are antisymmetric.
In other words,
\begin{gather}
\begin{gathered}
\gamma_{\mu}^{\intercal} = (\pm)_{\mu} \gamma_{\mu}, \\
(\pm)_{\mu} \defeq \begin{cases}
+ 1, & \text{for $0 \leq \mu \leq \frac{M(M+1)}{2} - 1$,} \\
- 1, & \text{for $\frac{M(M+1)}{2} \leq \mu \leq M^2 - 1$.}
\end{cases}
\end{gathered}
\end{gather}
The $\gamma_{\mu}$ are also taken to be orthonormal:
\begin{gather}
\tr \gamma_{\mu} \gamma_{\nu} = w \cdot \Kd_{\mu \nu}, \qquad
\gamma_{0,ab} = \sqrt{\frac{w}{M}} \cdot \Kd_{ab},
\end{gather}
with $a, b, c, d \in \{1, \ldots, M\}$, $\mu, \nu, \rho, \sigma \in \{0, \ldots, M^2-1\}$.

To treat the possible breaking of the $\Ugp(1)$ symmetry, let us introduce the following Nambu notation
\begin{align}
\tilde{\eta}_{a, +} &\equiv \eta_{a}, &
\tilde{\eta}_{a, -} &\equiv \eta_{a}^{*}
\end{align}
so that we can write
\begin{align}
\tilde{\eta}_{\tilde{a}} &= \tilde{\eta}_{a, s_a}
\end{align}
with $\tilde{a} \equiv (a, s_a)$ and $s \in \{+, -\}$.
We put tildes on all objects associated with the Nambu-space extension, i.e., which are in the Nambu-redundant basis.
We now introduce the bilinears
\begin{align}
\begin{aligned}
\tilde{\Psi}_{\mu, f}(x) &\defeq \sum_{ab} \sum_{s_1 s_2} \tilde{\eta}_{a, s_1}^{*}(x) \gamma_{\mu, ab} \Pauli_{f, s_1 s_2} \tilde{\eta}_{b, s_2}(x) \\
&= \sum_{\tilde{a} \tilde{b}} \tilde{\eta}_{\tilde{a}}^{*}(x) \mleft[\gamma_{\mu} \otimes \Pauli_{f}\mright]_{\tilde{a} \tilde{b}} \tilde{\eta}_{\tilde{b}}(x),
\end{aligned}
\end{align}
where $\Pauli_{f}$ are Pauli matrices in Nambu space and $f \in \{0, 1 \equiv x, 2 \equiv y, 3 \equiv z\}$.
We abbreviate
\begin{align}
\tilde{\Psi}_{\tilde{\mu}} &= \tilde{\Psi}_{\mu, f_{\mu}}
\end{align}
with $\tilde{\mu} \equiv (\mu, f_{\mu})$.
Because $\tilde{\eta}_{\tilde{a}}^{*} = [\Pauli_x \tilde{\eta}]_{\tilde{a}}$, it follows that the only allowed combinations $\tilde{\mu} = (\mu, f_{\mu})$ correspond to those that satisfy $\gamma_{\mu}^{\intercal} \Pauli_{f_{\mu}}^{\intercal} = \gamma_{\mu} \Pauli_x \Pauli_{f_{\mu}} \Pauli_x$.
Thus, the symmetric $\gamma_{\mu}$ matrices can only appear with $f = 0, x, y$, while the antisymmetric $\gamma_{\mu}$ matrices can only appear with $f = z$.
Physically, this is the statement that in the $2M$-component Nambu basis only overall symmetric matrices can be used to construct $\tilde{\Psi}_{\tilde{\mu}}$.

Alternatively, we could have introduced
\begin{align}
\tilde{\eta}_{a, 1}' &\equiv \sqrt{2} \Re \eta_{a}, &
\tilde{\eta}_{a, 2}' &\equiv \sqrt{2} \Im \eta_{a}
\end{align}
such that
\begin{align}
\tilde{\eta}_{\tilde{a}}' &= \tilde{\eta}_{a, i_a}'
\end{align}
with $\tilde{a} \equiv (a, i_a)$ and $i \in \{1, 2\}$.
Then, the previously defined $\tilde{\Psi}_{\tilde{\mu}}$ becomes
\begin{align}
\tilde{\Psi}_{\tilde{\mu}}(x) &= \sum_{\tilde{a} \tilde{b}} \tilde{\eta}_{\tilde{a}}'(x) \Gamma_{\tilde{\mu}, \tilde{a}\tilde{b}} \tilde{\eta}_{\tilde{b}}'(x),
\end{align}
where
\begin{align}
\Gamma_{\tilde{\mu}} \equiv \gamma_{\mu} \otimes U^{\dag} \Pauli_{f_{\mu}} U \label{eq:U1-theory-Gamma}
\end{align}
and $U = \frac{1}{\sqrt{2}} {\large \mleft(\begin{smallmatrix} 1 & \iu \\ 1 & - \iu \end{smallmatrix}\mright)}$.
As a result, $U^{\dag} (\Pauli_0, \Pauli_x, \Pauli_y, \Pauli_z) U = (\Pauli_0, \Pauli_z, -\Pauli_x, -\Pauli_y)$, respectively.

The number of independent quartic coefficients $g_{abcd}$, once they are symmetrized with respect to $a \leftrightarrow c$ and $b \leftrightarrow d$ exchange, is $(M (M+1) / 2)^2$.
Therefore, using the $g_{\mu \nu}$ components whose $\gamma_{\mu}$ are both symmetric and antisymmetric is redundant.
In the real OP formulation $\tilde{\eta}_{\tilde{a}}' \sim \sqrt{2} \, (\Re \eta_{a}, \Im \eta_{a})$, the symmetric $\Psi_{\mu} = \tilde{\Psi}_{\mu, 0}$ correspond to contractions with the Kronecker delta $\Kd_{i_1 i_2}$, while antisymmetric $\Psi_{\mu} = \tilde{\Psi}_{\mu, z}$ correspond to contractions with the Levi-Civita symbol $\LCs_{i_1 i_2}$.
Because $\Ugp(1) = \SO(2)$ does not include reflections, both tensors are individually invariant and the two sectors generally couple.
However, when there is a complex conjugation symmetry $\eta_a \mapsto \eta_a^*$, such as time-reversal, the two sectors decouple because it implies $g_{\mu \nu} = (\pm)_{\mu} (\pm)_{\nu} g_{\mu \nu}$.
In this case, because of $\LCs_{i_1 i_2} \LCs_{i_3 i_4} = \Kd_{i_1 i_3} \Kd_{i_2 i_4} - \Kd_{i_1 i_4} \Kd_{i_2 i_3}$ one can eliminate all antisymmetric components of $g_{\mu \nu}$ in favor of using only symmetric components of $g_{\mu \nu}$ without loss of generality.
We retain the antisymmetric components below to allow for systems without complex conjugation symmetry.

The quartic term can now be recast into
\begin{align}
\sum_{\mu\nu} \Psi_{\mu} g_{\mu\nu} \Psi_{\nu} = \sum_{\tilde{a} \tilde{b} \tilde{c} \tilde{d}} \tilde{g}_{\tilde{a} \tilde{b} \tilde{c} \tilde{d}} \tilde{\eta}_{\tilde{a}} \tilde{\eta}_{\tilde{b}} \tilde{\eta}_{\tilde{c}} \tilde{\eta}_{\tilde{d}},
\end{align}
with the $\Ugp(1)$ symmetry constraining the $\tilde{g}_{\tilde{a} \tilde{b} \tilde{c} \tilde{d}}$ so that only those components whose $s_a + s_b + s_c + s_d = 0$ can be finite.
Because the quartic term must be real, it also follows that  $\tilde{g}_{\tilde{a} \tilde{b} \tilde{c} \tilde{d}}^{*} = \tilde{g}_{\tilde{a}^* \tilde{b}^* \tilde{c}^* \tilde{d}^*}$ with $\tilde{a}^* \equiv (a, - s_a)$.
After symmetrization, we obtain
\begin{align}
\sum_{\mu\nu} \Psi_{\mu} g_{\mu\nu} \Psi_{\nu} = \sum_{\tilde{\mu} \tilde{\nu}} \tilde{\Psi}_{\tilde{\mu}} (\mathscr{S} \circ \tilde{g})_{\tilde{\mu} \tilde{\nu}} \tilde{\Psi}_{\tilde{\nu}}
\end{align}
where, we recall, $\mu, \nu, \rho, \sigma \in \{0, \ldots, M^2-1\}$, while $\tilde{\mu}, \tilde{\nu}, \tilde{\rho}, \tilde{\sigma}$ go over pairs $(\mu, f_{\mu})$ such that $\gamma_{\mu}^{\intercal} \Pauli_{f_{\mu}}^{\intercal} = \gamma_{\mu} \Pauli_x \Pauli_{f_{\mu}} \Pauli_x$.
The final result is
\begin{align}
(\mathscr{S} \circ \tilde{g})_{\tilde{\mu} \tilde{\nu}} &= \frac{g_{\mu \nu} + (\mathscr{F} \circ g)_{\mu \nu}}{3} (\Kd_{f_{\mu} 0} + \Kd_{f_{\mu} z}) (\Kd_{f_{\nu} 0} + \Kd_{f_{\nu} z}) \notag \\
&+ \frac{1}{3} \Re(\mathscr{L} \circ g)_{\mu \nu} (\Kd_{f_{\mu} x} \Kd_{f_{\nu} x} + \Kd_{f_{\mu} y} \Kd_{f_{\nu} y}) \label{eq:symmetrized-U1-gmunu} \\
&+ \frac{1}{3} \Im(\mathscr{L} \circ g)_{\mu \nu} (\Kd_{f_{\mu} x} \Kd_{f_{\nu} y} - \Kd_{f_{\mu} y} \Kd_{f_{\nu} x}), \notag
\end{align}
where
\begin{align}
(\mathscr{F} \circ g)_{\mu \nu} &\defeq \frac{1}{w^2} \sum_{\rho \sigma} \tr \gamma_{\mu} \gamma_{\rho} \gamma_{\nu} \gamma_{\sigma} \cdot g_{\rho \sigma}, \\
(\mathscr{L} \circ g)_{\mu \nu} &\defeq \frac{1}{w^2} \sum_{\rho \sigma} \tr \gamma_{\mu} \gamma_{\rho} \gamma_{\nu} \gamma_{\sigma}^{\intercal} \cdot g_{\rho \sigma}.
\end{align}
Note that the $w$ appearing above comes from the $\gamma_{\mu}$ normalization, $\tr \gamma_{\mu} \gamma_{\nu} = w \Kd_{\mu \nu}$.
The expressions above are obtained by inserting the $\Gamma_{\tilde{\mu}}$ of Eq.~\eqref{eq:U1-theory-Gamma} into Eq.~\eqref{eq:symmetrized-gmunu}.

The $f_{\mu} = 0, z$ components correspond to $\Ugp(1)$-invariant vestigial orders, while $f_{\mu} = x, y$ rotate into each other with a phase of $2 \varphi$ when $\eta_a \mapsto \Elr^{\iu \varphi} \eta_a$, i.e., they correspond to doubly-charged vestigial orders.
Thus, they correspond to the $m = 0$  and $m = \pm 2$ irreducible representations of $\Ugp(1)$, $m$ being the $z$ component of the angular momentum.
For any $g_{\mu\nu}^* = g_{\mu\nu} = g_{\nu\mu}$, it holds that $(\mathscr{F} \circ g)_{\mu \nu} = (\mathscr{F} \circ g)_{\nu \mu} = (\mathscr{F} \circ g)_{\nu \mu}^*$ and $(\mathscr{L} \circ g)_{\mu \nu} = (\pm)_{\mu} (\pm)_{\nu} (\mathscr{L} \circ g)_{\mu \nu} = (\mathscr{L} \circ g)_{\nu \mu}^*$ where $\gamma_{\nu}^{\intercal} = (\pm)_{\nu} \gamma_{\nu}$ and $\gamma_{\nu}^{\dag} = \gamma_{\nu}$.
When there is time-reversal symmetry, $g_{\mu \nu} = (\pm)_{\mu} (\pm)_{\nu} g_{\mu \nu}$ and so $\Im(\mathscr{L} \circ g)_{\mu \nu} = 0$.
Otherwise the $f = x, y$ degeneracy is lifted in favor of one of the time-reversal symmetry-breaking superpositions $\sim (1, \pm \iu)$.

The composite-field action is identical to the one given in Eq.~\eqref{eq:real-composite-action}, with the only difference being in the mean-field action, which is now given by
\begin{align}
\begin{aligned}
\actS_{m, \phi}[\eta] &= \int_{x_1 x_2} \sum_{ab} \eta_a^{*}(x_1) \chi_{ab}^{-1}(x_1-x_2) \eta_b(x_2) \\
&\phantom{=}\quad + \frac{1}{2} \int_x \sum_{\tilde{\mu}} \phi_{\tilde{\mu}}(x) \tilde{\Psi}_{\tilde{\mu}}(x).
\end{aligned} \label{eq:complex-meanfield-action}
\end{align}
The saddle-point equations are thus equivalent to Eq.~\eqref{eq:real-saddle-point-eq}:
\begin{align}
\begin{aligned}
\phi_{\tilde{\mu}}(x) &= \frac{1}{2} \sum_{\tilde{\nu}} (\mathscr{S} \circ \tilde{g})_{\tilde{\mu} \tilde{\nu}} \psi_{\tilde{\nu}}(x), \\
\psi_{\tilde{\mu}}(x) &= \ev{\tilde{\Psi}_{\tilde{\mu}}(x)}_{m,\phi}.
\end{aligned}
\end{align}
The large-$N$ limit is defined in an analogous manner as discussed in Sec.~\ref{sec:why-symm}, with only the symmetrized $\tilde{g}_{\tilde{\mu} \tilde{\nu}}$ appearing in the saddle-point equations.

\section{Magnetic primary order parameters \linebreak with $\mathrm{SO}(3)$ spin-rotation symmetry} \label{sec:spin-like-OPs}
Spin orders are common and their vestigial phases can be particularly rich when spin-orbit coupling is negligible and spin-rotation symmetry is present.
In this Appendix, we adapt the formalism of Part~\ref{sec:general-teo} to such systems.
An additional motivation is to study and compare the conventional large-$N$ limit, in which the spin $\SO(3)$ subgroup is extended to $\SO(N)$, with our large-$N$ limit (Sec.~\ref{sec:why-symm}), in which one adds a fictitious $\SO(N)$ symmetry along an extraneous dimension.

We consider a primary order that is described by a real-valued OP $\eta_{a, i}$ with $M L$ components, $a, b, c, d \in \{1, \ldots, M\}$, $i \in \{1, \ldots, L\}$, and that has an internal $\SO(L)$ symmetry $\eta_{a, i} \mapsto \sum_{i'} R_{ii'} \eta_{a, i'}$.
The fluctuations of $\eta_{a, i}$ are then described by the following general Euclidean action
\begin{align}
\actS[\eta] &= \frac{1}{2} \int_{x_1 x_2} \sum_{ab} \sum_{i_1 i_2} \eta_{a, i_1}(x_1) \chi_{ab}^{-1}(x_1-x_2) \Kd_{i_1 i_2} \eta_{b, i_2}(x_2) \notag \\
&\phantom{=}\quad + \frac{1}{8} \int_x \sum_{abcd} \sum_{i_1 i_2 i_3 i_4} g_{abcd} \Kd_{i_1 i_2} \Kd_{i_3 i_4} \eta_{a, i_1}(x) \notag \\[-4pt]
&\hspace{100pt} \times \eta_{b, i_2}(x) \eta_{c, i_3}(x) \eta_{d, i_4}(x) \notag \\[2pt]
&= \frac{1}{2} \int_{x_1 x_2} \sum_{ab} \sum_{i} \eta_{a, i}(x_1) \chi_{ab}^{-1}(x_1-x_2) \eta_{b, i}(x_2) \notag \\
&\phantom{=}\quad + \frac{1}{8} \int_x \sum_{\mu\nu} \Psi_{\mu}(x) g_{\mu\nu} \Psi_{\nu}(x). \label{eq:spin-primary-action}
\end{align}
The main assumption in writing this action is that $\Kd_{i_1 i_2}$ and $\Kd_{i_1 i_2} \Kd_{i_3 i_4}$ are, up to permutations, the only $\SO(L)$-invariant tensors of rank $2$ and $4$.
This is not the case for $L = 2$, where we can also have $\LCs_{i_1 i_2}$ and $\Kd_{i_1 i_2} \LCs_{i_3 i_4}$, and for $L = 4$, where we can have $\LCs_{i_1 i_2 i_3 i_4}$.
To eliminate these options, we shall assume an additional internal reflection symmetry and enlarge $\SO(2)$ and $\SO(4)$ to $\Ogp(2)$ and $\Ogp(4)$, respectively.
Note that, in the current case, we are not allowed to assume that $g_{\mu\nu}$ is symmetrized, $g_{\mu\nu} \neq (\mathscr{S} \circ g)_{\mu\nu}$, because this is not the most general possibility, as shown in  Eq.~\eqref{eq:real-primary-action-large-N}.
This is to be contrasted with the case of complex order parameters, as discussed after Eq.~\eqref{eq:U1-theory-Gamma} of the previous Appendix.

The form of the action~\eqref{eq:spin-primary-action} implies that the most general quartic self-interaction can be expressed as a coupling between $\SO(L)$-invariant composite OPs ($g_{\mu\nu}^* = g_{\mu\nu} = g_{\nu\mu}$) of the form:
\begin{align}
\Psi_{\mu}(x) \defeq \vb{\eta}^{\intercal}(x) \gamma_{\mu} \otimes \one \, \vb{\eta}(x).
\end{align}
Here $\gamma_{\mu}$ are Hermitian $M \times M$ matrices, $\gamma_{\mu}^{\dag} = \gamma_{\mu}$.
We order the $M^2$ Hermitian $\gamma_{\mu}$ so that the first $\frac{M(M+1)}{2}$ are symmetric, while the last $\frac{M(M-1)}{2}$ are antisymmetric.
They are normalized in the usual way ($w > 0$):
\begin{align}
\tr \gamma_{\mu} \gamma_{\nu} &= w \cdot \Kd_{\mu \nu}, &
\gamma_{0,ab} &= \sqrt{\frac{w}{M}} \cdot \Kd_{ab}.
\end{align}
Here, $a, b, c, d \in \{1, \ldots, M\}$, $\mu, \nu, \rho, \sigma \in \{0, \ldots, M^2-1\}$.
Clearly, $\Psi_{\mu}$ are nonzero only for the symmetric $\gamma_{\mu}$.

Next, we consider composite OPs which are nontrivial in spin space.
To do so, let us recall that there are three irreducible representations (irreps) of $\SO(L)$ associated with rank-$2$ tensors $T_{ij}$~\cite{Hamermesh1989}: the scalar irrep $(\tr T) \Kd_{ij}$, the symmetric traceless irrep $T_{ij} + T_{ji} - \frac{2}{L} (\tr T) \Kd_{ij}$, and the antisymmetric irrep $T_{ij} - T_{ji}$.
For $L = 3$, these correspond to angular momentum $\ell = 0$, $2$, and $1$, respectively.
To make the notation less cumbersome, we shall refer to the general $\SO(L)$ irreps as $\ell = 0$, $2$, and $1$ as well, but we emphasize that $\ell$ can only be associated with angular momentum for $L = 3$.
We now introduce a basis of $L \times L$ matrices $\Lambda_f$ that are Hermitian ($\Lambda_f^{\dag} = \Lambda_f$), orthonormal ($\tr \Lambda_{f_1} \Lambda_{f_2} = L \Kd_{f_1 f_2}$), and ordered so that the zeroth one is the identity, $\Lambda_0 = \one$, the next $\frac{L(L+1)}{2} - 1$ are symmetric, while the last $\frac{L(L-1)}{2}$ are antisymmetric.
For $L = 3$, the Gell-Mann matrices $\GellMann_f$ of App.~\ref{sec:GM-mat} are a possible choice of $\Lambda_f$.
Clearly, with these conventions, $\{\Lambda_0\}$ is the basis of the scalar $\ell = 0$ irrep, $\{\Lambda_1, \ldots, \Lambda_{L(L+1)/2-1}\}$ is the basis of the symmetric traceless $\ell = 2$ irrep, and $\{\Lambda_{L(L+1)/2}, \ldots, \Lambda_{L^2-1}\}$ is the basis of the antisymmetric $\ell = 1$ irrep.

The corresponding composite OP is then given by
\begin{align}
\tilde{\Psi}_{\tilde{\mu}}(x) = \tilde{\Psi}_{\mu, f_{\mu}}(x) \defeq \vb{\eta}^{\intercal}(x) \gamma_{\mu} \otimes \Lambda_{f_{\mu}} \vb{\eta}(x),
\end{align}
where $\tilde{\mu} \equiv (\mu, f_{\mu})$.
The indices $\mu$ and $f_{\mu}$ are paired up so that they are both either symmetric or antisymmetric under transposition, i.e., so that $\gamma_{\mu}^{\intercal} \Lambda_{f_{\mu}}^{\intercal} = \gamma_{\mu} \Lambda_{f_{\mu}}$.
Denoting
\begin{align}
\Gamma_{\tilde{\mu}} \equiv \gamma_{\mu} \otimes \Lambda_{f_{\mu}},
\end{align}
and applying Eq.~\eqref{eq:symmetrized-gmunu} with $w \to w L$ in $\mathscr{F}$, we find that the symmetrized $(\mathscr{S} \circ \tilde{g})_{\tilde{\mu} \tilde{\nu}}$ appearing in
\begin{align}
\sum_{\mu\nu} \Psi_{\mu} g_{\mu\nu} \Psi_{\nu} = \sum_{\tilde{\mu} \tilde{\nu}} \tilde{\Psi}_{\tilde{\mu}} (\mathscr{S} \circ \tilde{g})_{\tilde{\mu} \tilde{\nu}} \tilde{\Psi}_{\tilde{\nu}}
\end{align}
is given by
\begin{align}
(\mathscr{S} \circ \tilde{g})_{\tilde{\mu} \tilde{\nu}} &= \frac{1}{3} g_{\mu \nu} \Kd_{f_{\mu} 0} \Kd_{f_{\nu} 0} + \frac{2}{3} (\mathscr{F} \circ g)_{\mu \nu} \frac{\Kd_{f_{\mu} f_{\nu}}}{L}, \label{eq:symmetrized-spin-gmunu}
\end{align}
where
\begin{align}
(\mathscr{F} \circ g)_{\mu \nu} &\defeq \frac{1}{w^2} \sum_{\rho, \sigma=0}^{\frac{M(M+1)}{2}-1} \tr \gamma_{\mu} \gamma_{\rho} \gamma_{\nu} \gamma_{\sigma} \cdot g_{\rho \sigma}.
\end{align}
This is the main result of the current Appendix.
$(\mathscr{S} \circ \tilde{g})_{\tilde{\mu} \tilde{\nu}}$ is diagonal in $f_{\mu}, f_{\nu}$, as one would expect: there is no mixing of different $\SO(L)$ irreps.
The $(\mathscr{S} \circ \tilde{g})_{\mu 0, \nu 0} \equiv (\mathscr{S} \circ g)_{\mu, \nu}^{\ell=0}$ with symmetric $\mu, \nu$ corresponds to the scalar channel; the $(\mathscr{S} \circ \tilde{g})_{\mu 1, \nu 1} \equiv (\mathscr{S} \circ g)_{\mu, \nu}^{\ell=2}$ with symmetric $\mu, \nu$ corresponds to the traceless spin-symmetric channel; and the $(\mathscr{S} \circ \tilde{g})_{\mu \frac{L(L+1)}{2}, \nu \frac{L(L+1)}{2}} \equiv (\mathscr{S} \circ g)_{\mu, \nu}^{\ell=1}$ with antisymmetric $\mu, \nu$ corresponds to the spin-antisymmetric channel.

The composite field action is introduced in the usual way (as in Sec.~\ref{sec:SPA}), with the saddle-point equations having the same form as Eq.~\eqref{eq:real-saddle-point-eq}:
\begin{align}
\begin{aligned}
\phi_{\tilde{\mu}}(x) &= \frac{1}{2} \sum_{\tilde{\nu}} (\mathscr{S} \circ \tilde{g})_{\tilde{\mu} \tilde{\nu}} \psi_{\tilde{\nu}}(x), \\
\psi_{\tilde{\mu}}(x) &= \ev{\tilde{\Psi}_{\tilde{\mu}}(x)}_{m,\phi}.
\end{aligned}
\end{align}

Finally, let us discuss the conventional large-$L$ (``large-$N$'') limit in which the spin $\SO(3)$ subgroup is enlarged to $\SO(L)$.
To make the large-$L$ limit well-defined, one needs to rescale $g_{\mu \nu} \to (3 / L) g_{\mu \nu}$ in Eq.~\eqref{eq:spin-primary-action}.
The first issue in this approach is that the spin-nontrivial composite OPs are not even considered, as the Hubbard-Stratonovich transformations are carried out using only spin-trivial bilinears whose condensation is then studied via the saddle-point equations.
This makes the treatment blind to vestigial orders that are spin-nontrivial, such as the vestigial spin loop currents found in the example of Sec.~\ref{sec:M-SDW}.
The second issue is that enlarging the spin subgroup biases the results against spin-nontrivial channels.
As we demonstrated in Sec.~\ref{sec:Hartree-Fock}, the Hartree-Fock self-consistency equations are governed by the symmetrized coupling constants $\mathscr{S} \circ \tilde{g}$.
According to Eq.~\eqref{eq:symmetrized-spin-gmunu}, the spin-nontrivial channels of $\mathscr{S} \circ \tilde{g}$ come with an overall $L^{-1}$ factor when compared to the spin-trivial channels.
Thus taking $L \to \infty$ suppresses spin-nontrivial channels at weak coupling, and possibly beyond.
In contrast, in our large-$N$ approach of Sec.~\ref{sec:why-symm}, we leave the original group structure intact and do not bias the vestigial couplings in favor or against spin-nontrivial channels.

\section{Relationship between Gaussian variational and Hartree-Fock approaches} \label{sec:comp-variational}
Here we show that the self-consistent Hartree-Fock approximation of Sec.~\ref{sec:Hartree-Fock}~\cite{How2023, How2024} is completely equivalent to the Gaussian variational approximation of Refs.~\cite{Fischer2016, Nie2017, Hecker2023} for homogeneous solutions of the composite fields.

To formulate the variational approximation, let us start by writing down the free energy:
\begin{align}
\beta F &= - \log \partZ,
\end{align}
where $\beta \equiv 1 / (k_B T)$ and $\partZ = \int \DD{\eta} \Elr^{- \actS[\eta]}$, as usual.
Given an arbitrary reference action $\actS_{\text{ref}}[\eta]$, we may now re-express $\partZ$ as
\begin{align}
\begin{aligned}
\partZ &= \int \DD{\eta} \Elr^{- \actS_{\text{ref}}[\eta]} \Elr^{- \mleft(\actS[\eta] - \actS_{\text{ref}}[\eta]\mright)} \\
&= \partZ_{\text{ref}} \ev{\Elr^{- \mleft(\actS[\eta] - \actS_{\text{ref}}[\eta]\mright)}}_{\text{ref}},
\end{aligned}
\end{align}
where $\partZ_{\text{ref}} = \int \DD{\eta} \Elr^{- \actS_{\text{ref}}[\eta]}$.
Hence
\begin{align}
\beta F &= - \log \partZ_{\text{ref}} - \log \ev{\Elr^{- \mleft(\actS[\eta] - \actS_{\text{ref}}[\eta]\mright)}}_{\text{ref}}.
\end{align}
Within the variational approximation, we now exploit Jensen's inequality $- \log \ev{\Elr^{-x}} \leq \ev{x}$, which follows from convexity, to bound the free energy from above with
\begin{align}
\beta F &\leq \beta F_{\text{var}} \defeq - \log \partZ_{\text{ref}} + \ev{\actS[\eta] - \actS_{\text{ref}}[\eta]}_{\text{ref}}.
\end{align}
In the Gaussian variational approach~\cite{Nie2017, Fischer2016, Hecker2023}, for $\actS_{\text{ref}}[\eta]$ one uses the mean-field action $\actS_{m, \phi}[\eta]$ of Eq.~\eqref{eq:real-meanfield-action}, with the main difference from Sec.~\ref{sec:SPA} being that $\phi_{\mu}(x)$ are now treated as variational parameters.
$\beta F_{\text{var}}$ is readily evaluated to be
\begin{align}
\begin{aligned}
\beta F_{\text{var}}[\phi] &= \frac{1}{2} \Tr \log G_{\phi}^{-1} - \frac{1}{2} \Tr G_{\phi} \mleft(G_{\phi}^{-1} - \chi^{-1}\mright) \\
&\phantom{=}\quad + \ev{\actS_{\text{int}}[\eta]}_{m, \phi},
\end{aligned}
\end{align}
which is identical to the Baym-Kadanoff functional~\eqref{eq:Baym-Kadanoff} within the Hartree-Fock approximation $\Upphi_{\text{LW}} = \ev{\actS_{\text{int}}[\eta]}$.

By extremizing $\beta F_{\text{var}}$ with respect to $\phi_{\mu}(x)$, one obtains the variational extremum equation
\begin{gather}
\begin{gathered}
\Tr\mleft(G_{\phi}^{-1} - \chi^{-1}\mright) \fdv{G_{\phi}}{\phi_{\rho}(x')} = \\
= \frac{1}{2} \int_x \sum_{\mu\nu} 3 (\mathscr{S} \circ g)_{\mu\nu} \ev{\Psi_{\mu}(x)}_{m, \phi} \tr \Gamma_{\nu} \fdv{G_{\phi}(x,x)}{\phi_{\rho}(x')}.
\end{gathered}
\end{gather}
For homogeneous $\phi_{\mu}(x) = \bar{\phi}_{\mu}$, $\var{G_{\phi}(x_1, x_2)} / \var{\phi_{\rho}(x')}$ can be factored out, yielding Eqs.~\eqref{eq:Hartree-Fock-self-energy} and~\eqref{eq:Hartree-Fock-equation}.

\section{Gell-Mann matrices} \label{sec:GM-mat}
We use the following non-standard choice for the nine Gell-Mann matrices:
\begin{align}
\begin{aligned}
\GellMann_0 &= \begin{pmatrix}
1 & 0 & 0 \\
0 & 1 & 0 \\
0 & 0 & 1
\end{pmatrix}, & \GellMann_1 &= \frac{1}{\sqrt{2}} \begin{pmatrix}
1 & 0 & 0 \\
0 & 1 & 0 \\
0 & 0 & -2
\end{pmatrix}, \\
\GellMann_2 &= \sqrt{\frac{3}{2}} \begin{pmatrix}
1 & 0 & 0 \\
0 & -1 & 0 \\
0 & 0 & 0
\end{pmatrix}, & \GellMann_3 &= \sqrt{\frac{3}{2}} \begin{pmatrix}
0 & 0 & 0 \\
0 & 0 & 1 \\
0 & 1 & 0
\end{pmatrix},
\end{aligned}
\end{align}
\newpage
\noindent and
\begin{align}
\begin{aligned}
\GellMann_4 &= \sqrt{\frac{3}{2}} \begin{pmatrix}
0 & 0 & 1 \\
0 & 0 & 0 \\
1 & 0 & 0
\end{pmatrix}, & \GellMann_5 &= \sqrt{\frac{3}{2}} \begin{pmatrix}
0 & 1 & 0 \\
1 & 0 & 0 \\
0 & 0 & 0
\end{pmatrix}, \\
\GellMann_6 &= \sqrt{\frac{3}{2}} \begin{pmatrix}
0 & 0 & 0 \\
0 & 0 & -\iu \\
0 & \iu & 0
\end{pmatrix}, & \GellMann_7 &= \sqrt{\frac{3}{2}} \begin{pmatrix}
0 & 0 & \iu \\
0 & 0 & 0 \\
-\iu & 0 & 0
\end{pmatrix}, \\
\GellMann_8 &= \sqrt{\frac{3}{2}} \begin{pmatrix}
0 & -\iu & 0 \\
\iu & 0 & 0 \\
0 & 0 & 0
\end{pmatrix}.
\end{aligned}
\end{align}
They are normalized so that $\tr \GellMann_{\mu} \GellMann_{\nu} = 3 \Kd_{\mu \nu}$.

In addition, it is convenient to introduce
\begin{align}
\begin{aligned}
\tilde{\GellMann}_1 &= - \frac{1}{2} \GellMann_1 + \frac{\sqrt{3}}{2} \GellMann_2 = \frac{1}{\sqrt{2}} \begin{pmatrix}
1 & 0 & 0 \\
0 & -2 & 0 \\
0 & 0 & 1
\end{pmatrix}, \\
\tilde{\GellMann}_2 &= - \frac{\sqrt{3}}{2} \GellMann_1 - \frac{1}{2} \GellMann_2 = \sqrt{\frac{3}{2}} \begin{pmatrix}
-1 & 0 & 0 \\
0 & 0 & 0 \\
0 & 0 & 1
\end{pmatrix}
\end{aligned}
\end{align}
which are orthonormal alternatives to $\GellMann_1$ and $\GellMann_2$.

\bibliography{REFS.bib}

\end{document}